\documentclass[onecolumn,notitlepage,showkeys,superscriptaddress]{revtex4-1}

\usepackage{graphicx}
\usepackage{amsmath,amsfonts,amssymb}
\usepackage{color}

\begin{document}

\title{Learning dynamics explains human behavior in Prisoner's Dilemma on networks}

\author{Giulio Cimini}
\email{g.cimini@math.uc3m.es}
\affiliation{Grupo Interdisciplinar de Sistemas Complejos (GISC), Departamento de Matem\'{a}ticas, Universidad Carlos III de Madrid, 28911 Legan\'{e}s, Madrid, Spain}
\author{Angel S\'{a}nchez}
\affiliation{Grupo Interdisciplinar de Sistemas Complejos (GISC), Departamento de Matem\'{a}ticas, Universidad Carlos III de Madrid, 28911 Legan\'{e}s, Madrid, Spain}
\affiliation{Instituto de Biocomputaci\'{o}n y F\'{i}sica de Sistemas Complejos (BIFI), Universidad de Zaragoza, 50018 Zaragoza, Spain}

\begin{abstract}
Cooperative behavior lies at the very basis of human societies, yet its evolutionary origin remains a key unsolved puzzle. 
Whereas reciprocity or conditional cooperation is one of the most prominent mechanisms proposed to explain the emergence of cooperation in social dilemmas, 
recent experimental findings on networked Prisoner's Dilemma games suggest that conditional cooperation also depends on the previous action of the player---namely 
on the `mood' in which the player currently is. Roughly, a majority of people behaves as conditional cooperators if they cooperated in the past, 
while they ignore the context and free-ride with high probability if they did not. However, the ultimate origin of this behavior represents a conundrum itself. 
Here we aim specifically at providing an evolutionary explanation of moody conditional cooperation. To this end, we perform an extensive analysis of different evolutionary dynamics 
for players' behavioral traits---ranging from standard processes used in game theory based on payoff comparison to others that include non-economic or social factors. 
Our results show that only a dynamic built upon reinforcement learning is able to give rise to evolutionarily stable moody conditional cooperation, 
and at the end to reproduce the human behaviors observed in the experiments.
\end{abstract}
\keywords{evolutionary game theory, prisoner's dilemma, social networks, moody conditional cooperation, reinforcement learning}
\maketitle

Cooperation and defection are at the heart of every social dilemma \cite{Dawes1980}. 
While cooperative individuals contribute to the collective welfare at a personal cost, defectors choose not to. 
Due to the lower individual fitness of cooperators arising from that cost of contribution, selection pressure acts in favor of defectors, 
thus making the emergence of cooperation a difficult puzzle. 
Evolutionary game theory \cite{MaynardSmith1973} provides an appropriate theoretical framework to address the issue of cooperation among selfish and unrelated individuals. 
At the most elementary level, many social dilemmas can be formalized as two-person games where each player can either cooperate (C) or defect (D). 
The {\it Prisoner's Dilemma} game (PD) \cite{Axelrod1984} has been widely used to model a situation in which mutual cooperation leads to the best outcome in social terms, 
but defectors can benefit the most individually. In mathematical terms, this is described by a payoff matrix (entries correspond to the row player's payoffs)
$$\begin{array}{c|cc}
	& \mbox{C}	& \mbox{D}	\\
\hline
\mbox{C}	& R	& S	\\
\mbox{D}	& T	& P	
\end{array}$$
where mutual cooperation yields the reward $R$, mutual defection leads to punishment $P$, 
and the mixed choice gives the cooperator the sucker's payoff $S$ and the defector the temptation $T$. 
The essence of the dilemma is captured by $T>R>P>S$: both players prefer any outcome in which the opponent cooperates, but the best option for both is to defect. 
In particular, the temptation to cheat ($T>R$) and the fear of being cheated ($S<P$) can put cooperation at risk,
and according to the principles of Darwinian selection, cooperation extinction is inevitable \cite{Hofbauer1998}.

Despite the conclusion above, cooperation is indeed observed in biological and social systems alike \cite{MaynardSmith1995}. 
The evolutionary origin of such cooperation hence remains a key unsolved issue, 
particularly because the manner in which individuals adapt their behavior---which is usually referred to as evolutionary dynamics or strategy update---is unknown a priori. 
Traditionally, most of the theoretical studies in this field have built on update rules based on payoff comparison \cite{Hofbauer2003,Szabo2007,Roca2009a} \cite{foot0}. 
While such rules fit in the framework of biological evolution, where payoff is understood as fitness or reproductive success, they are also questionable, especially from an economic perspective, as it is often the case that individuals perceive the others' actions but not how much they benefit from them. 
Indeed, experimental observations \cite{Fischbacher2001,Grujic2010,Gracia2012b} (with some exceptions \cite{Traulsen2010}, but see also the reanalysis of those data in \cite{Grujic2013}) point out that human subjects playing PD or Public Good games do not seem to take payoffs into consideration. 
Instead, they respond to the cooperation that they observe in a reciprocal manner, being more prone to contribute  the more their partners do. 

Reciprocity \cite{Trivers1971} has been studied in 2-player games through the concept of reactive strategies \cite{Sigmund2010}, 
the most famous of which is {\it Tit-For-Tat} \cite{Axelrod1981} (given by playing what the opponent played in the previous run). 
Reactive strategies generalize this idea by considering that players choose their action with probabilities that depend on the opponent's previous action. 
A further development was to consider memory-one reactive strategies \cite{Sigmund2010}, in which the probabilities depend on the previous action of both the focal player and her opponent. 
In multiplayer games, conditional cooperation, {\it i.e.}, the dependence of the chosen strategy on the amount of cooperation received, 
had been reported in related experiments \cite{Fischbacher2001} and observed also for the spatial iterated PD \cite{Traulsen2010} (often along with a large percentage of free-riders). 
The analysis of the two largest-scale experiment to date with humans playing an iterated multiplayer PD game on a network \cite{Grujic2010,Gracia2012b} 
extended this idea by including the dependence on the focal player's previous action, giving rise to the so-called {\it moody conditional cooperation} (MCC). 

The MCC strategy can be described as follows \cite{Grujic2012}: if in the previous round the player defected, she will cooperate with probability $p_D=q$ 
(approximately independently of the observed cooperation), whereas, if she cooperated, she will cooperate again with a probability $p_C(x)=p\,x+r$ (subject to the constraint $p_C(x)\leq 1$), 
where $x$ is the fraction of cooperative neighbors in the previous round. 
There is ample evidence supporting this aggregate behavior, as it has been observed in at least five independent experiments: 
the two already quoted \cite{Grujic2010,Gracia2012b}; another one on multiplayer PD \cite{Grujic2012b}; a lab-in-the-field experiment with people attending a fair in Barcelona, 
where participants in the age range 17-87 behaved consistently according to the MCC strategy \cite{Gutierrez-Roig2013}, and finally, in \cite{Traulsen2010}, 
as revealed by a recent meta-analysis of those experimental results \cite{Grujic2013}. On the other hand, it could be argued that MCC behavior arises from learning processes 
experienced by the players. In this respect, it is true that when a number of iterations of the PD is regarded as a single 'supergame', repetitions of such supergame 
show changes in behavior \cite{DalBo2011}. This is in agreement with the observations in \cite{Grujic2010}, where two repetitions of the supergame were carried out 
with the same players (Experiments 1 and 2 in the reference), and it was found that the initial behavior was indeed different in both. However, analysis that exclude the first few rounds 
of those experiments show clear evidence for MCC behavior which, if anything, becomes even more marked in the second one. Similar analysis were carried out in all other experiments, 
precisely to check for the effects of learning, finding in all cases strong evidence in support of the MCC strategy, even in \cite{Grujic2012b}, where 100 iterations of the PD were played. 
Therefore, we are confident that the observation of MCC behavior is reproducible and correctly interpreted, and we believe it is a good framework to study the problem as we propose here.  However, from the viewpoint of ultimate origins and evolutionary stability of this kind of behavior, conditional cooperation and its moody version are a puzzle themselves. 
For instance, theoretical results based on replicator dynamics show that the coexistence of moody conditional cooperators with free-riders 
is not possible beyond very small groups \cite{Grujic2012}. Additionally, whereas the strategies reported in \cite{Grujic2010,Gracia2012b} are aggregate behaviors, 
it is not clear how individual MCC behavioral profiles $\{q,p,r\}$ evolve in time and how many evolutionarily stable profiles can exist among the players.

Here we aim precisely at addressing these issues by developing and studying a model for the evolutionary dynamics of MCC behavioral traits. 
To this end, we perform agent-based simulations of a population consisting of $N$ differently-parameterized moody conditional cooperators, 
either on a well-mixed population or placed on the nodes of a network, who play an iterated PD game with their neighbors 
(which is the same setting used in recent experiments \cite{Traulsen2010,Grujic2010,Gracia2012b}) and whose behavioral parameters $\{q,p,r\}$ are subject to a strategy update process. 
Specifically, during each round $t$ of the game each player selects which action to take (C or D) according to her MCC traits, 
then plays a PD game with their neighbors---the chosen action being the same with all of them---and  collects the resulting payoff $\pi^t$. Subsequently, 
every $\tau$ rounds players may update their MCC parameters according to a given evolutionary rule. 

The key and novel point in this study is that we explore a large set of possible update rules for the MCC parameters, whose details are given in SI Materials and Methods. 
To begin with, the first set of rules that we consider are of imitative nature, in which players simply copy the parameters from a selected counterpart. Imitation has been related 
to bounded rationality or to a lack of information that forces players to copy the strategies of others \cite{Schlag1998}. 
The rules that we consider here cover different aspects of imitation. Thus, we study the classical imitative dynamics that are based on payoff comparison: 
stochastic rules as {\it Proportional Imitation} \cite{Helbing1992} (equivalent, for a large and well-mixed population, to the replicator dynamics \cite{Hofbauer2003}), 
the {\it Fermi rule} \cite{Szabo1998} (featuring a parameter $\beta$ that controls the intensity of selection, and that can be understood as the inverse of temperature 
or noise in the update rule \cite{Blume1993,Traulsen2006}) and the {\it Death-Birth rule} (inspired on Moran dynamics \cite{Moran1962}), as well as the deterministic dynamics 
given by {\it Unconditional Imitation} (also called ``Imitate the Best'') \cite{Nowak1992}. 
In all these cases, players decide to copy one of their neighbors with a probability (that may be 1, {\it i.e.}, with certainty) that depends in a specific manner on the payoffs 
that they and their partners obtained in the previous round of the game. To widen the scope of our analysis, we also analyze 
another imitative mechanism that is not based on payoff comparison, namely the {\it Voter model} \cite{Holley1975}, 
in which players simply follow the social context without any strategic consideration \cite{Fehr2000}. Finally, in order to go beyond pure imitation, 
we also consider another two evolutionary dynamics which are innovative, meaning that they allow extinct strategies to be reintroduced in the population 
(whereas imitative dynamics cannot do that). The first one is {\it Best Response} \cite{Matsui1992,Blume1993}, a rule that has received a lot of attention in the literature, 
especially in economics, and that represents a situation in which each player has enough cognitive abilities to compute an optimum strategy given what her neighbors did in the previous round. The second one is {\it Reinforcement Learning} \cite{Bush1955,Macy2002,Izquierdo2008}, which instead embodies the condition of a player that 
uses her experience to choose or avoid certain actions based on their consequences: actions that met or exceeded aspirations in the past tend to be repeated in the future, 
whereas choices that led to unsatisfactory experiences are avoided. Note that neither of these two last rules relies on the use of information on others' payoffs. 

With the different update schemes that we have summarized above, we have an ample spectrum of update rules representing most of the alternatives that have been proposed 
to implement evolutionary dynamics. The point of considering such comprehensive set is directly related to our aim: finding how evolution, in a broad sense, 
can give rise to situations that are compatible with the ones seen in the experiments \cite{Grujic2010,Gracia2012b}, in terms of values and stationarity of the MCC parameters, 
as well as of the final level of cooperation achieved. Additionally, we study different spatial structures determining the interactions among the players: 
the simple setup of a well-mixed population (modeled by a random graph of average degree $\bar{k}=m$, rewired after each round of the game), as well as 
more complex structures---such as Barab\'{a}si-Albert scale free network \cite{Barabasi1999} (with degree distribution $P(k)\sim 2\,m^2/k^3$) 
and regular lattices with periodic boundary conditions (where each node is connected to its $k\equiv m$ nearest neighbors) as used in the available experimental results. 
In so doing, we add another goal to our research, namely to check whether evolution can also explain the observed lack of  {\it network reciprocity} \cite{Nowak2006}, 
which is another important experimental outcome \cite{Grujic2010,Gracia2012b}. Indeed, experimental results show very clearly that, 
when it comes to human behavior, the existence of an underlying network of contacts does not have any influence on the final level of cooperation. 
Therefore, any evolutionary proposal to explain the way subjects behave in the experiments must also be consistent with this additional observation.

\section*{Results and Discussion}

We have carried out an extensive simulation program on the set of update rules and underlying networks that we have introduced above. 
In what follows, we separate the discussion of the corresponding results in two main groups: imitative and non-imitative strategies. 
Additional aspects of our numerical approach are described in SI Results.

\subsection*{Imitative updates} The five topmost sets of plots of Fig.\ \ref{fig.C} show the evolution of the level of cooperation $c$ (defined as the percentage of players 
who cooperate in each round of the game), as well as the stationary probability distribution of the individual MCC parameters among the population, 
when different evolutionary dynamics are employed to update players' behavioral traits. Note that all the plots refer to the case $\tau=1$ (meaning that the update takes place after each round). 
We will show only results for this choice below, because we have observed that the value of $\tau$ basically influences only the convergence rate of the system to its stationary state, 
but not its characteristic features. As can be seen from the plots, the final level of cooperation here is, generally, highly dependent on the population structure, 
and often the final outcome is a fully defective state (especially for a well-mixed population) \cite{foot1}. 
Then, as expected from non-innovative strategies, the number of profiles $\{q,p,r\}$ that survive at the end of the evolution is always very low and, in general, 
only one profile is left for every individual realization of the system. 
Notwithstanding, the surviving profiles are very different among independent realizations (except when the final outcome is full defection, where $q\rightarrow0$ irrespectively of $p$ and $r$), 
indicating the absence of a stationary distribution for MCC parameters, {\it i.e.}, the lack of evolutionarily stable profiles.
The only case in which the parameters $q$ and $r$  tend to concentrate around some stationary non-trivial values is given by games played on lattices 
and with Unconditional Imitation updating. Finally, we note that, when the update rule is the Voter model, the surviving profile is just picked randomly among the population 
(as expected from a rule that is not meant to improve payoffs), and hence the cooperation level remains close on average to the value set by the initial distribution of MCC parameters.
A similar behavior is observed with the Fermi rule for low $\beta$, where $\beta$ is the parameter that controls the intensity of the selection. 
Whereas for high $\beta$ (low temperature) errors are unlikely to occur and players always choose the parameters 
that enhance their payoffs, resulting in full defection as final outcome, for low $\beta$ (high temperature) errors are frequent, 
so that MCC parameters basically change randomly and $c$ remains close to its initial value. 
It is also worth noting that Proportional Imitation and the Fermi Rule lead to very similar results, except for the parameter $q$, which makes sense in view that they are very similar 
unless $\beta$ is very small. The fact that both the Fermi rule and the Death-Birth update lead also to similar outcomes is probably related to those two dynamics being both error-prone, 
with specific features of either one showing, for instance, in the different results on lattices. Nonetheless, beyond all these peculiarities of each imitative dynamics, 
the main conclusion of our simulation program is that this type of update schemes is not compatible with the experimental observations. 

Note that it is not our goal to explain in detail the effects of a particular updating rule on a given population structure. 
However, it is possible to gain qualitative insights on the behavior of the system from rather naive considerations. Take for instance scale-free networks, which feature hubs 
(players with high degree) that thus get higher payoff than average players do. If the dynamics is of imitative nature, hubs' strategy is stable and tends to spread over the network: 
there is the possibility for a stable subset of cooperators to form around hubs \cite{GomezGardenes2007}. This behavior (which cannot occur in random or regular graphs, 
where the degree distribution is more homogeneous) is clearly visible when the updating rule is Proportional Imitation. Notably, the stability of the subset of cooperators is destroyed 
when mistakes are possible (as with the Fermi rule); on the other hand, it is enhanced when the updating selects preferentially individuals with high payoffs 
(as with the Death-Birth rule or Unconditional Imitation). In these two latter cases cooperation becomes sustainable also in lattices, as these structures naturally allow clusters 
of mutually connected cooperators to emerge. Instead, the independence on the network observed---as we shall see---in the case of Reinforcement Learning 
is easily explained by players not even looking at each other, which makes the actual population structure irrelevant.

\subsection*{Non-imitative updates} 
\begin{figure*}
\includegraphics[width=0.24\textwidth]{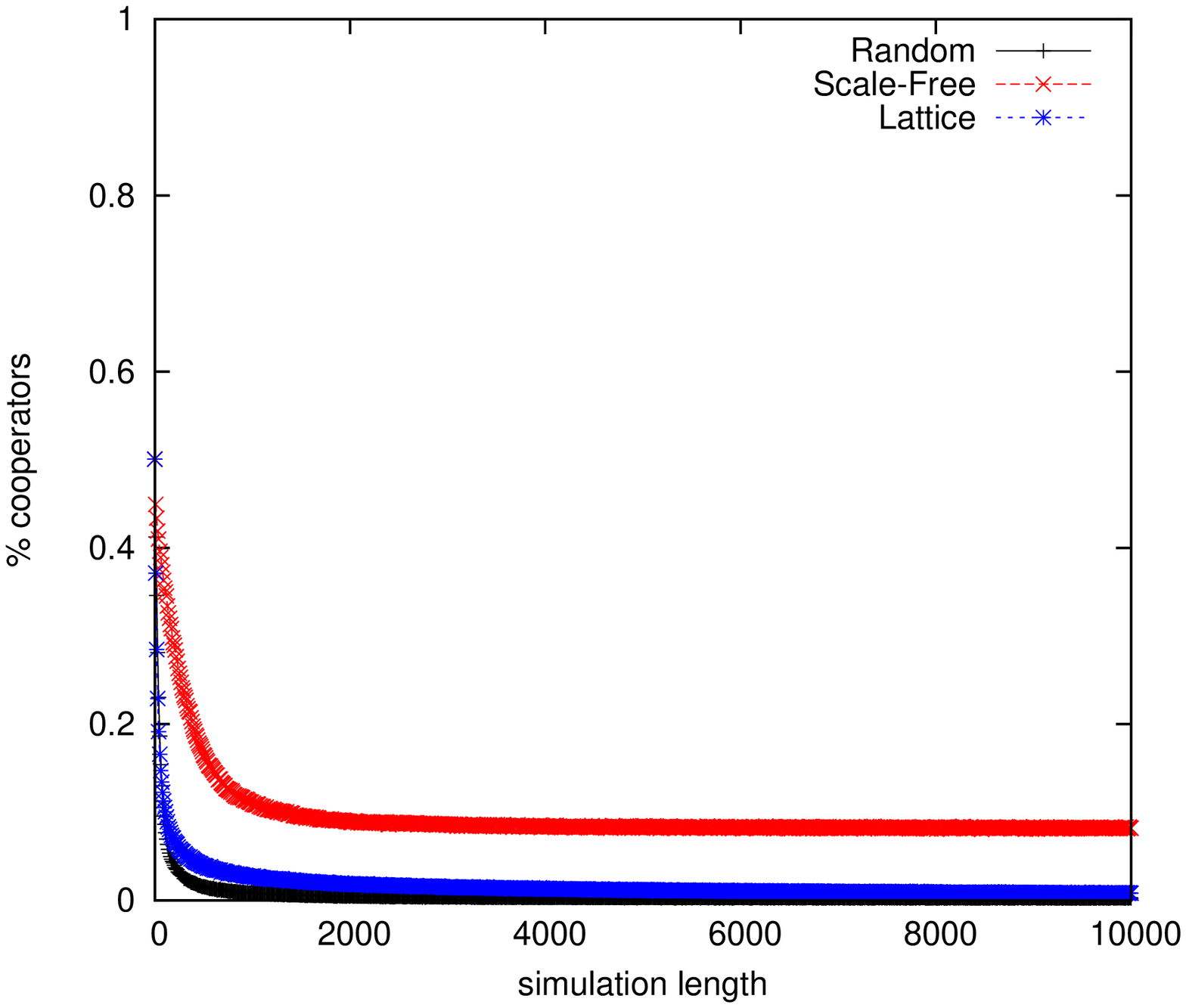}
\includegraphics[width=0.24\textwidth]{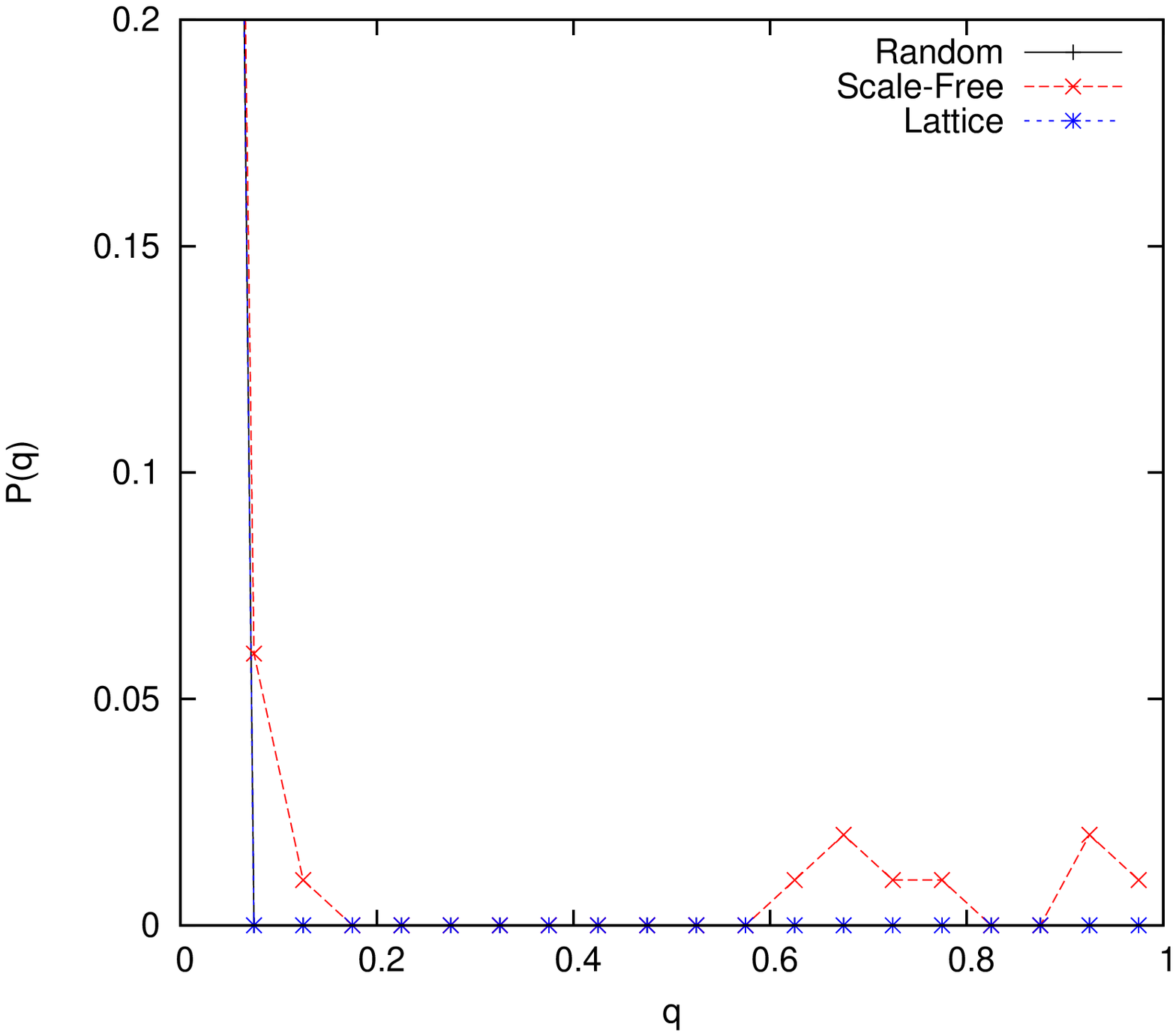}
\includegraphics[width=0.24\textwidth]{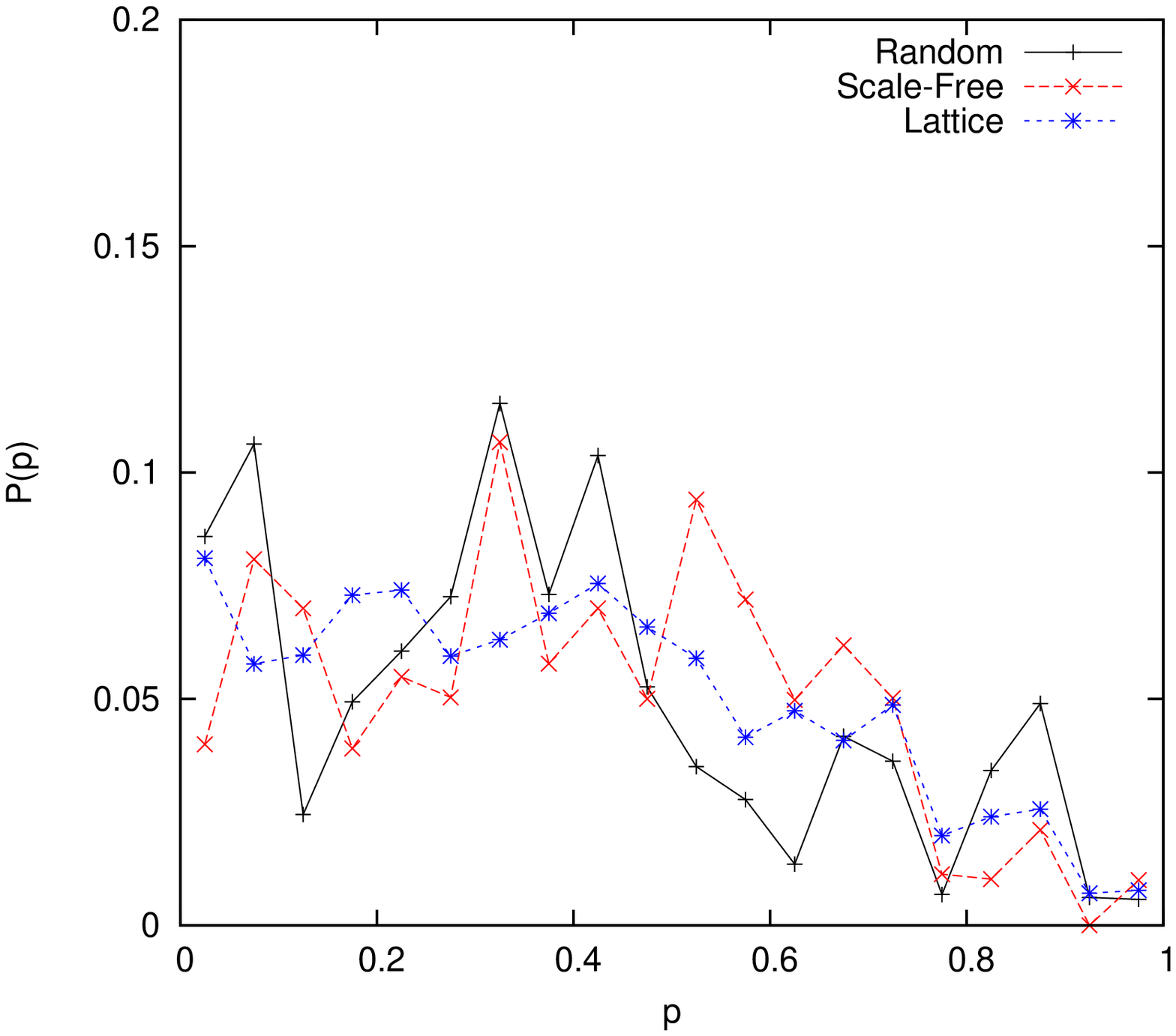}
\includegraphics[width=0.24\textwidth]{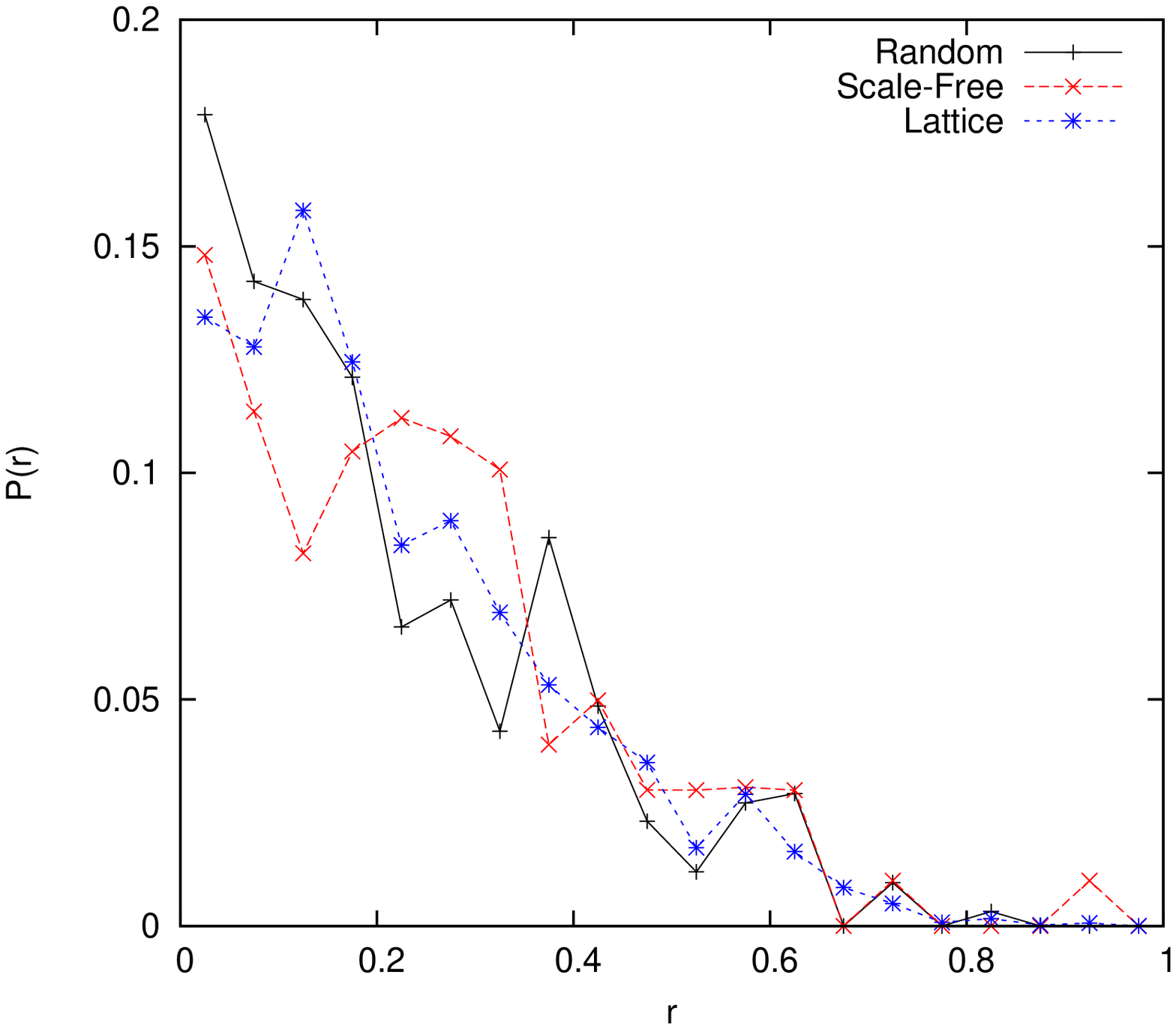}
\newline
\includegraphics[width=0.24\textwidth]{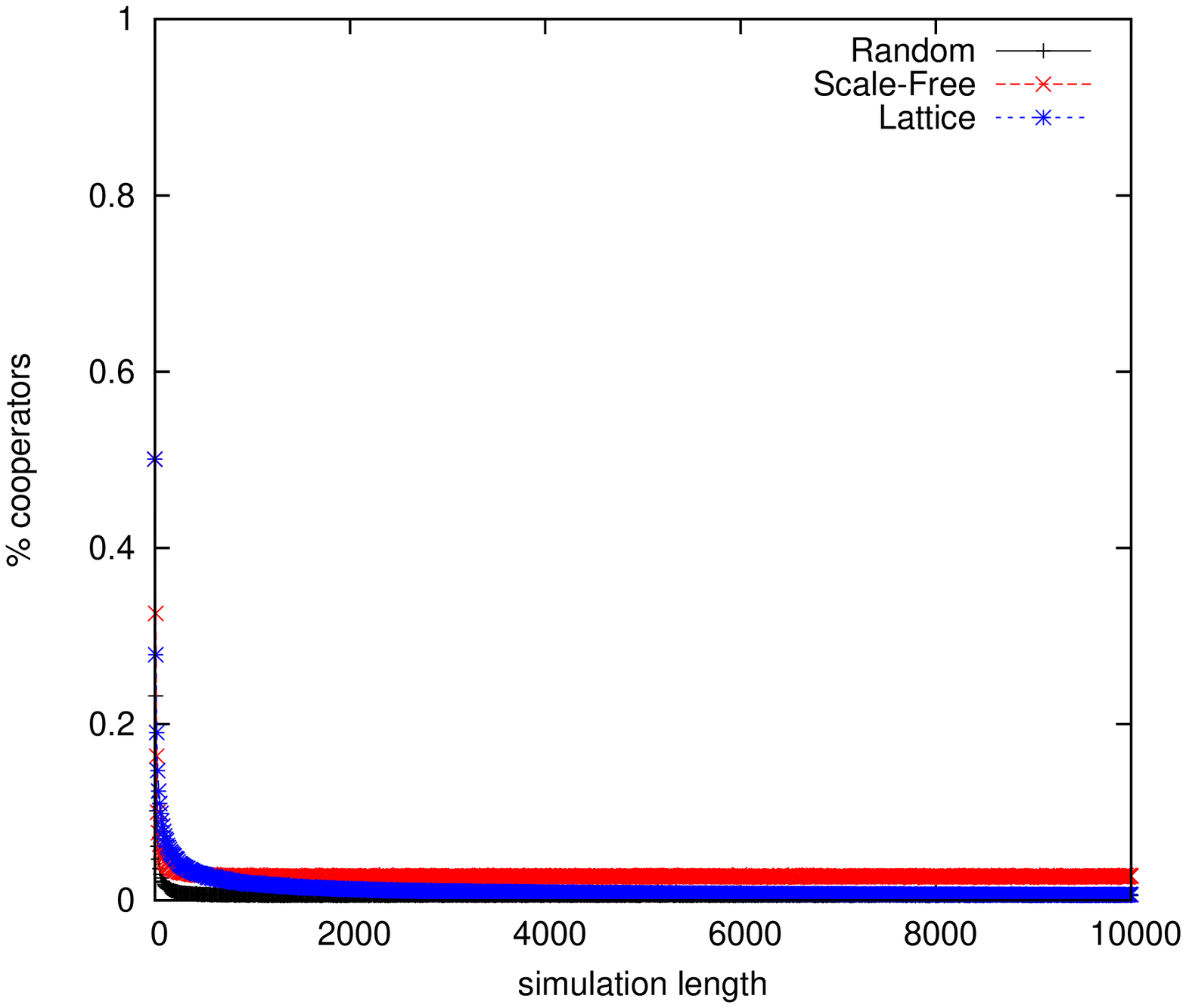}
\includegraphics[width=0.24\textwidth]{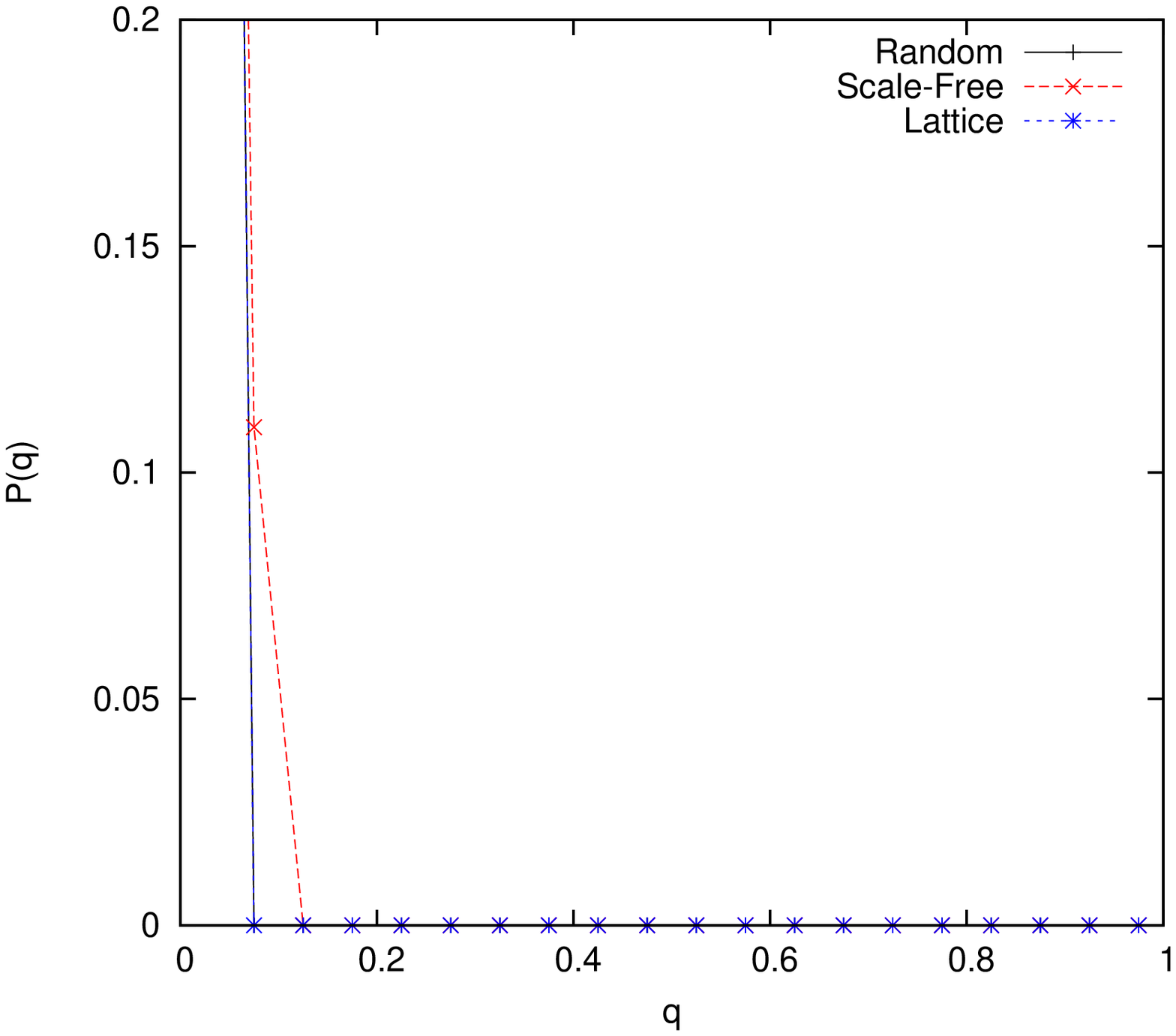}
\includegraphics[width=0.24\textwidth]{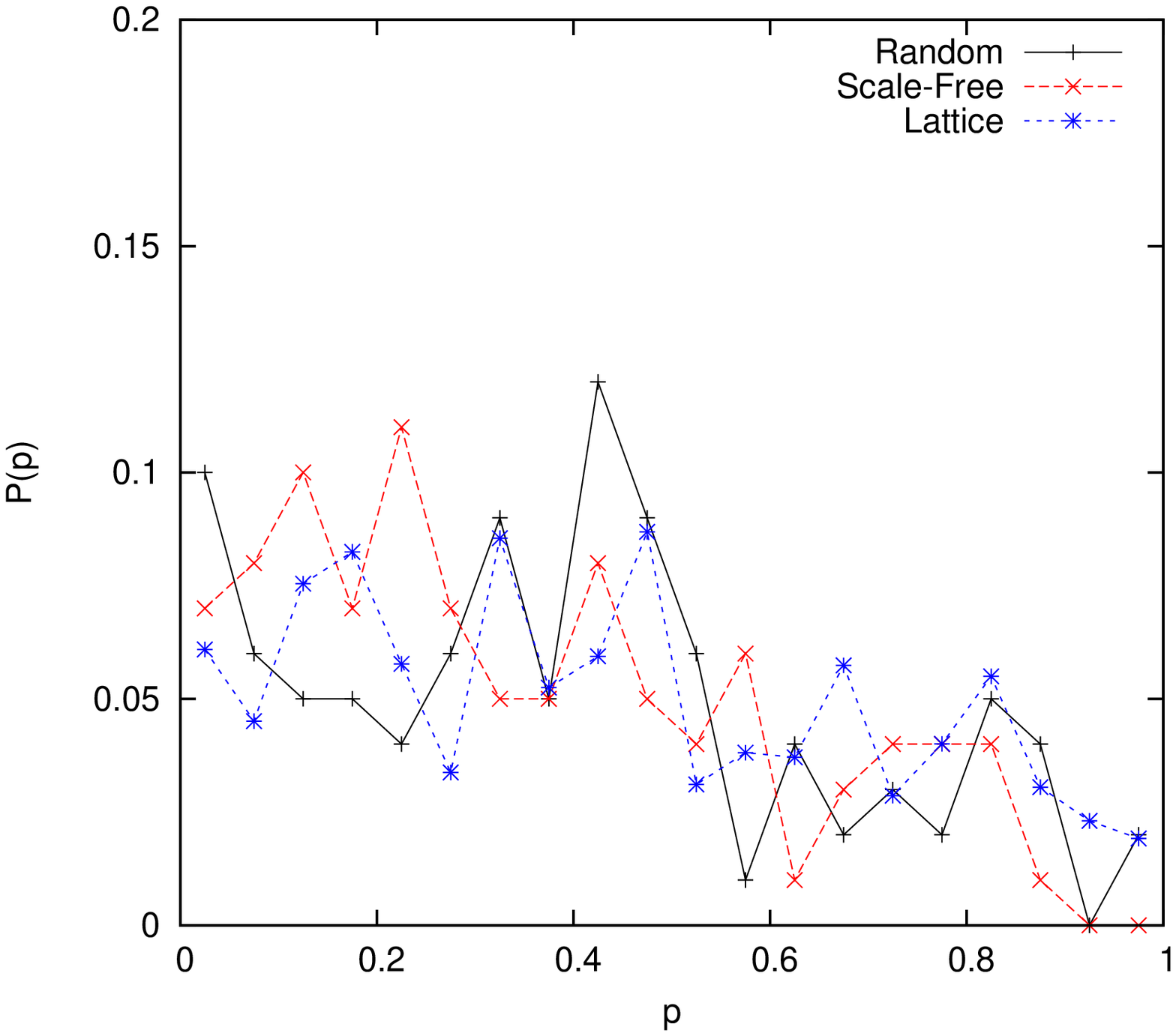}
\includegraphics[width=0.24\textwidth]{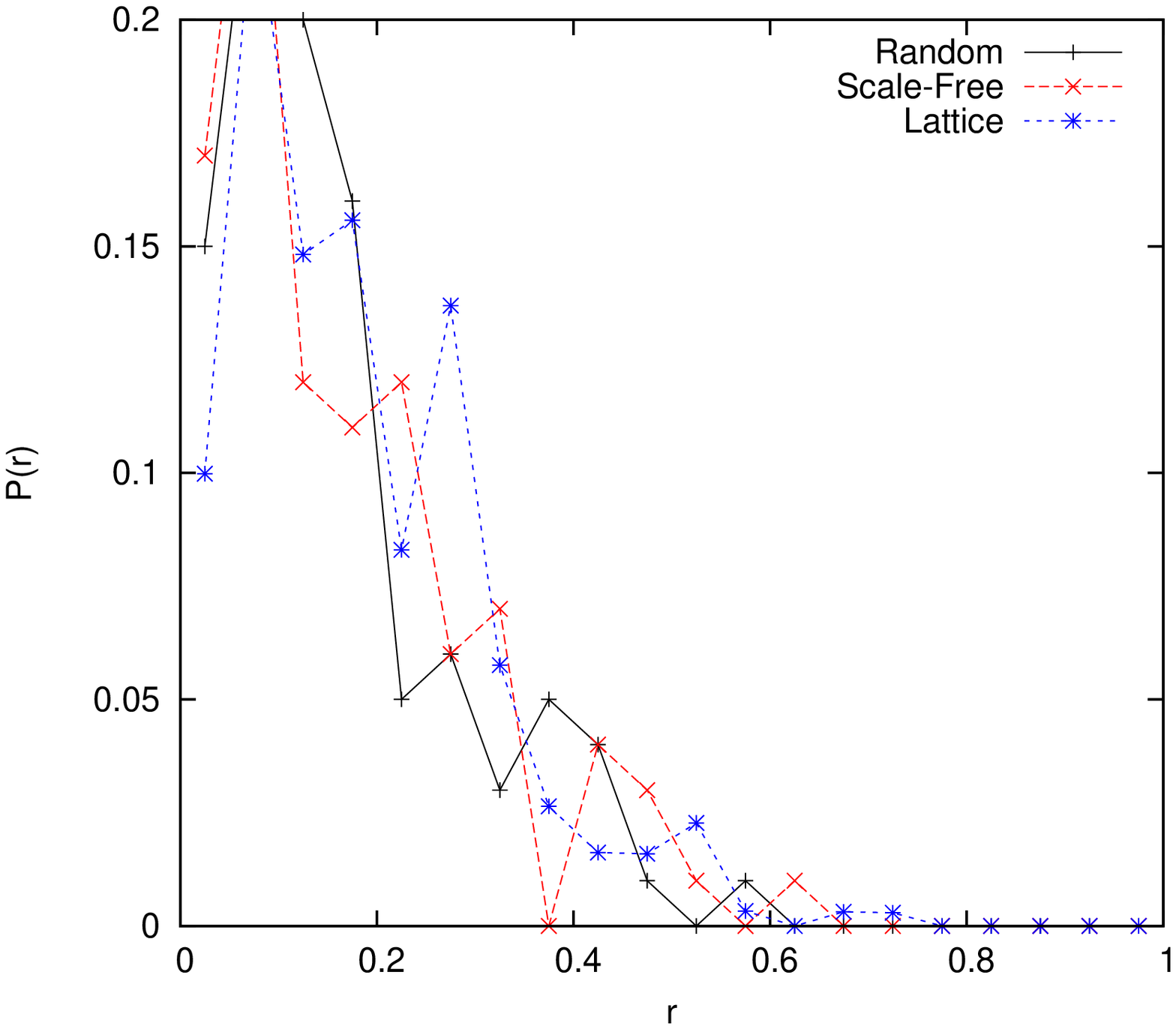}
\newline
\includegraphics[width=0.24\textwidth]{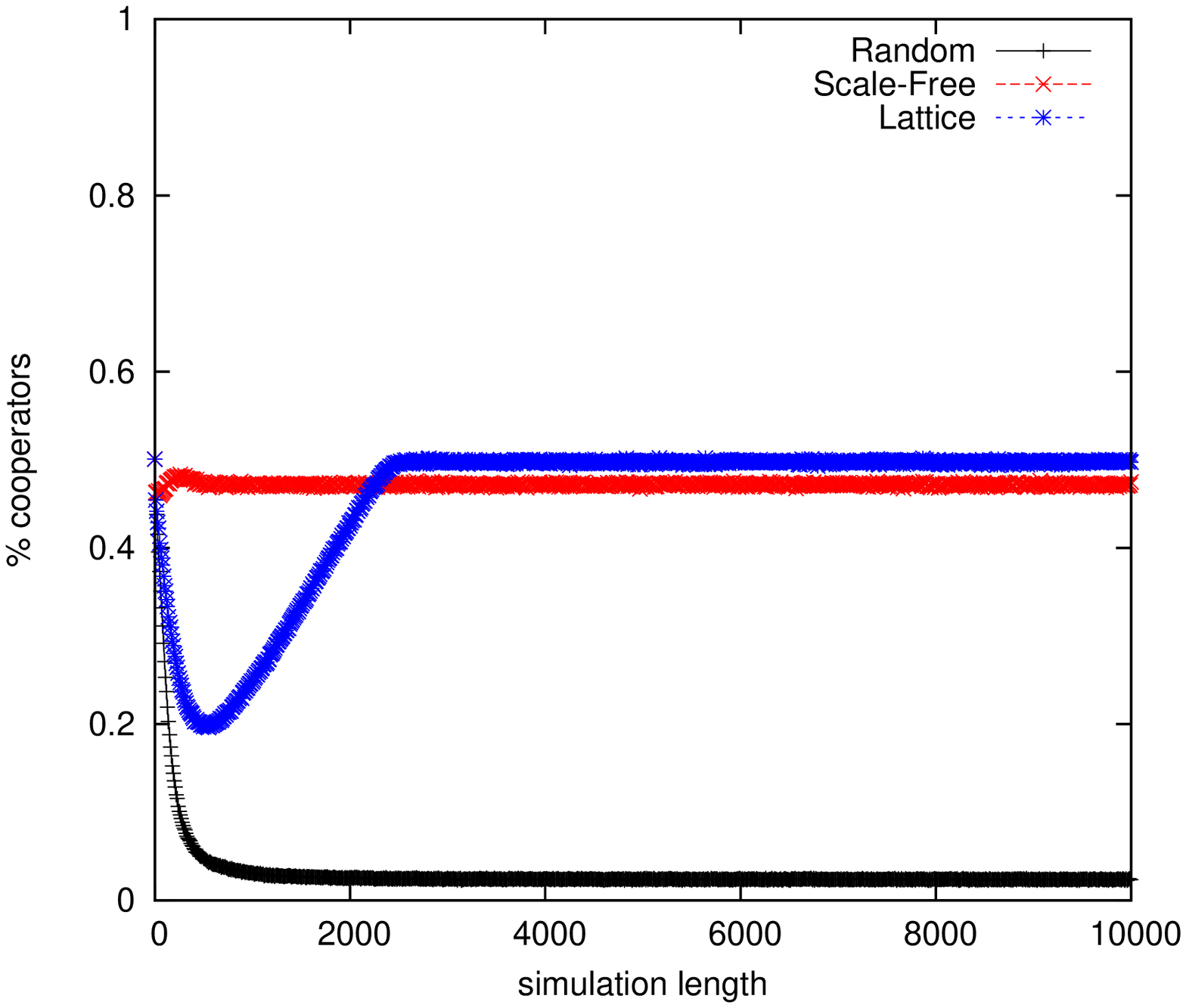}
\includegraphics[width=0.24\textwidth]{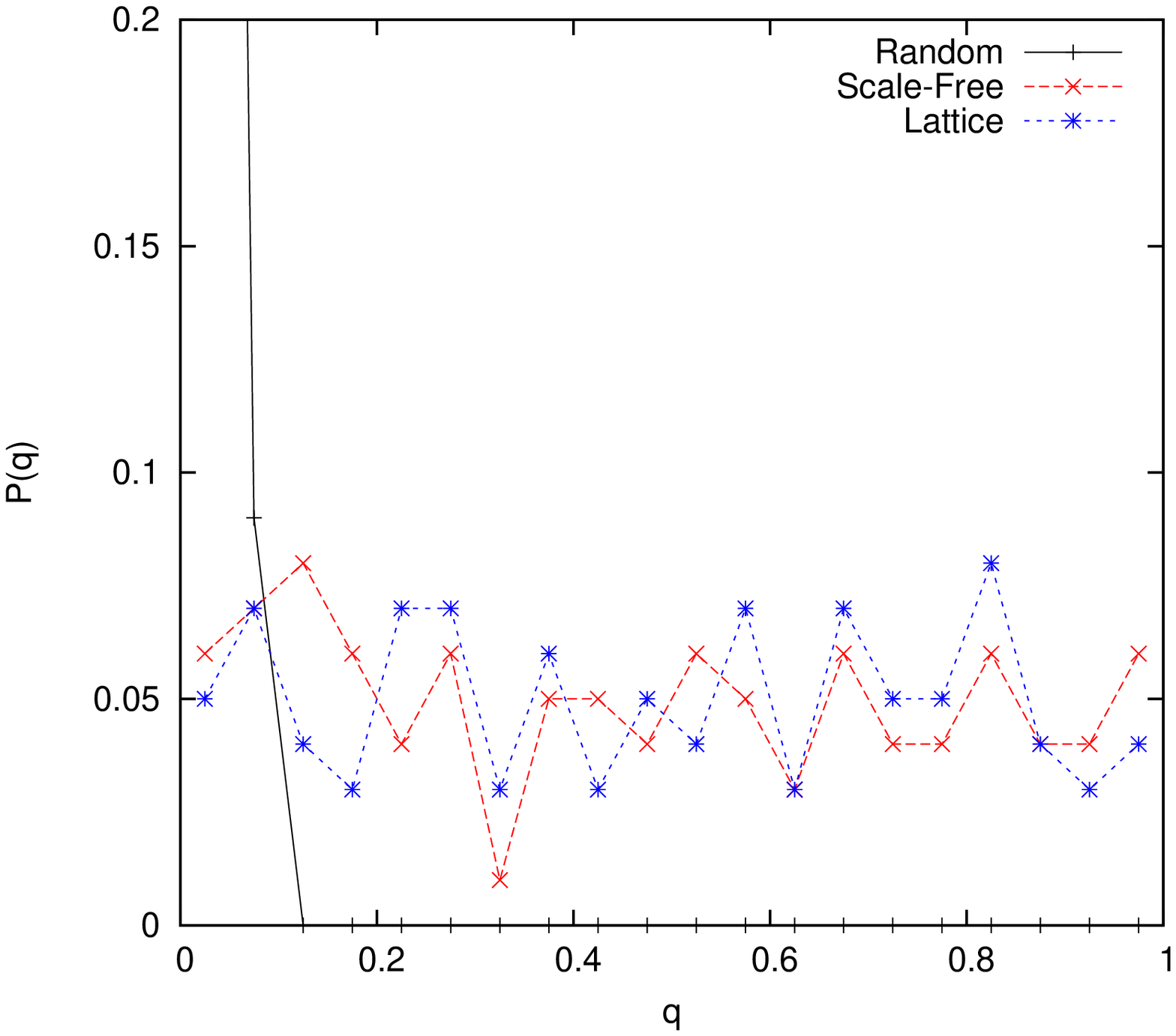}
\includegraphics[width=0.24\textwidth]{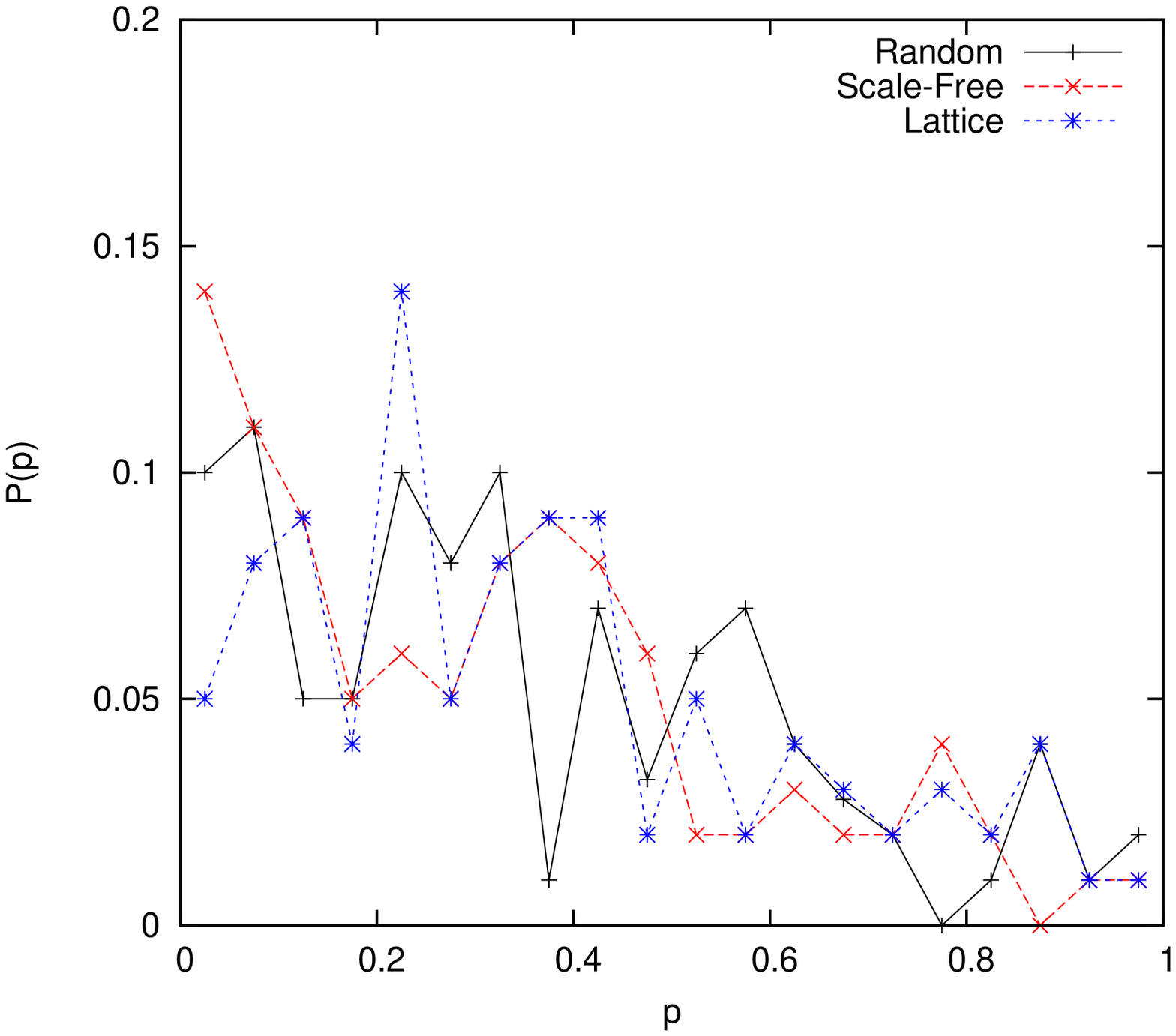}
\includegraphics[width=0.24\textwidth]{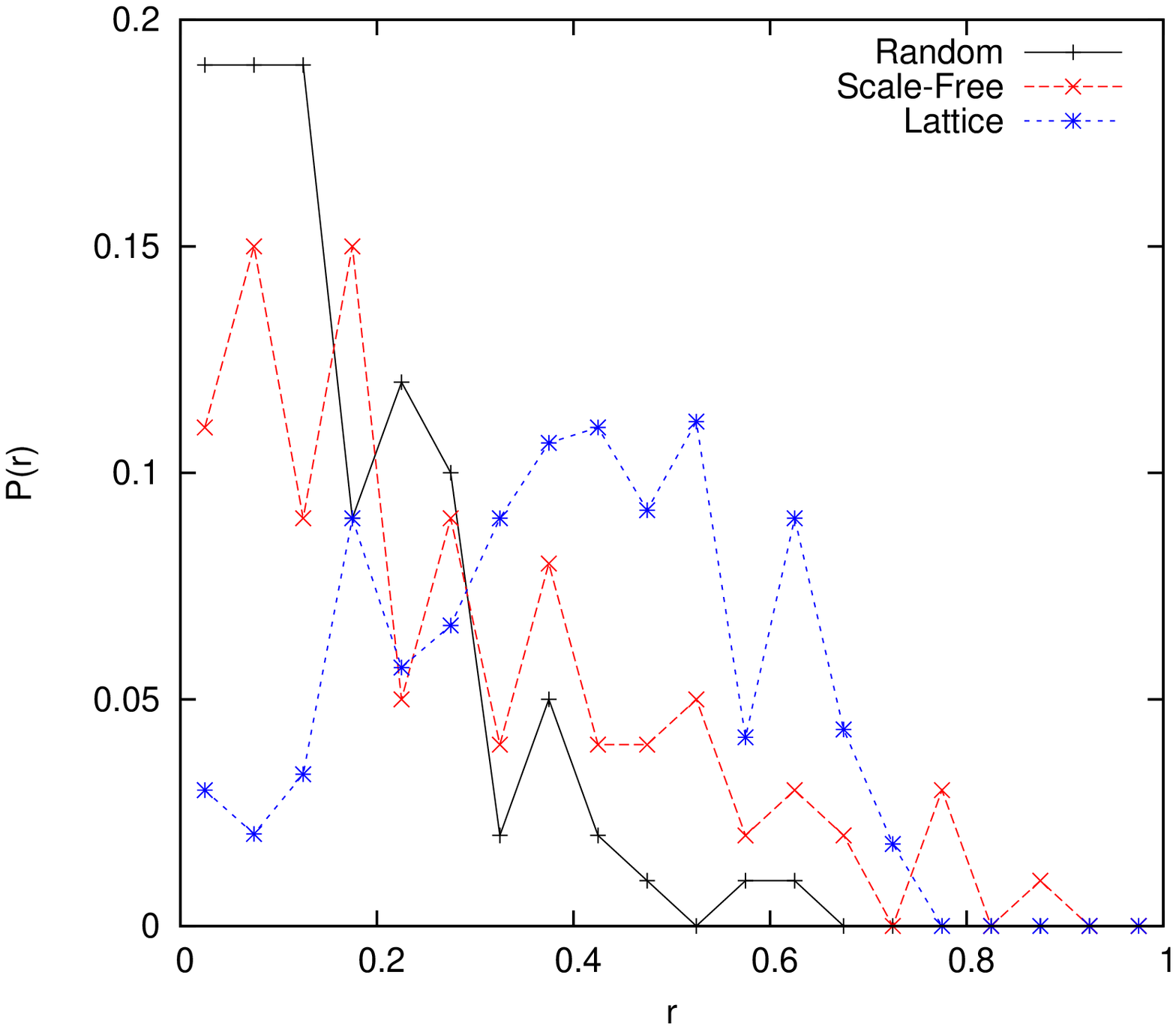}
\newline
\includegraphics[width=0.24\textwidth]{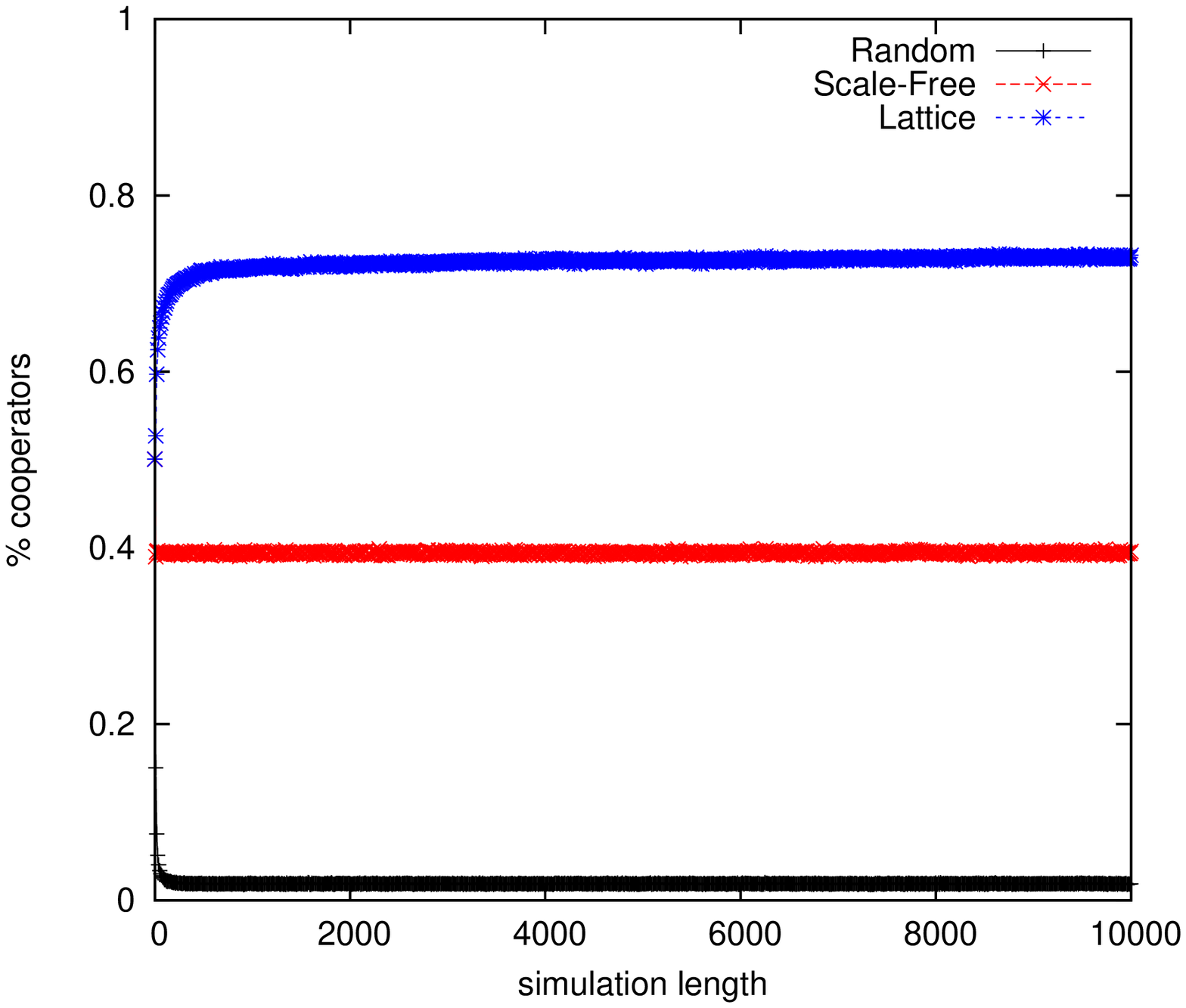}
\includegraphics[width=0.24\textwidth]{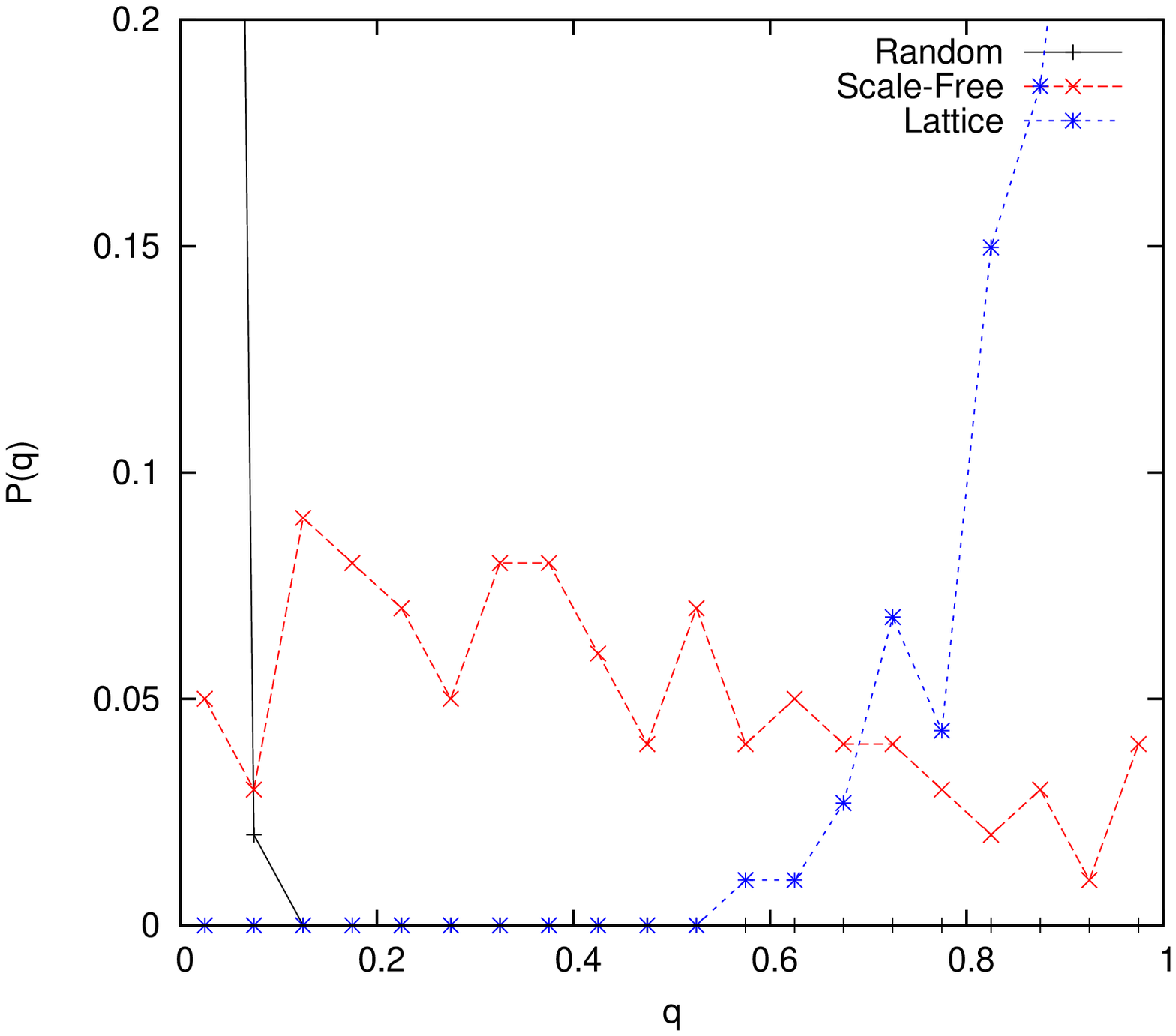}
\includegraphics[width=0.24\textwidth]{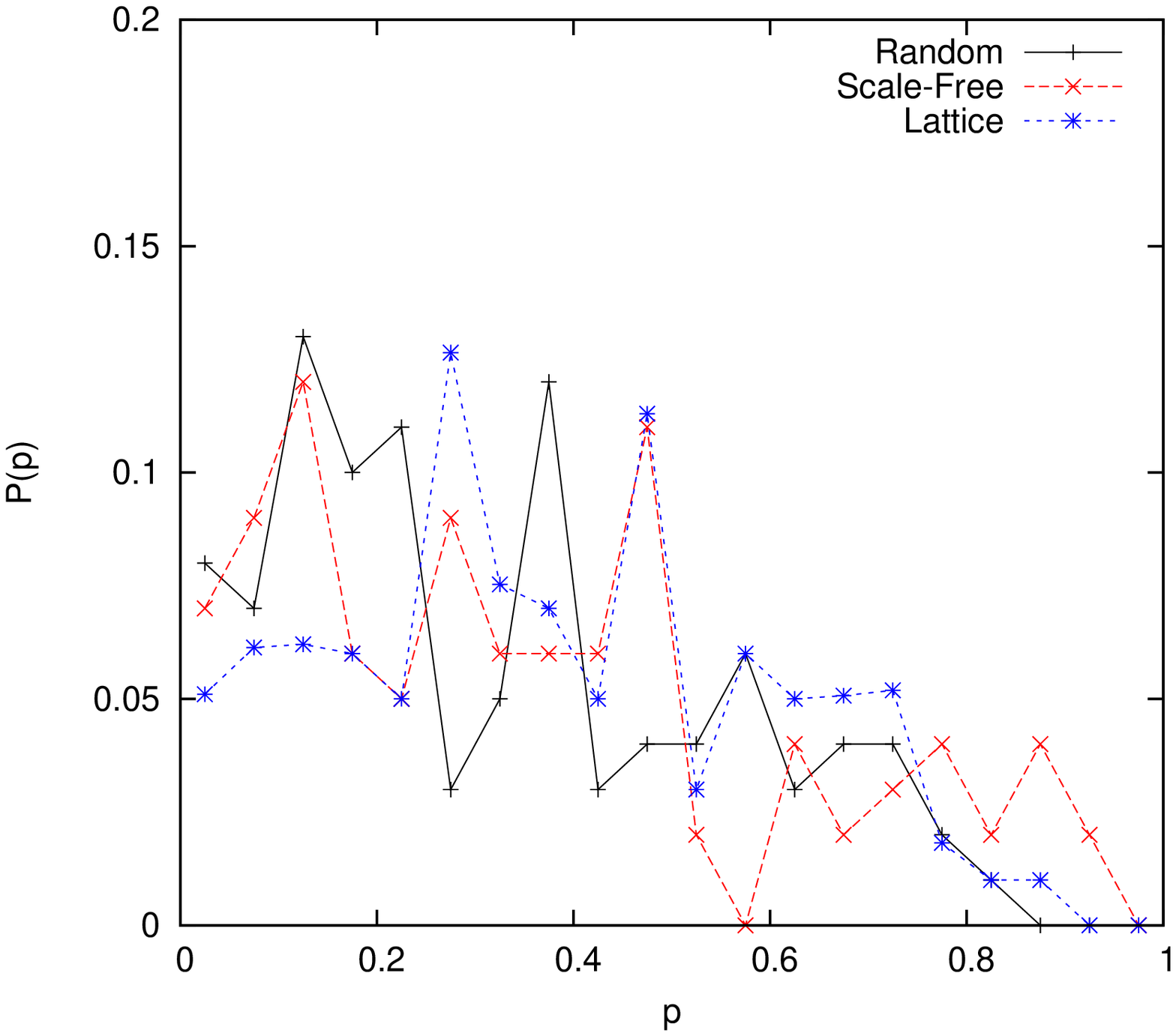}
\includegraphics[width=0.24\textwidth]{vd_tau1_4_dyn_r_10000.ps}
\newline
\includegraphics[width=0.24\textwidth]{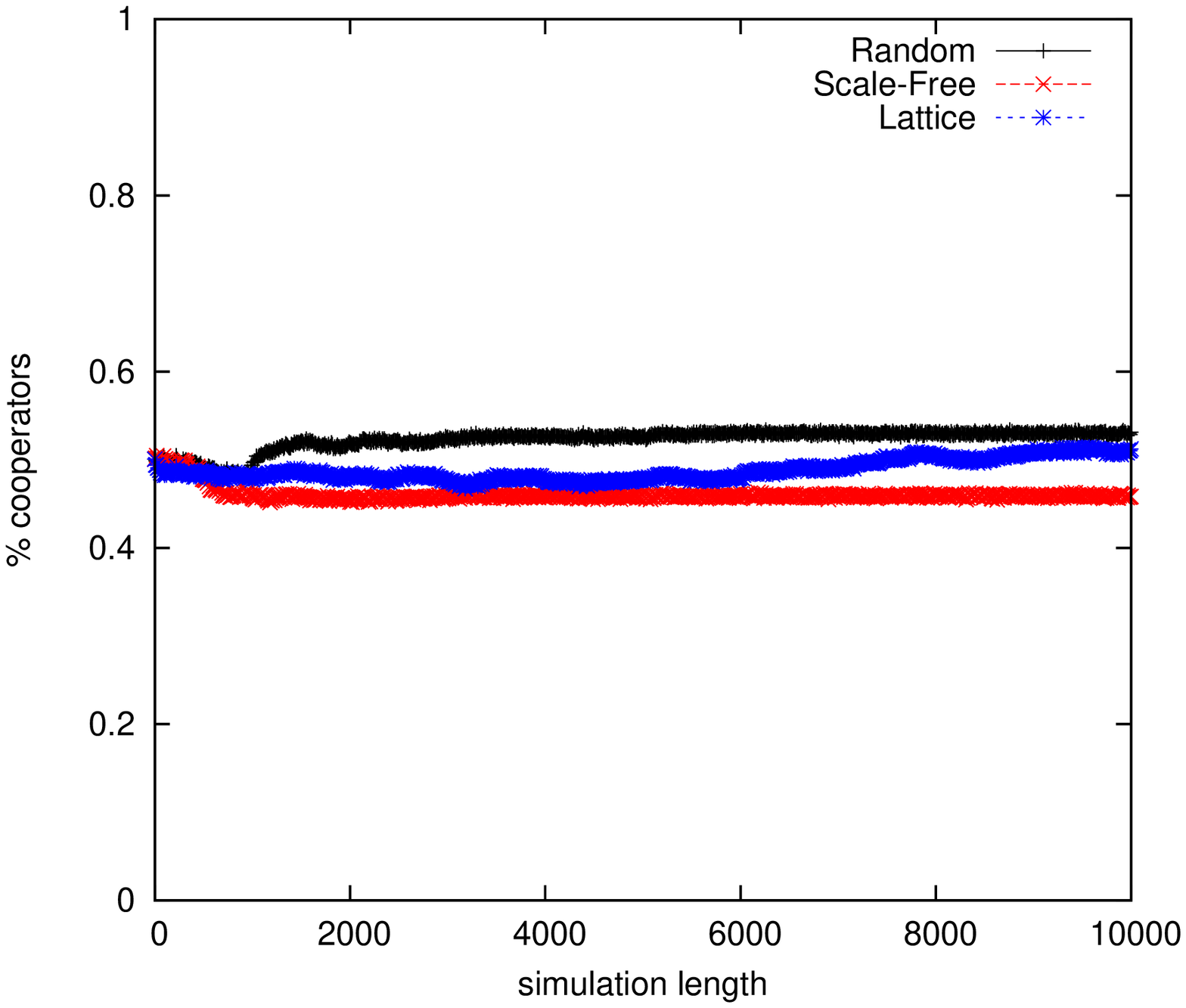}
\includegraphics[width=0.24\textwidth]{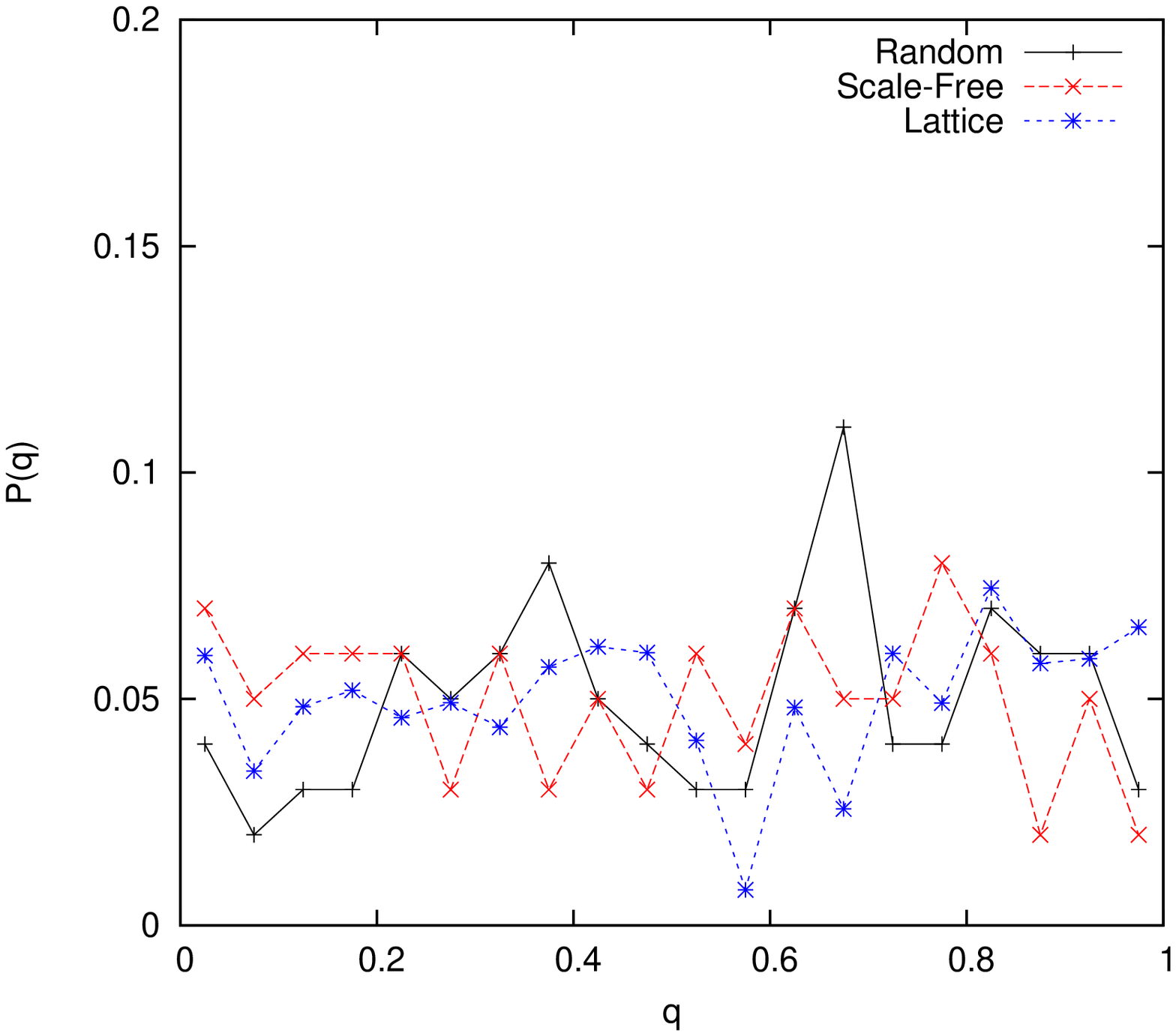}
\includegraphics[width=0.24\textwidth]{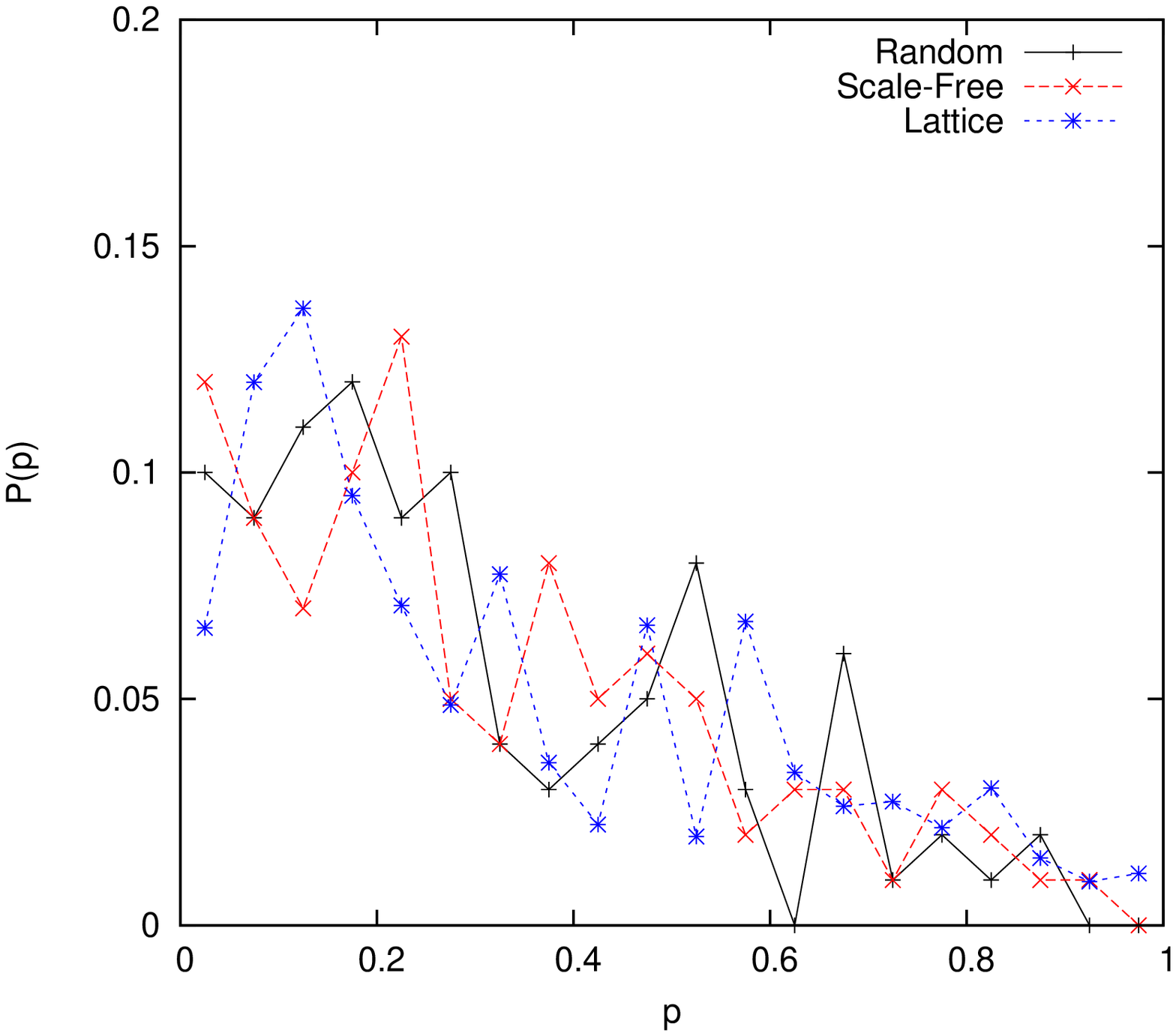}
\includegraphics[width=0.24\textwidth]{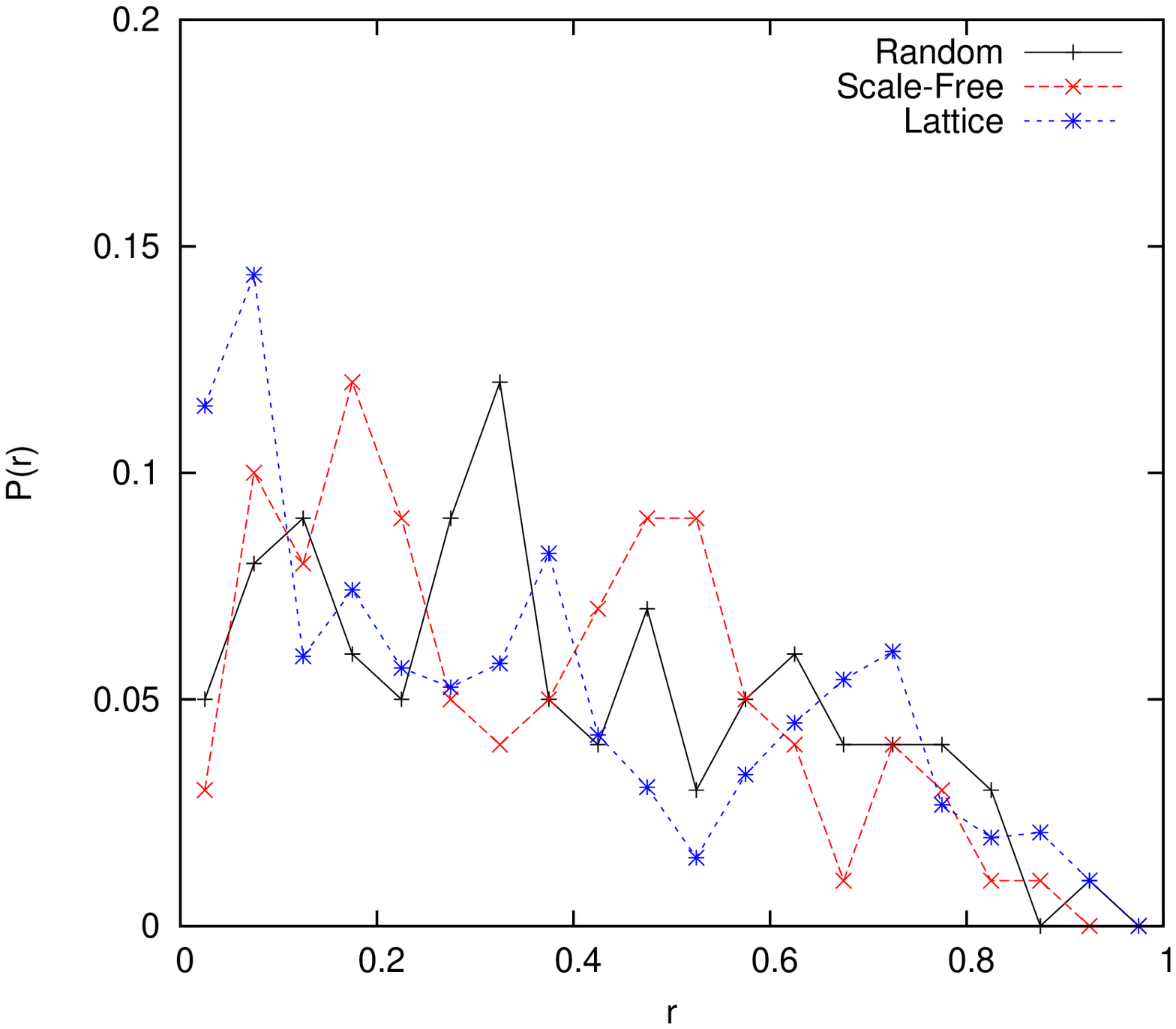}
\newline
\includegraphics[width=0.24\textwidth]{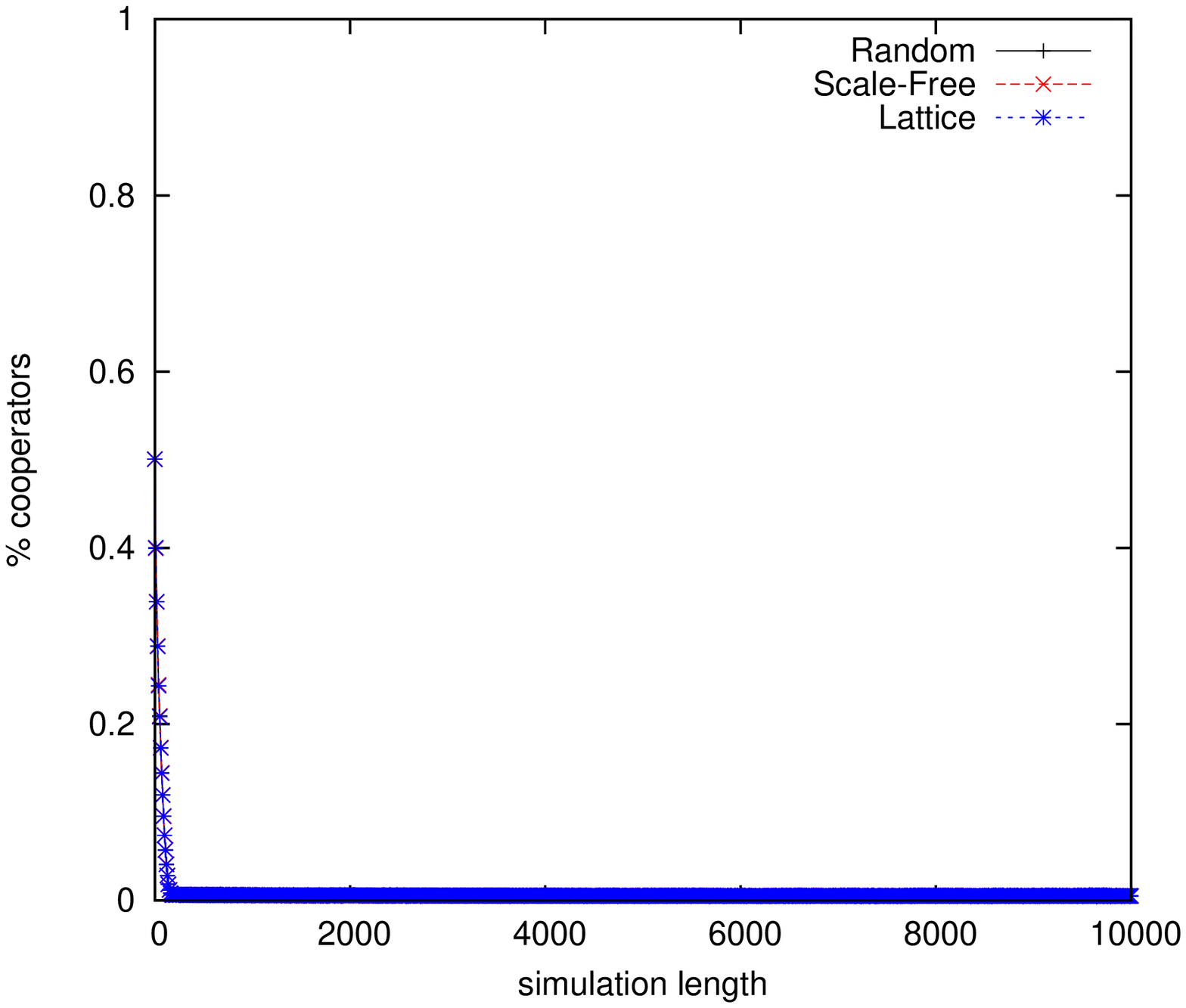}
\includegraphics[width=0.24\textwidth]{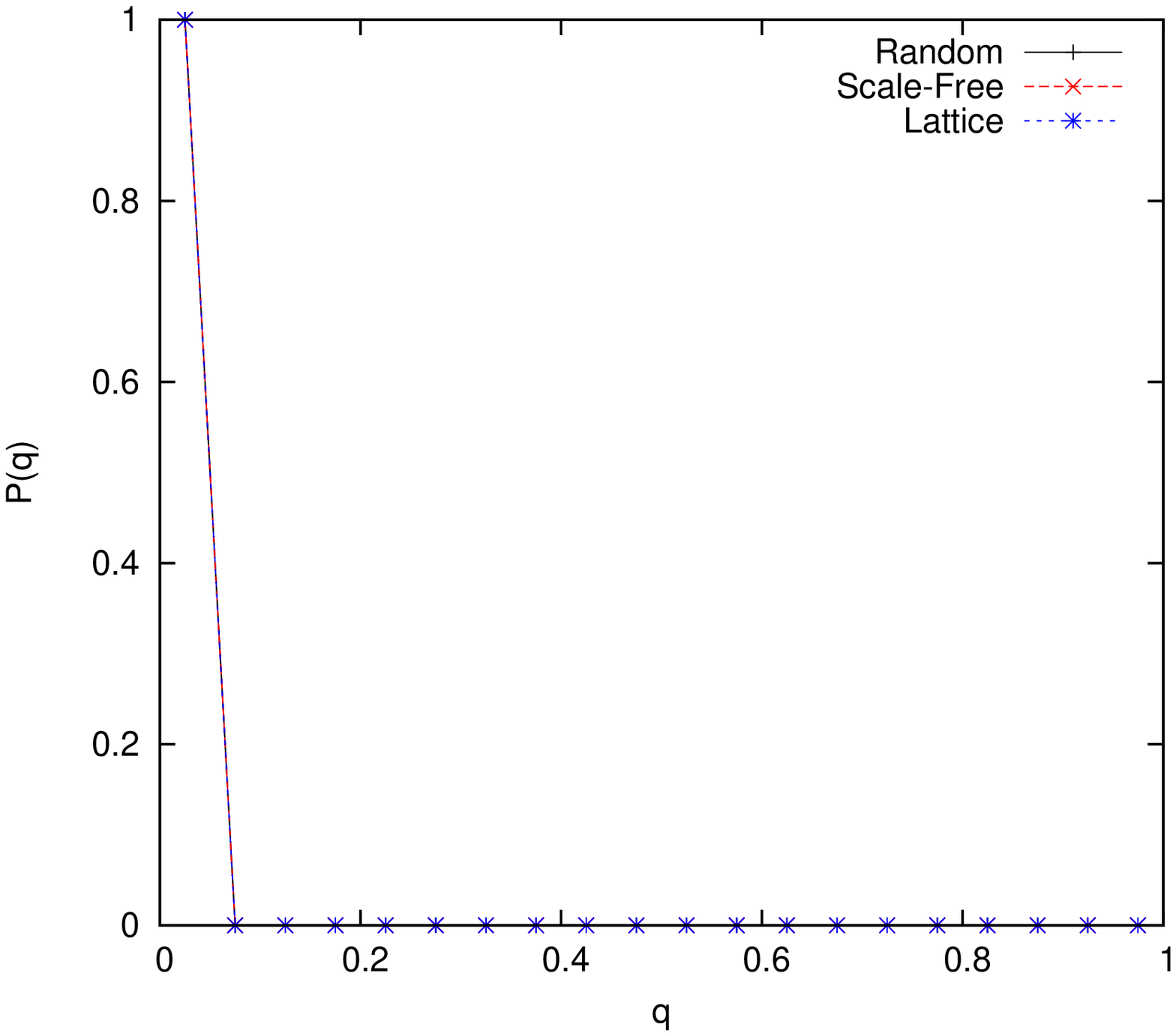}
\includegraphics[width=0.24\textwidth]{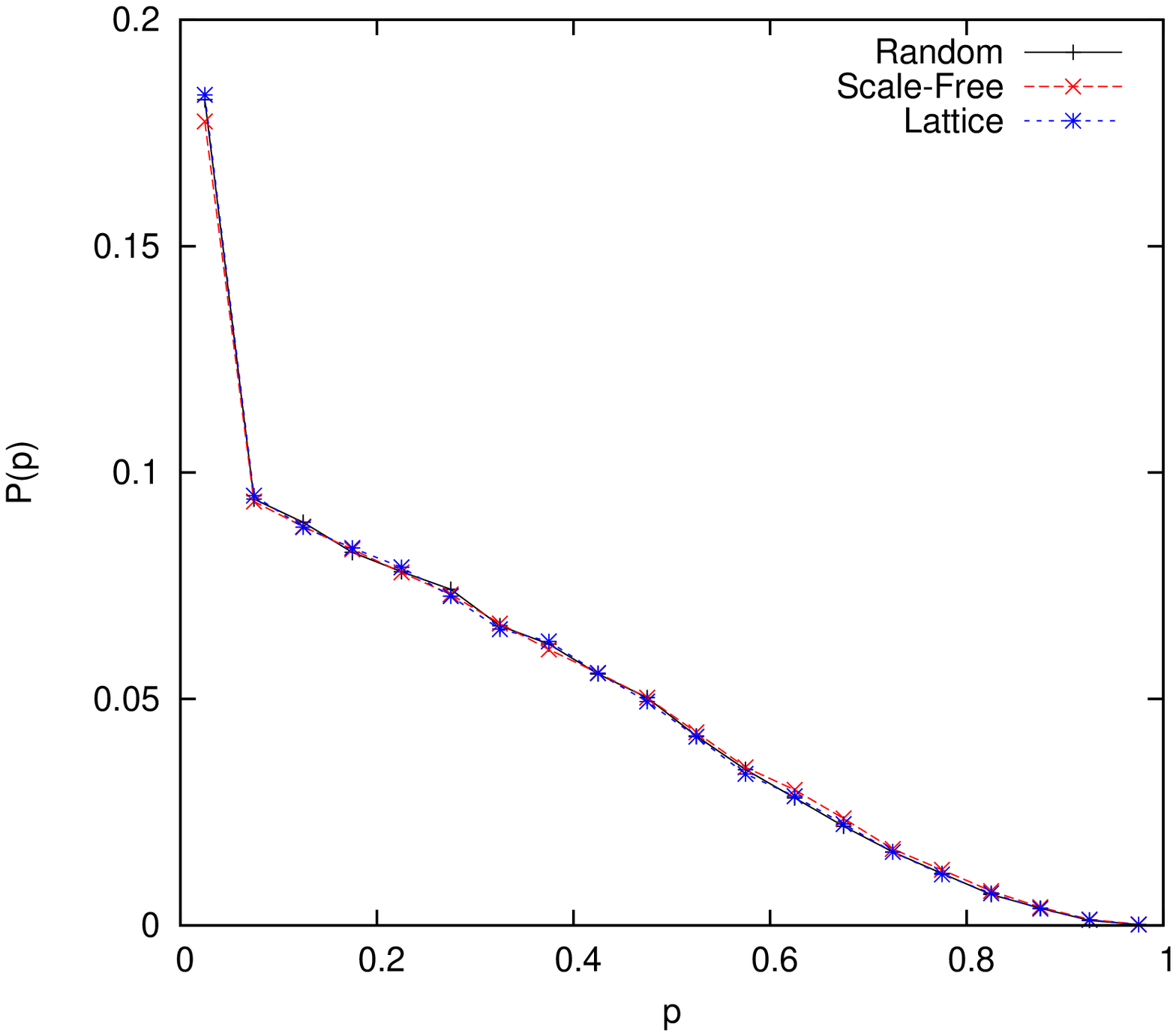}
\includegraphics[width=0.24\textwidth]{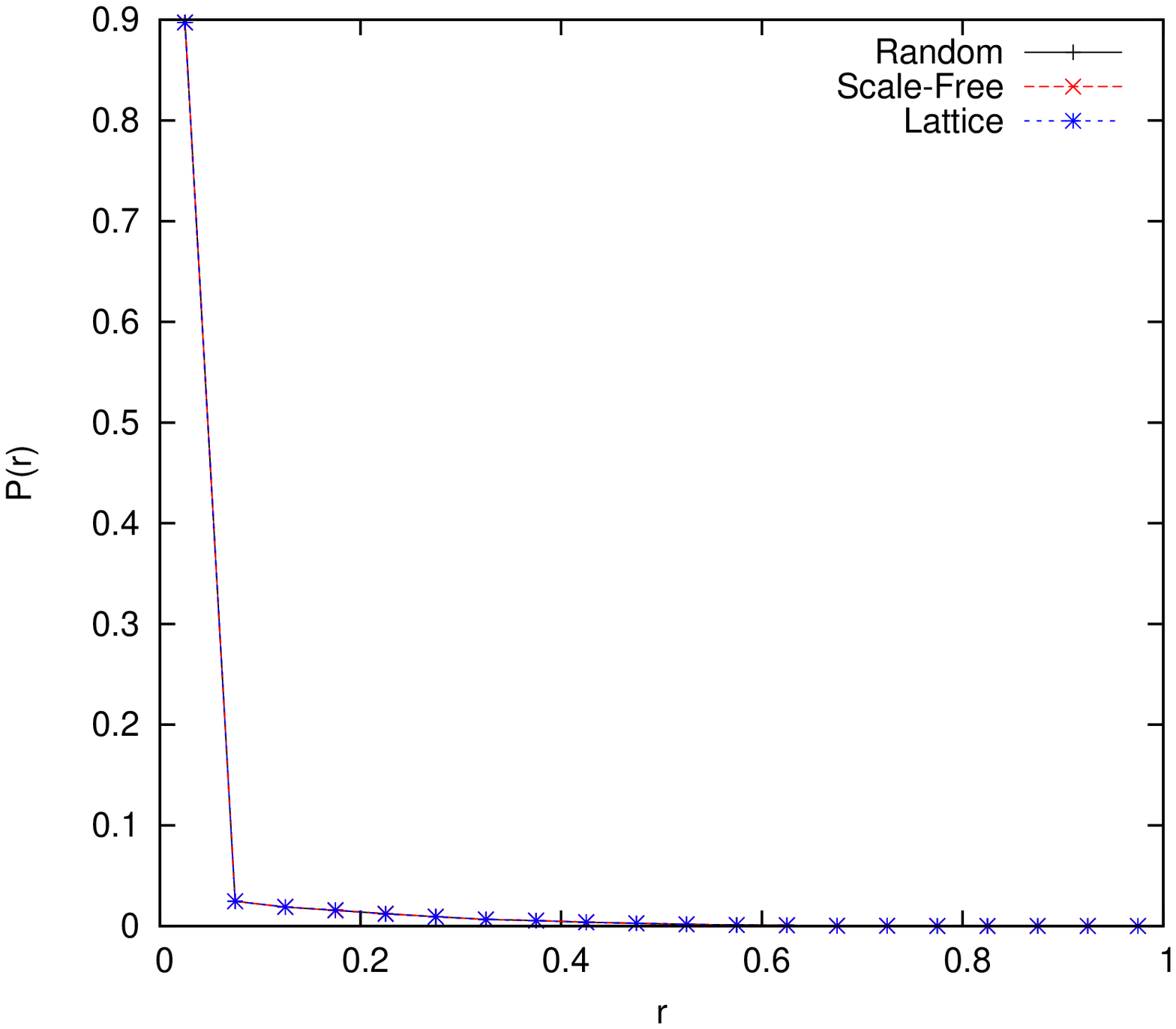}
\caption{Evolution of the level of cooperation (left column) and stationary distributions of MCC parameters (from left to right: $q$, $p$ and $r$) when the evolutionary dynamics is 
(from top to bottom): Proportional Imitation, Fermi rule with $\beta=1/2$, Death-Birth rule, Unconditional Imitation, Voter model and Best Response with $\delta=10^{-2}$. 
Results are averaged over 100 independent realizations.}\label{fig.C}
\end{figure*}

A first general finding about this type of evolutionary rules is that, because of their own nature, they allow for a very large number of 
surviving MCC profiles ($\sim N$), even when the parameters tend to concentrate around specific values. 
The bottom set of plots of Fig.\ \ref{fig.C} summarizes our results for the Best Response dynamics, which is the most ``rational'' of the ones that we are studying here. 
For this choice, the system always ends up in a fully defective state, irrespectively of the network's structure, which is the outcome 
that would be obtained by global maximization of the individual payoffs. In this sense, the amount $\delta$ by which parameters are shifted at each update influences only 
the convergence rate of the system: higher $\delta$ arrives faster to full defection ($q=r=0$). We then see that evolution by Best Response fails completely 
to explain any of the main experimental results. 

\begin{figure*}
\includegraphics[width=0.24\textwidth]{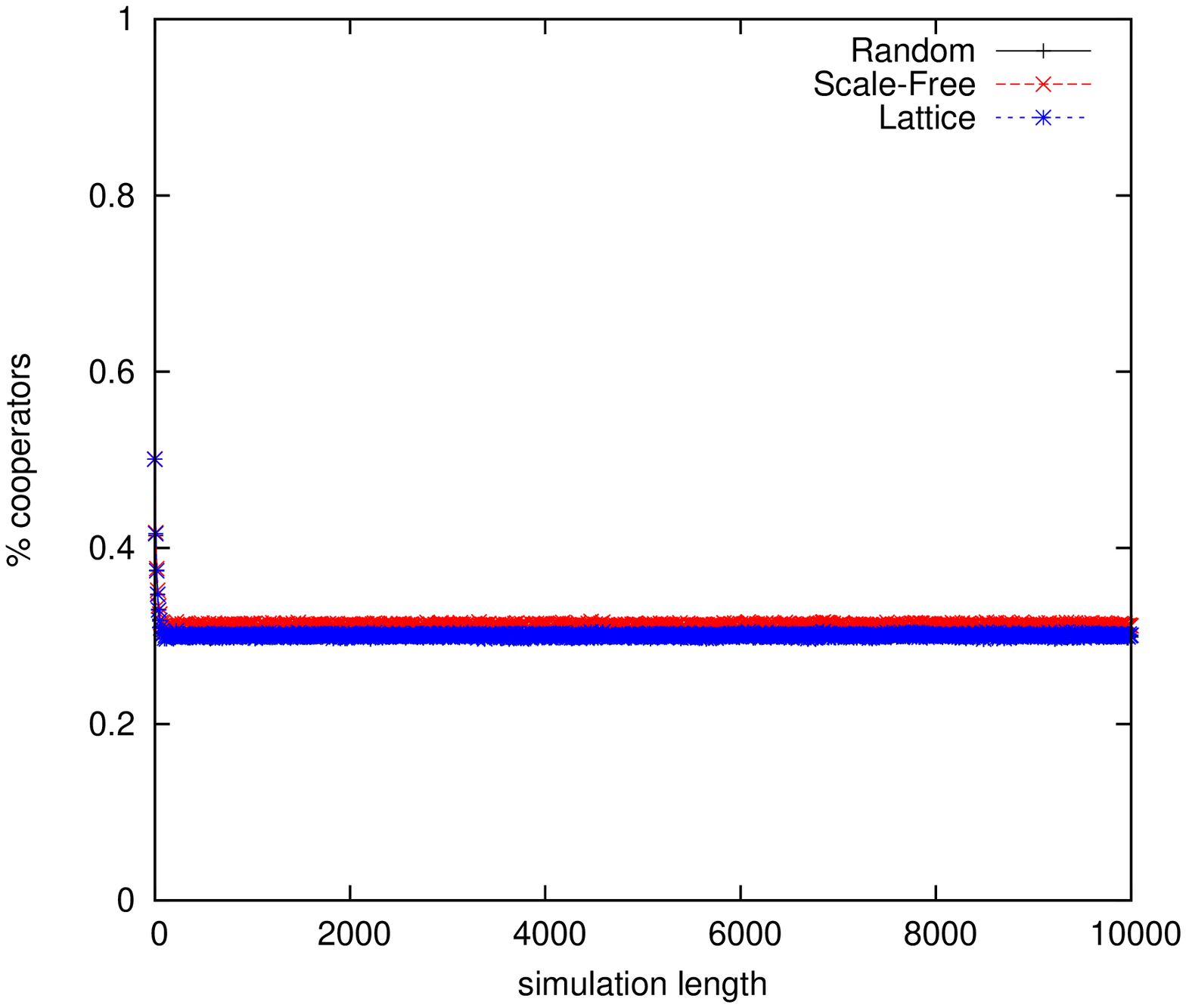}
\includegraphics[width=0.24\textwidth]{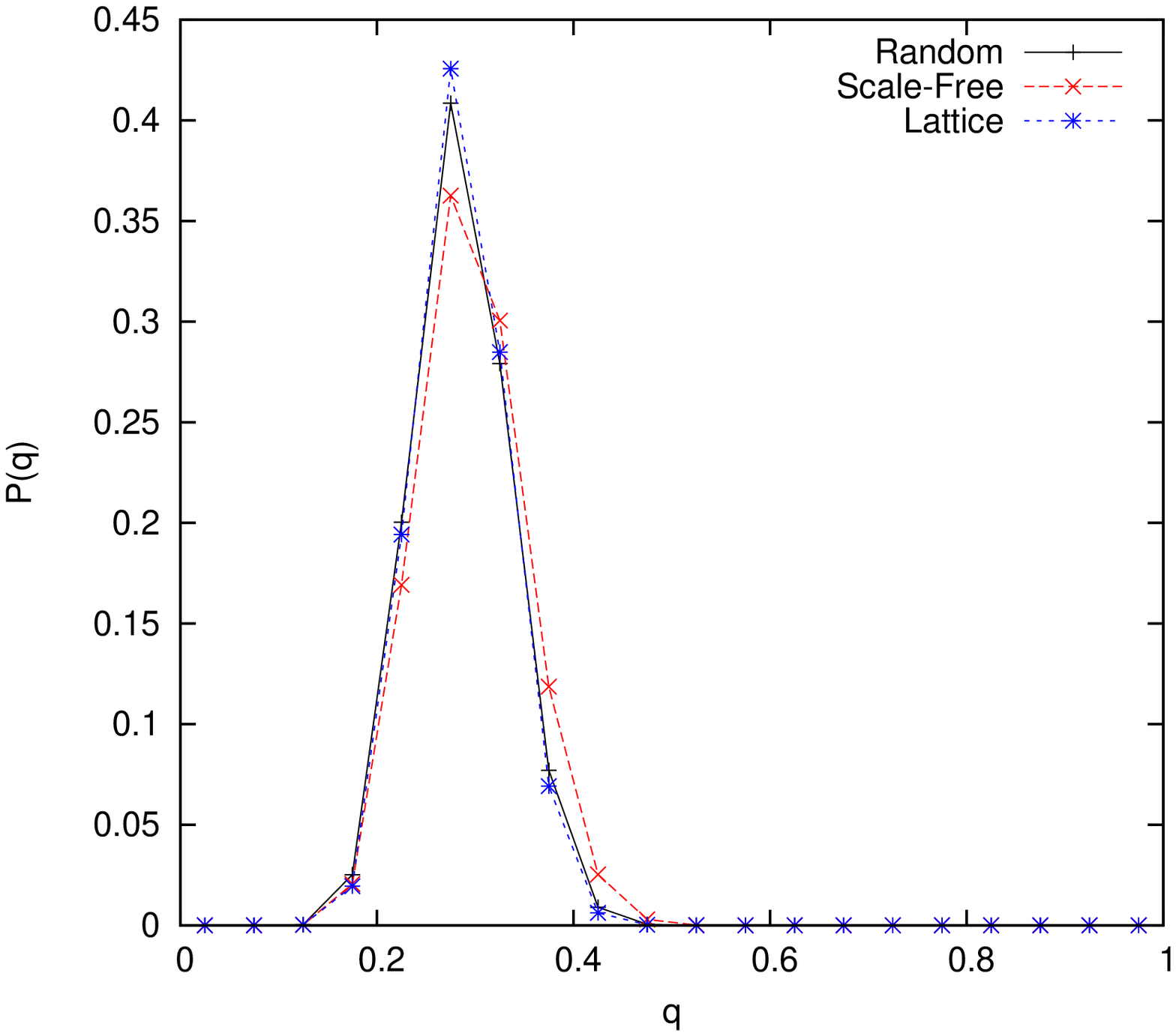}
\includegraphics[width=0.24\textwidth]{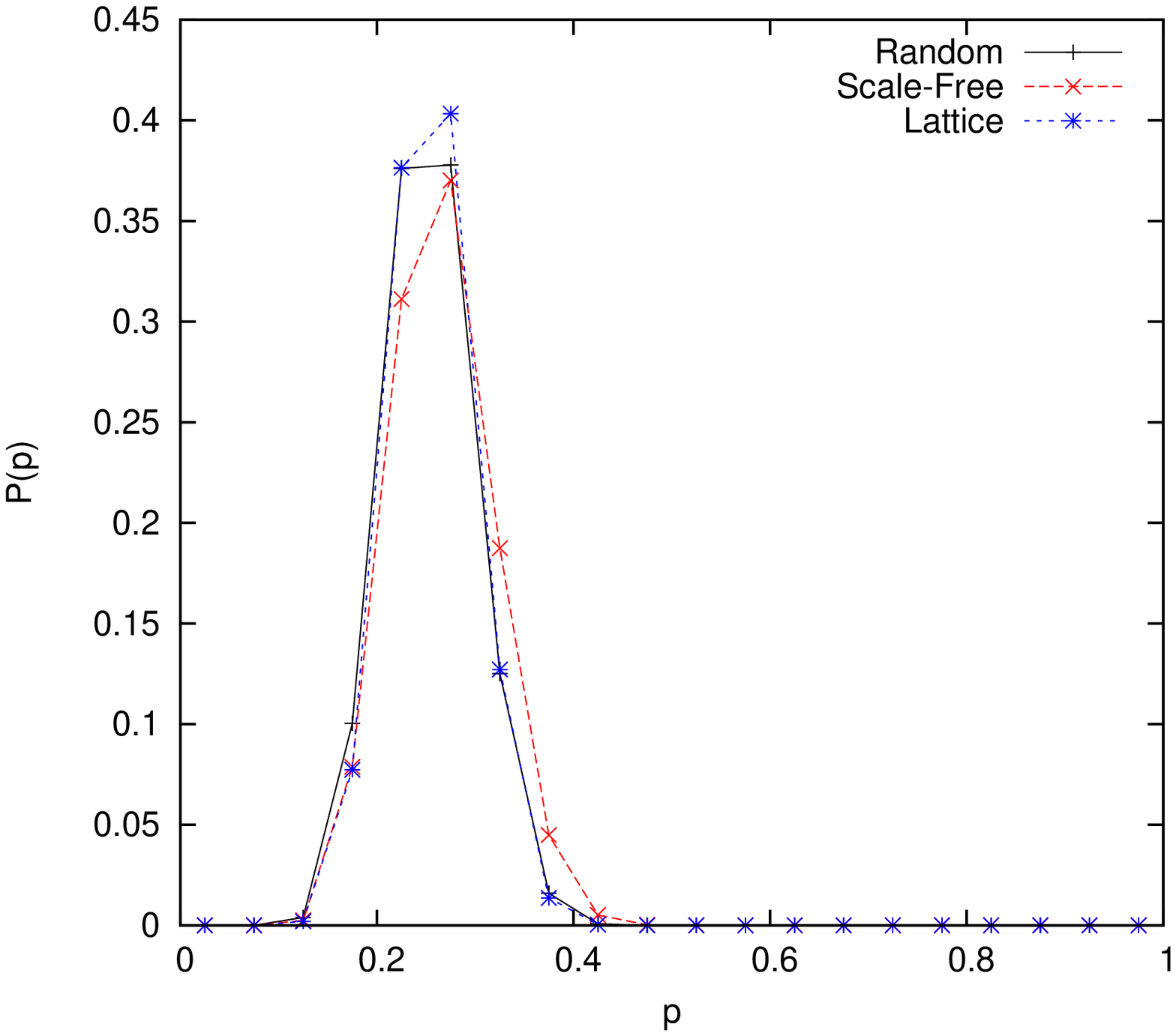}
\includegraphics[width=0.24\textwidth]{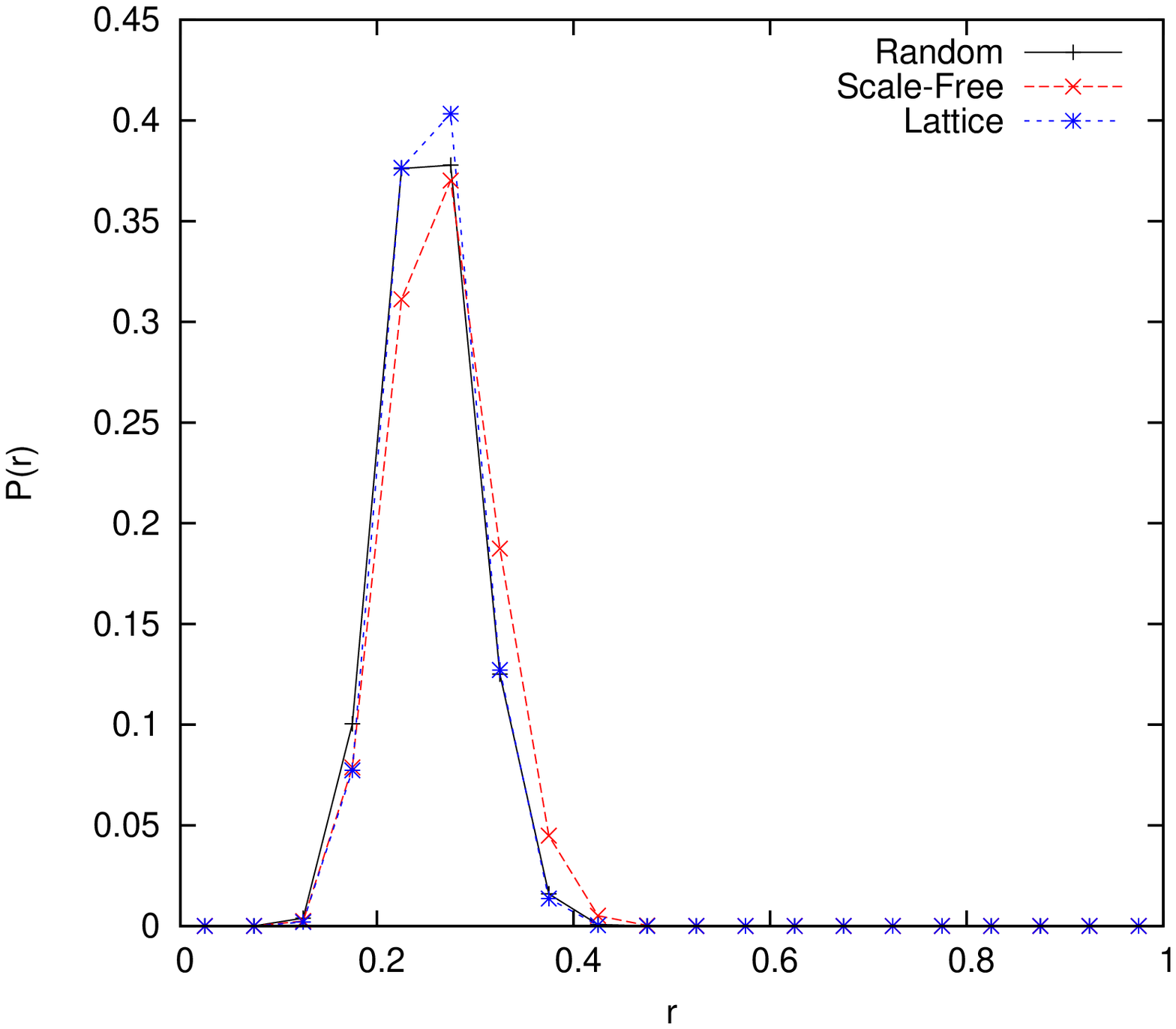}
\newline
\includegraphics[width=0.24\textwidth]{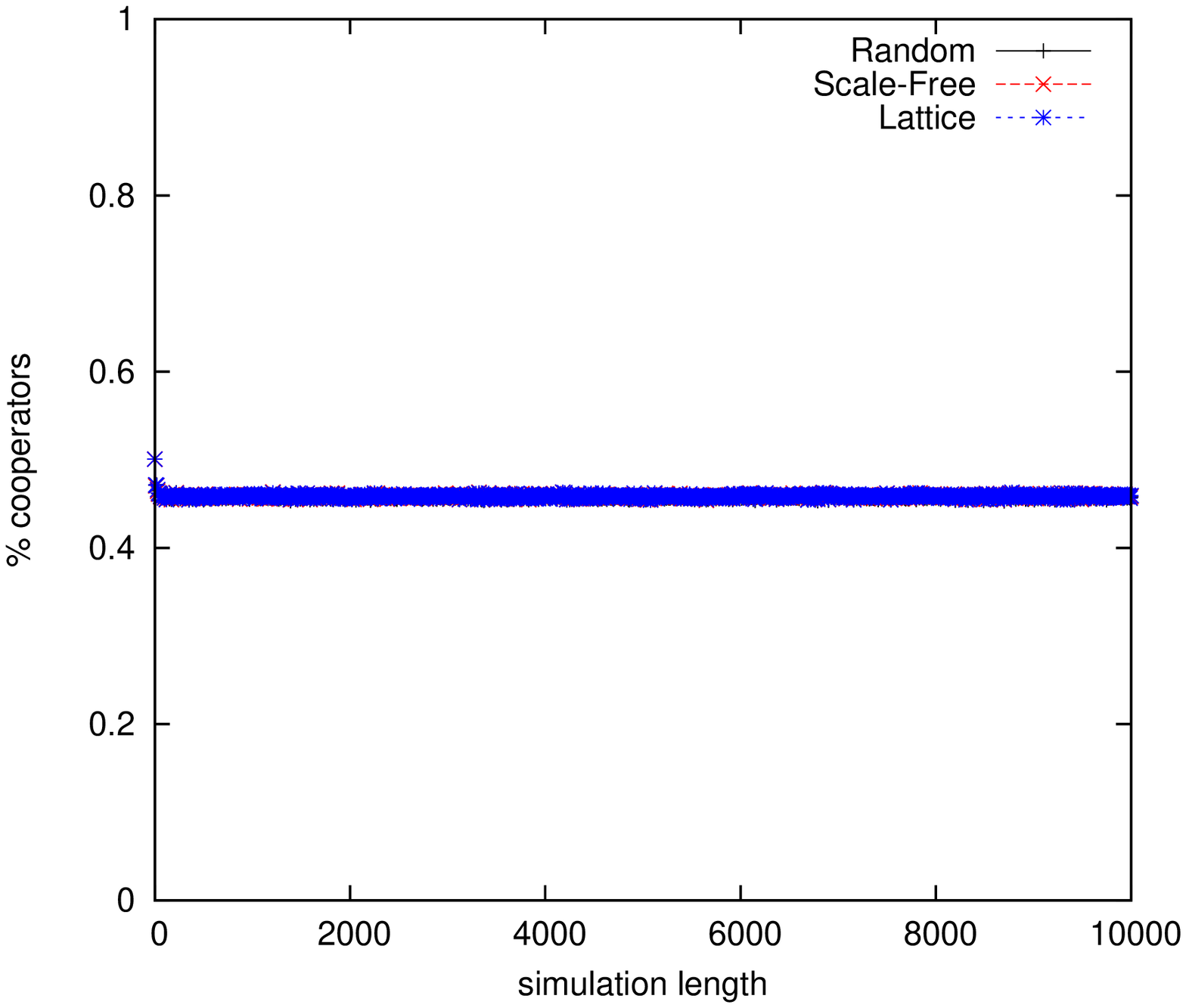}
\includegraphics[width=0.24\textwidth]{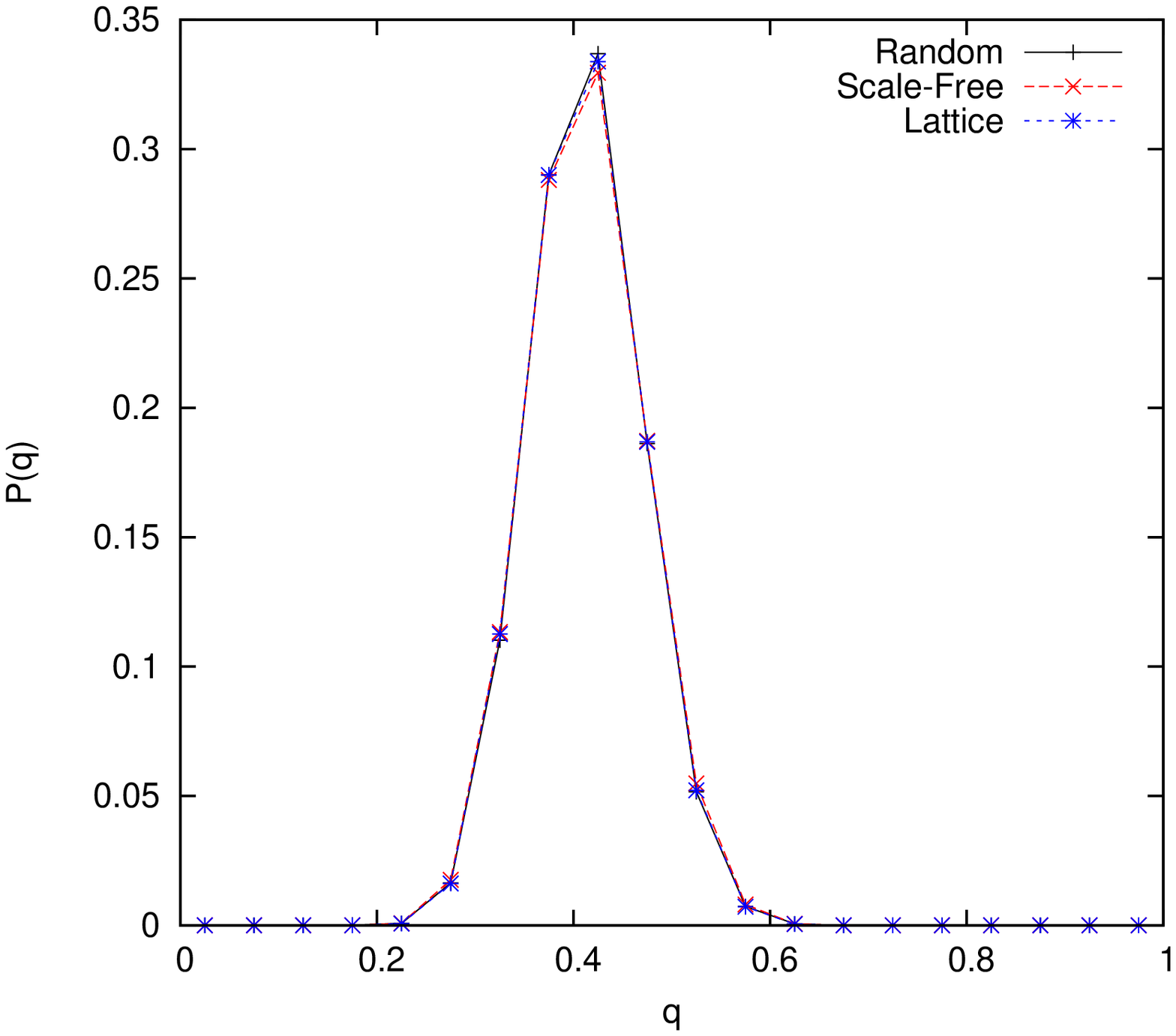}
\includegraphics[width=0.24\textwidth]{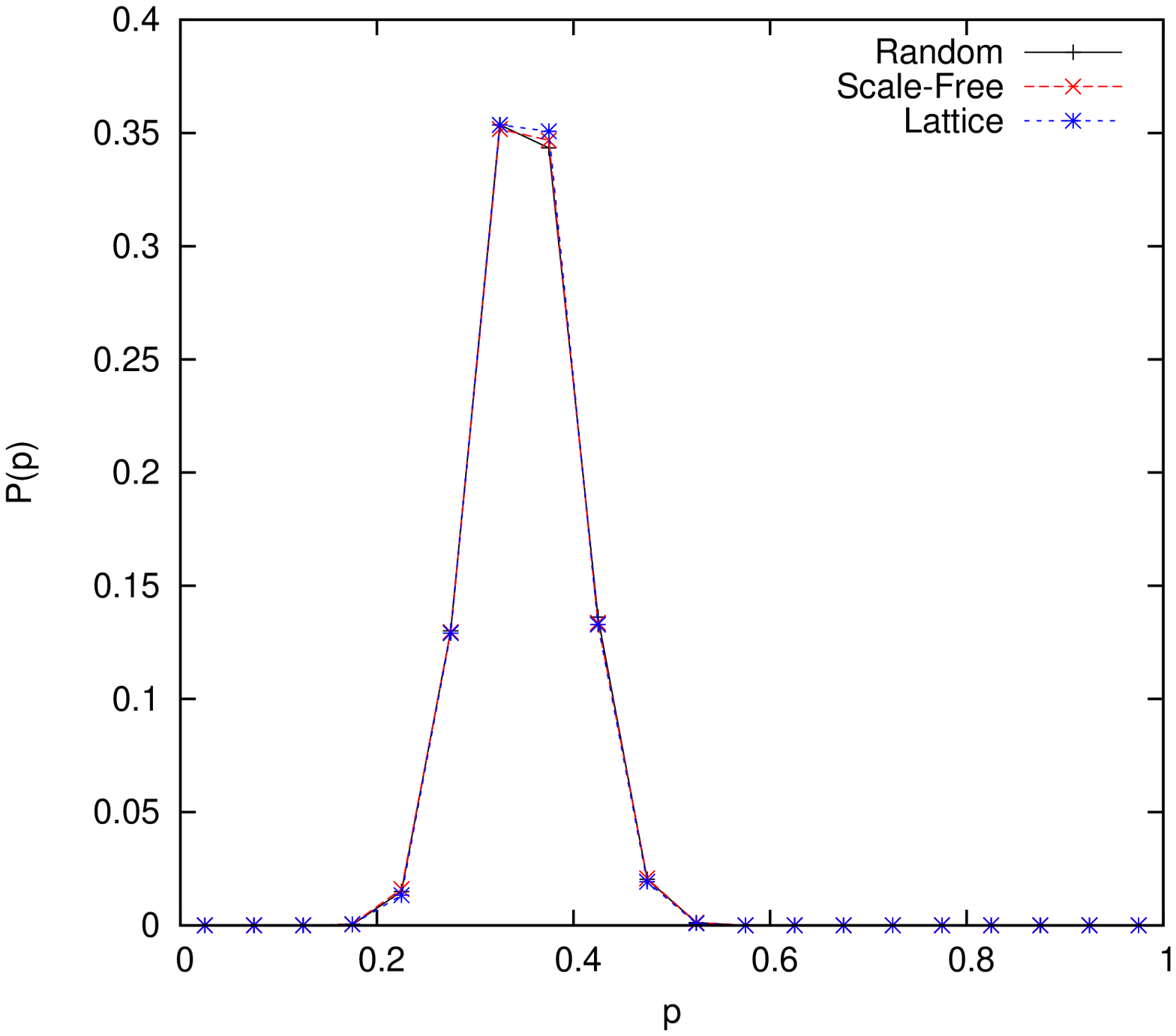}
\includegraphics[width=0.24\textwidth]{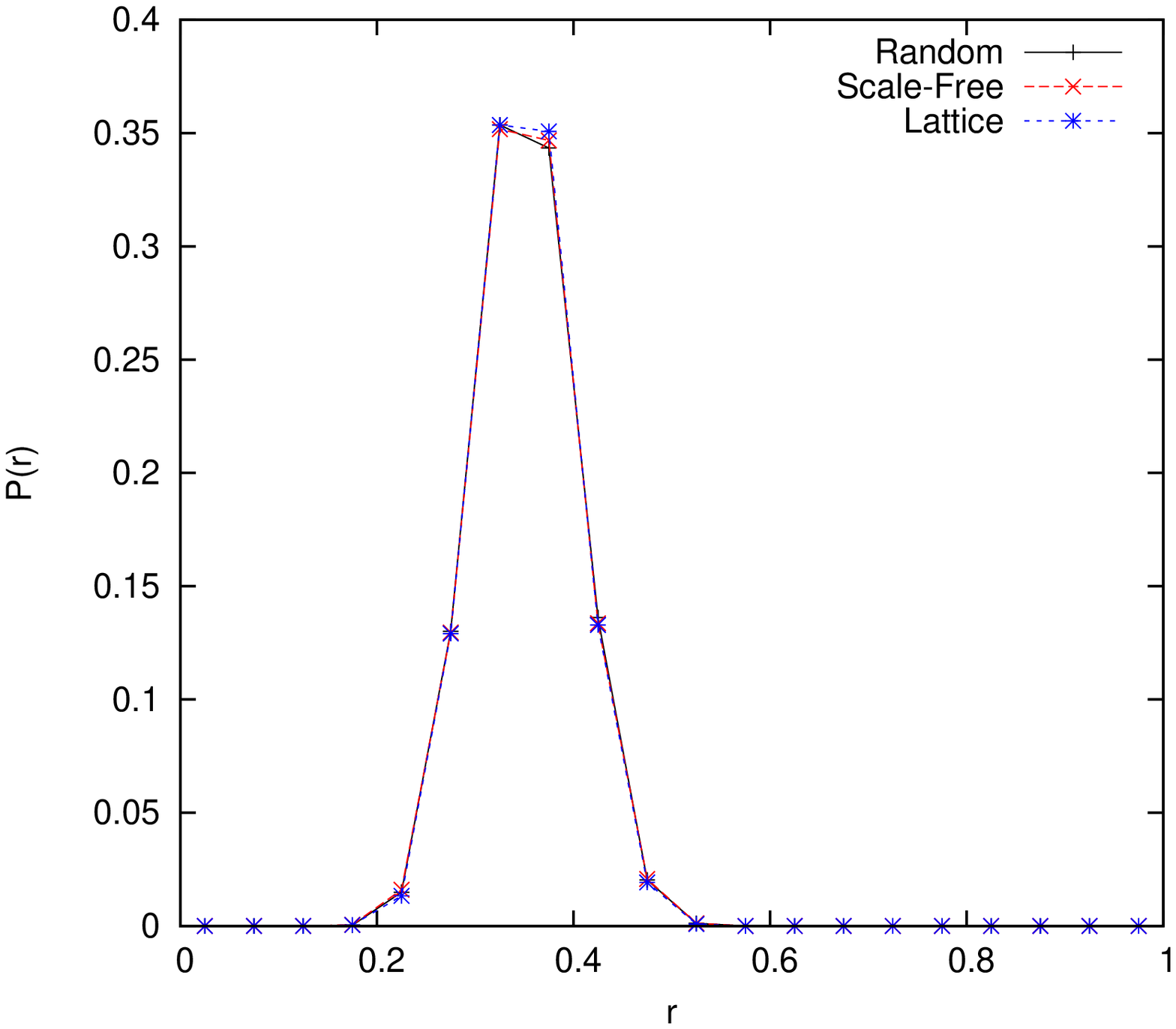}
\newline
\includegraphics[width=0.24\textwidth]{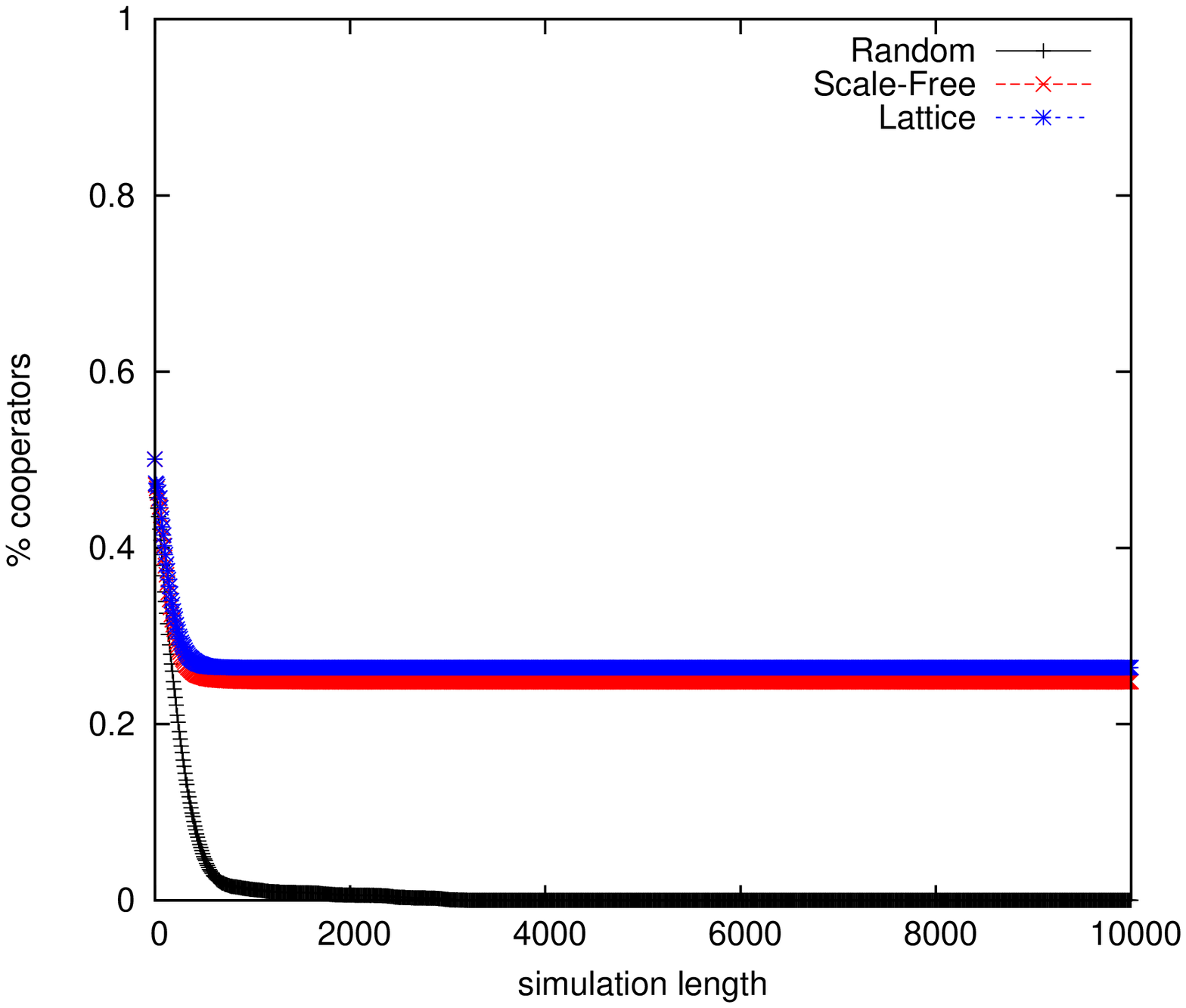}
\includegraphics[width=0.24\textwidth]{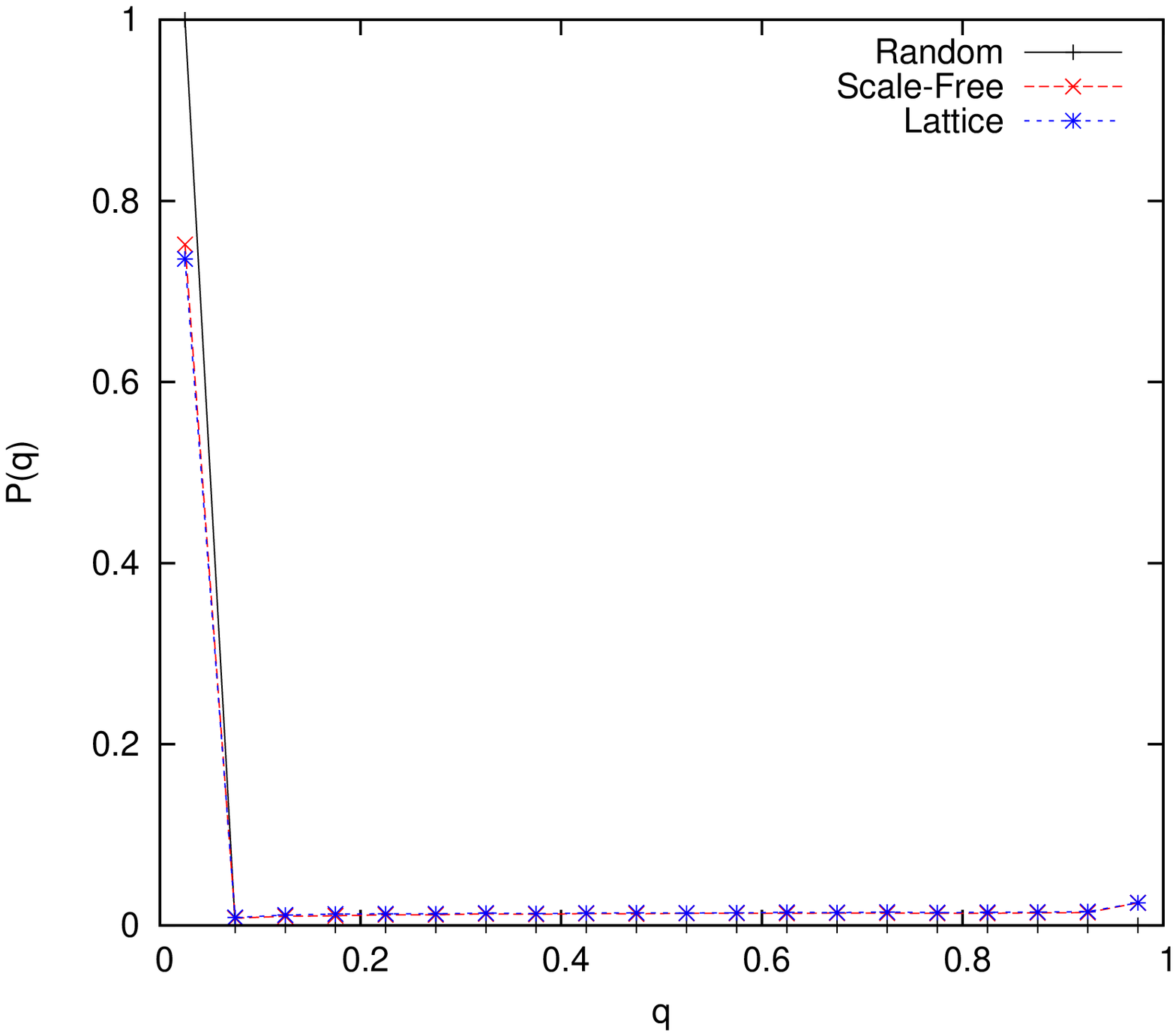}
\includegraphics[width=0.24\textwidth]{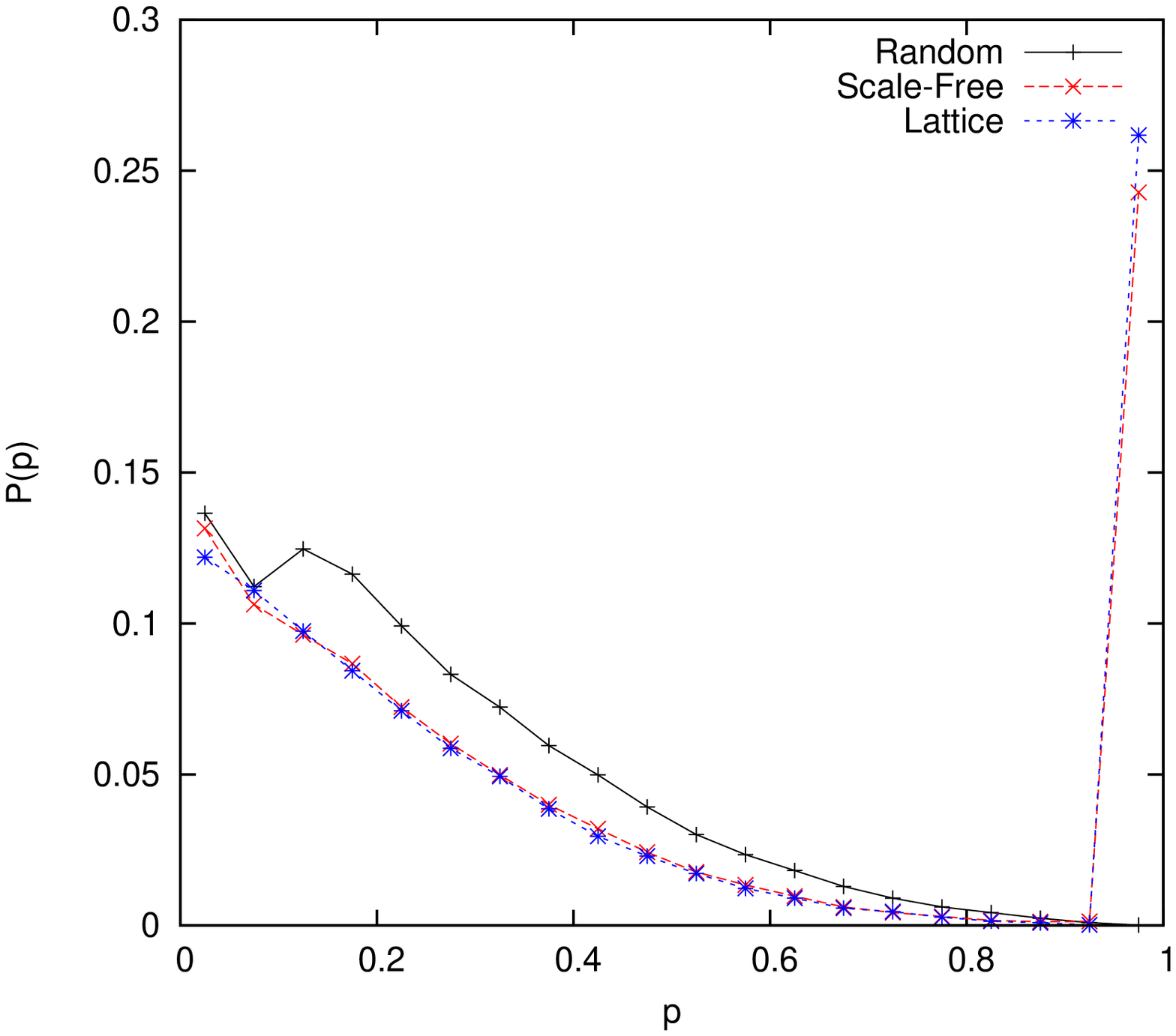}
\includegraphics[width=0.24\textwidth]{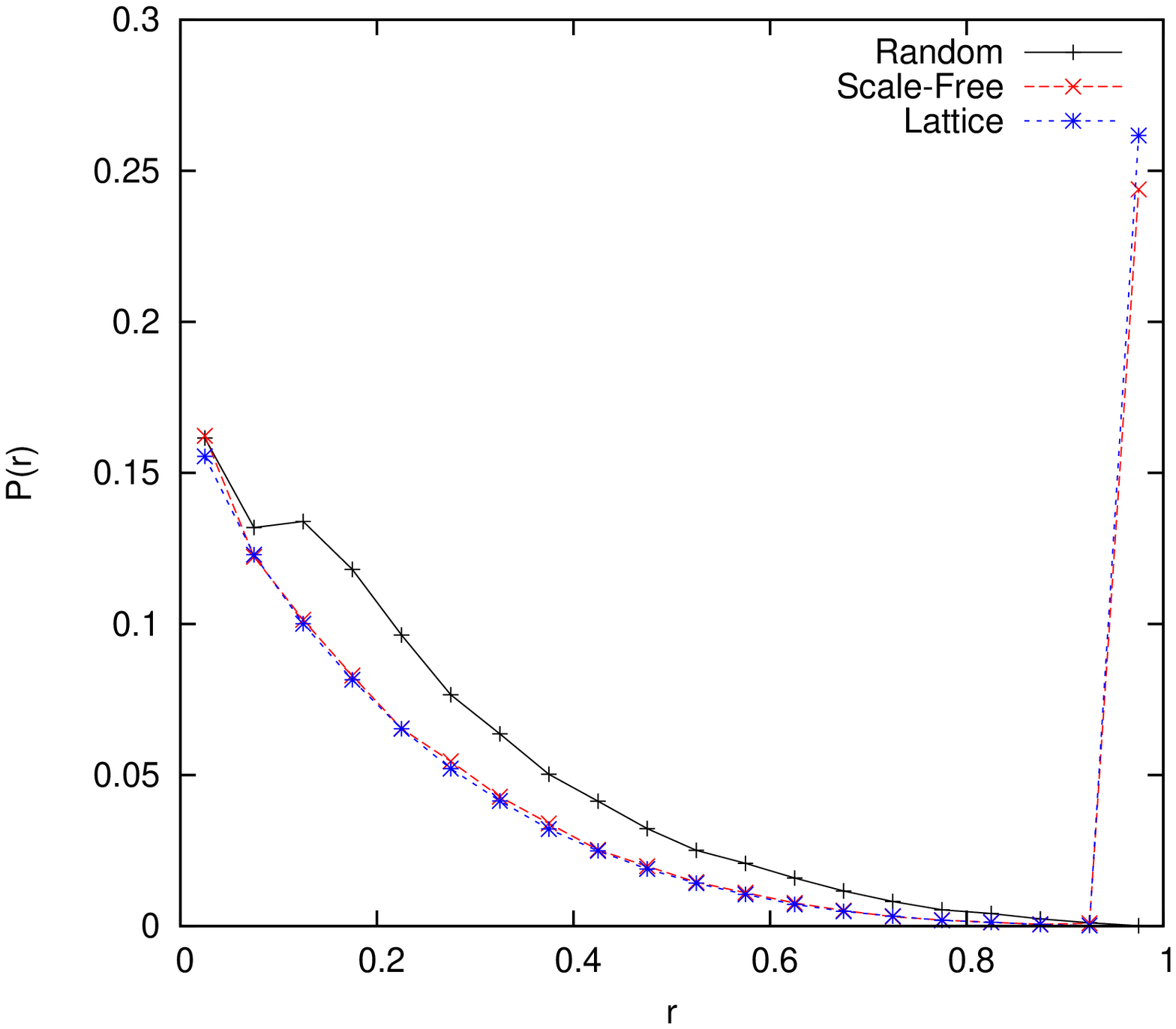}
\newline
\includegraphics[width=0.24\textwidth]{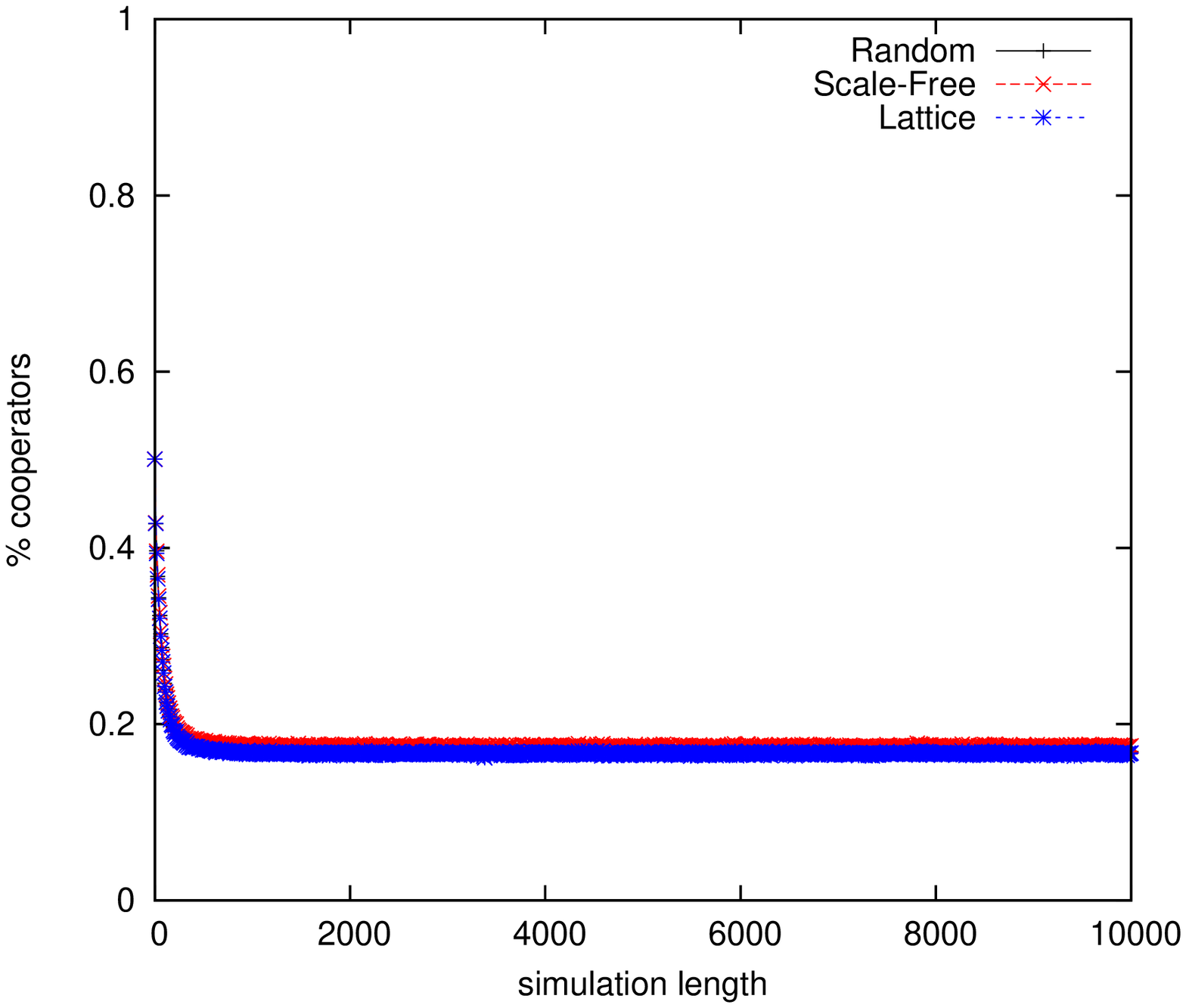}
\includegraphics[width=0.24\textwidth]{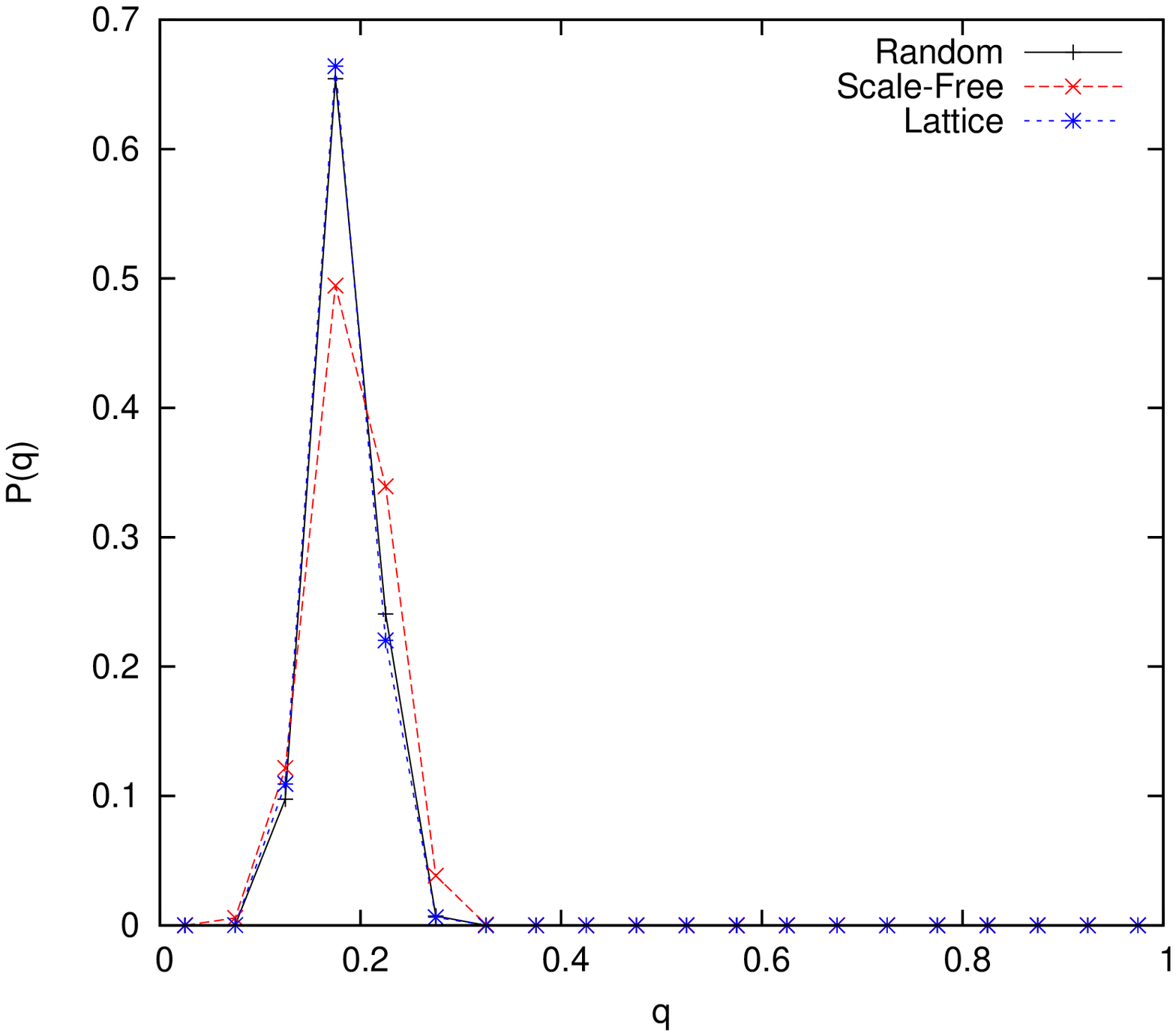}
\includegraphics[width=0.24\textwidth]{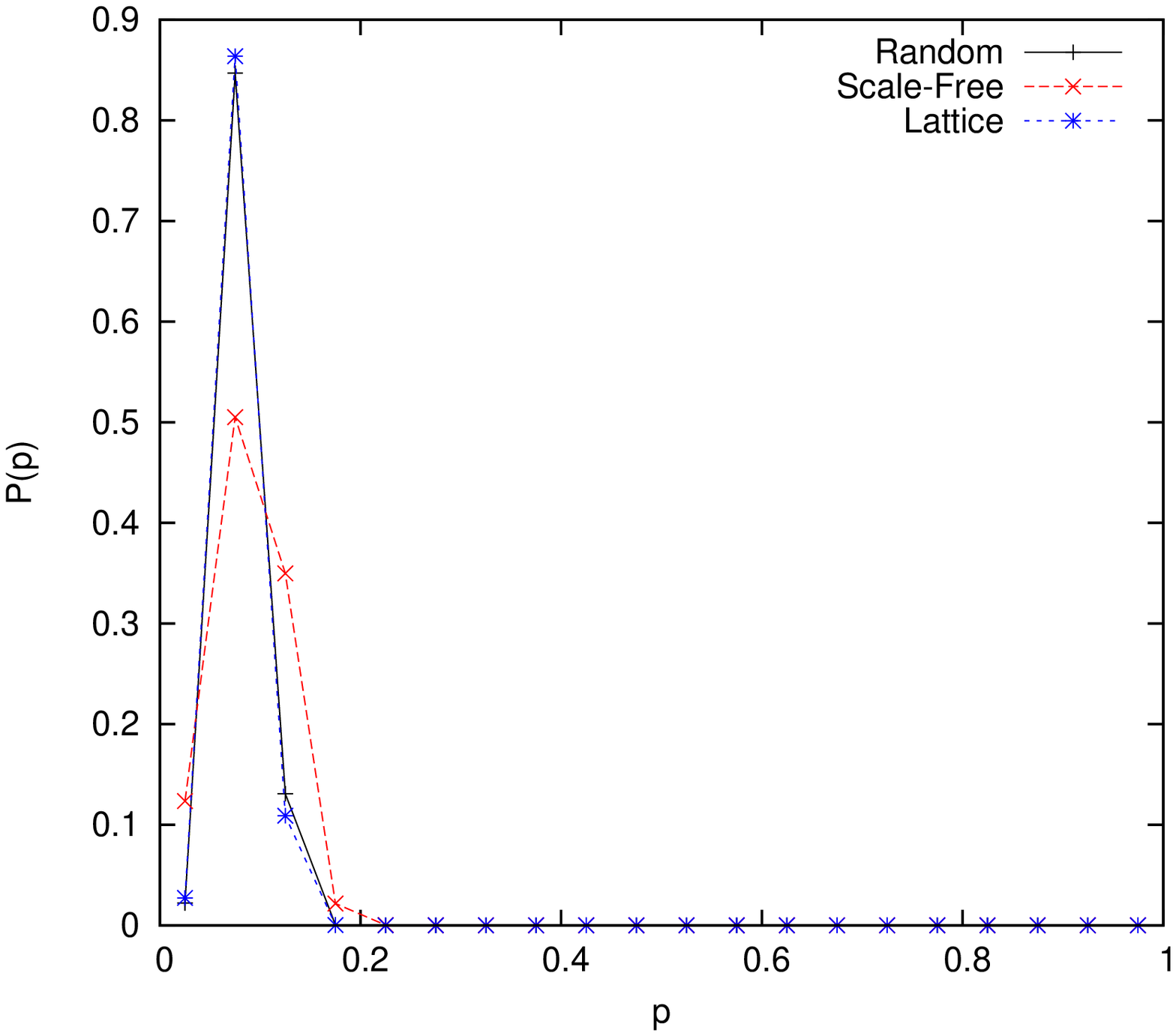}
\includegraphics[width=0.24\textwidth]{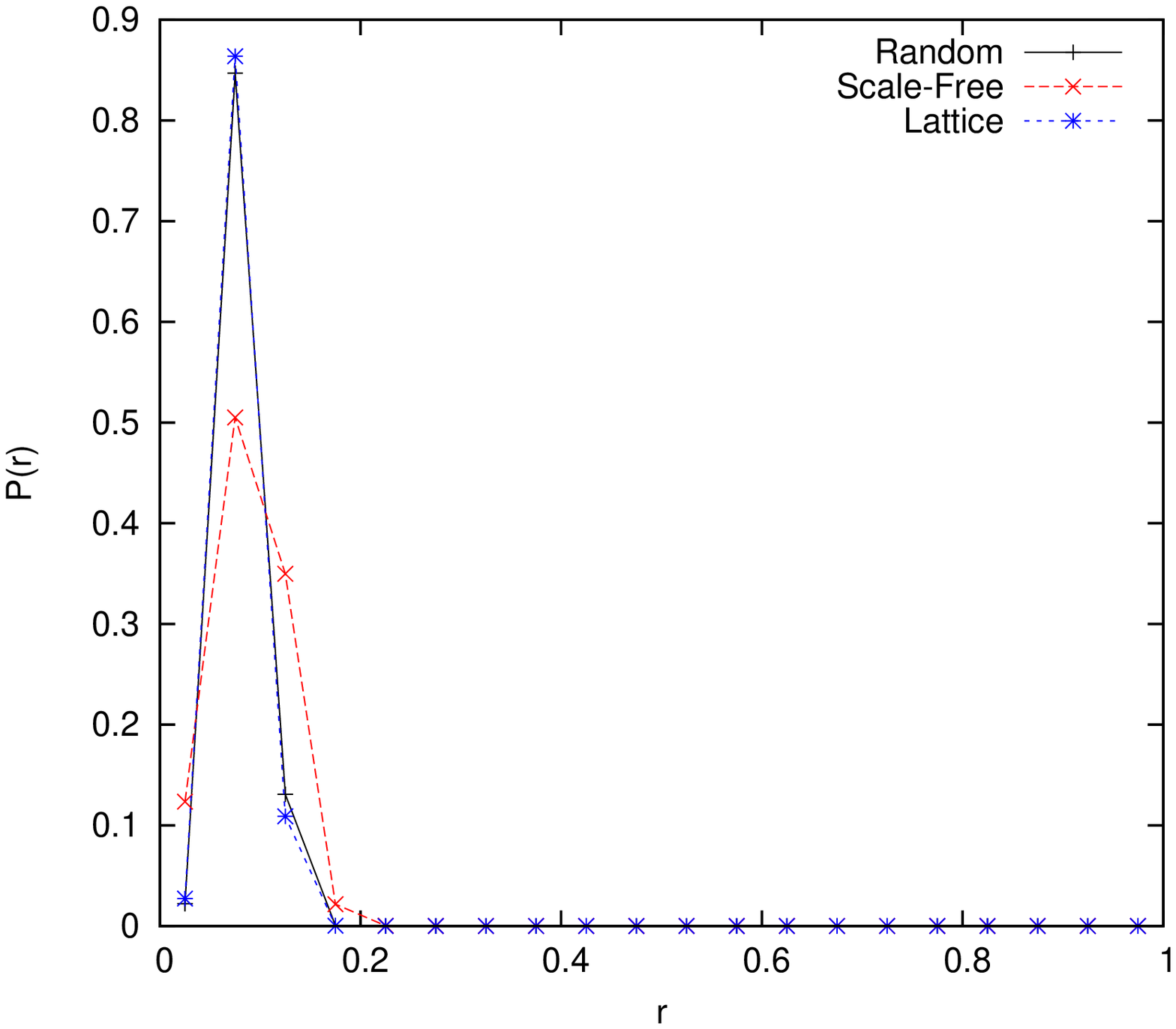}
\caption{Evolution of the level of cooperation (left column) and stationary distributions of MCC parameters (from left to right: $q$, $p$ and $r$) when the evolutionary dynamics is 
Reinforcement Learning with learning rate $\lambda=10^{-1}$. From top to bottom: $A=1/2$, $A=5/4$, $A=-1/4$ and adaptive $A$ with $h=0.2$ and $A^0=1/2$. 
Results are averaged over 100 independent realizations.}\label{fig.C_RL}
\end{figure*}

Our other rule of choice in this type is Reinforcement Learning. We will begin by assuming that aspiration levels $A$ remain fixed in time.  
Our results regarding this rule are presented in Fig.\ \ref{fig.C_RL}. When $A$ is midway between the punishment and reward payoffs ($P<A<R$) we observe a stationary, 
non vanishing level of cooperation around 30\% that does not depend on the population structure. 
This behavior, that is robust with respect to the learning rate $\lambda$, is  in good qualitative agreement with the experimental observations \cite{Grujic2010,Gracia2012b}. 
However the most remarkable outcome of this dynamic is that, contrary to all other update procedures that we have discussed so far, the values of the MCC parameters $\{q,p,r\}$ 
concentrate around some stationary, non-trivial values which are independent on the population structure and on the initial conditions of the system. 
Indeed, we have checked that the stationary values of $\{q,p,r\}$ do not depend on the initial form of their distributions, and also that fixing one of these three parameters 
does not influence the stationary distributions of the others. More importantly, these values are compatible with the ones obtained by linear fits of the aggregate MCC behavior 
extracted from the experiments \cite{Grujic2010,Gracia2012b}. Reinforcement learning thus represents the only mechanism (among those considered here) 
which is able to give rise to evolutionarily stable moody conditional cooperators, while at the same time reproducing the cooperation level and the lack of network reciprocity 
(note that, as we already said, the type of network on which the population sits does not affect the cooperation level). 
It is worth mentioning two other features of this dynamics. First, we have checked that the value of $\lambda$  influences only the convergence rate of the system; 
however, if players learn too rapidly ($\lambda\sim1$) then the parameters change too quickly and too much to reach stationary values---a phenomenon typical 
of this kind of learning algorithms. Second, if we introduce in the system a fraction $d$ of players who always defect (recall that full defectors coexist with moody conditional cooperators 
in the experiments), what happens is that the final cooperation level changes---it drops to 25\% for $d=0.2$ and to 20\% for $d=0.4$---but the stationary distributions 
of MCC parameters are not affected. This means that Reinforcement Learning is able to account for the heterogeneity of the behaviors observed in the experimental populations, 
which is consistent with the fact that this update rule does not take into account either the payoffs or the actions of the rest of the players. 

Further evidence for the robustness of the Reinforcement Learning evolutionary dynamics arises from extending our study to other aspiration levels, including dynamic ones. 
In general, what we observe is that the higher $A$, the higher the final level of cooperation achieved. 
When $R<A<T$ players are not satisfied with the reward of mutual cooperation; however an outcome of mutual defection leads to a great stimulus towards cooperation in the next round. 
This is why players' parameters tend to concentrate around values that allow for a strategy which alternates cooperation and defection, and that brings to stationary cooperation levels 
around 50\%. Instead if $S<A<P$, then defection-defection is a satisfactory outcome for each pair of players. In this case cooperation may thrive only on stationary networks 
(where clusters of cooperator may form). However for a well-mixed population the final state is necessarily fully defective ($q\rightarrow 0$). 
Hence we observe in this case a dependence on the network structure which is not observed in the experiments; nonetheless, setting an aspiration level below punishment 
is at least questionable. Therefore, unless players make very strange decisions on their expectations from the game, we find behaviors that agree qualitatively with the experiments. 
Finally, we consider the case in which players adapt their aspiration level after each round: $A^{t+1}\leftarrow (1-h)A^t+h\pi^t/k$, where $h$ is the adaptation 
(or habituation) rate and $P<A^0<R$. What we observe now is that the stationary level of cooperation lies around 20\%, the absence of network reciprocity is recovered, 
and players' average aspiration levels remain in the range $P<\bar{A}<R$. Thus this case is again compatible with experimental observations, and the fact that aspiration levels 
of an intermediate character are selected (corresponding to the case that better describes them) provides a clear rationale for this choice in the preceding paragraph. 

A final important validation of Reinforcement Learning comes from studying the \emph{EWA} (experience-weighted attraction) updating \cite{Camerer1999}, 
an evolutionary dynamics that combines aspects of Belief Learning models (to which Best Response belongs) and of Reinforcement Learning. 
Results for this choice of the updating scheme (which are reported in SI EWA) confirm in fact that Reinforcement Learning is the determinant contribution 
which allows to achieve situations matching with empirical outcomes.

\section*{Conclusion}

Understanding cooperation is crucial because all major transitions in evolution involve the spreading of some sort of cooperative behavior \cite{MaynardSmith1995}. 
In addition, the archetypical  tensions that generate social dilemmas are present in fundamental problems of the modern world: resource depletion, pollution, overpopulation, 
and climate change. This work, inspired by experiments \cite{Grujic2010,Gracia2012b}, aimed at finding an evolutionary framework capable of modeling and justifying 
real people behavior in an important class of social dilemmas---namely Prisoner's Dilemma games.
To this end, we have studied the evolution of a population of differently-parameterized MCC whose parameters can evolve. 
We have considered several rules for parameters' changes---both of imitative nature and innovative mechanisms, as well as rules based on payoff comparison 
and others based on non-economic or social factors. 
Our research shows that Reinforcement Learning with a wide range of learning rates is the only mechanism able to explain the evolutionary stability of moody conditional cooperation, 
leading to situations that are in agreement with the experimental observations in terms of the stationary level of cooperation achieved, 
average values and stationary distributions of the MCC parameters, and absence of network reciprocity. 
Note that we have considered only PD games; however, given that in our setup players have to play the same action with all their neighbors, 
it is clear that our results should be related to Public Goods experiments (where conditional cooperation was first observed \cite{Fischbacher2001}). 
Our findings thus suggest that MCC can also arise and be explained through reinforcement learning dynamics in repeated Public Goods games. 

We stress that this is a very relevant result, as for the first time to our knowledge we are providing a self-consistent picture of how people behave in PD games on networks. 
Indeed, starting from the observation that players do not take others' payoffs into account, we find that if this behavior is to be explained in an evolutionary manner, 
it has to be because people learn from what they experience, and not from the information they may gather on their neighbors. Such a learning process is in turn very sensible 
in the heavily social framework in which we as humans are embedded, and compatible with the knowledge that we have on the effects of our choices on others. 
On the other hand, the evolutionary dynamics that our work eliminates as possible responsible for how we behave are, in fact, difficult to justify in the same social context, 
either because they would require a larger cognitive effort (Best Response) or, on the contrary, because they assume a very limited rationality that only allows to imitate 
without reflecting on how we have been affected by our choices. Our work thus provides independent evidence that, at least in the context of human subjects interacting in 
PD, the observed behaviors arise mostly from learning. Of course, this does not mean that other ways to update one's strategy are not possible: 
indeed, a large fraction of people have been observed to be full defectors, a choice they may have arrived at by considering the PD game from a purely rational viewpoint. 
In addition, specific individuals may behave in idiosyncratic manners that are not described within our framework here. Still, as we have seen, our main result, namely that 
Reinforcement Learning explains the behavior of a majority of people and its macro-consequences (level of cooperation, lack of network reciprocity) would still hold true 
in the presence of these other people. 

Although a generalization of our results to other classes of social dilemma beyond PD and Public Goods is not straightforward, our 
conclusions here should guide further research on games on networks. We believe that the experimental results, to which the present work provides a firm theoretical support, 
allow to conclude that many of the evolutionary dynamics used in theory and in simulations simply do not apply to the behavior of human subjects and, therefore, their use should be avoided. 
As a matter of fact, much of the research published in the last decade by using all these update schemes is only adding confusion to an already very complicated problem. 
Even so, our findings do not exclude the plausibility of other strategy updating in different contexts. For instance, analytical results with imitative dynamics \cite{Wu2010} 
display an agreement with experimental outcomes on dynamical networks \cite{Fehl2011}, where it was also shown that selection intensity (which can be thought as a measure of players' rationality) can dramatically alter the evolutionary outcome \cite{VanSegbroeck2011}. It is also important to stress that our findings here relate to human behavior, and other species could behave differently; for instance, it has been recently reported that bacteria improve their cooperation on a spatial structure \cite{Hol2013} and this could arise because of more imitative 'strategies'. Finally, a promising line of research could be to compare the distribution of values for the MCC parameters that we have obtained here with the observations 
on single individuals, thus going beyond the check agains aggregate data to address the issue of reproducing whole histograms. Unfortunately, the data that we currently have is not good 
in terms of individual behavior, as observations are noisy and statistics is insufficient to assign significant values of the parameters to single participants. In this respect, 
experiments specifically designed to overcome this difficulty could be a very relevant contribution to further verifying our claims.

Another important suggestion arising from our research is the relevance of theoretical concepts derived within Reinforcement Learning to the study of games on networks. 
In this respect, it is very interesting to recall that a theoretical line of work based on Reinforcement Learning models for 2-player repeated games has received quite 
some attention recently \cite{Erev2001,Bendor2001b}. In this context, a generalized equilibrium concept has been introduced in order to explain the findings in simulations 
of 2-player PD \cite{Macy2002,Izquierdo2008}, called self-correcting equilibrium: it obtains when the expected change of parameters is zero but there is a positive probability 
to incur into a negative as well as positive stimulus. The extension of the Reinforcement Learning dynamics to multiplayer PD that we have presented here points to the explanatory power 
of such equilibrium concepts in the framework of network games, as the level of cooperation observed in experiments is in close agreement with the predicted equilibrium. 
Importantly, it has recently been shown that behavioral rules with intermediate aspiration levels, as the ones we find here to be relevant, are the most successful ones among 
all possible reactive strategies in a wide range of 2-player games \cite{Vaquero2012}. This suggests that this type of evolutionary dynamics may indeed be relevant in general. 
It would therefore be important to study whether or not the associated equilibrium concept is also the most important one when other types of games are played on an underlying network. 
If that is the case, we would have a very powerful tool to understand and predict human behavior in those situations. 
\newline\newline

\textbf{Materials and Methods} --- Agent-based simulations of the model were carried out using the following parameters:
$c_0=0.5$ (initial fraction of cooperators) \cite{foot2}, 
$R=1$, $P=0$, $S=-1/2$, $T=3/2$ (entries of the PD's payoff matrix, such that $T>R$, $S<P$ and $2R>T+S$) \cite{foot3},
$N=1000$ and $m=10$ (network parameters) \cite{foot4}.
The MCC behavioral parameters $\{q,p,r\}$ are all drawn for each player before the first round of the game from a uniform distribution $\mathcal{U}[0,1]$, 
with the additional constraint $p+r\le1$ to have $0\le p_C(x)\le1$. 
Note that the particular form of the initial distribution as well as the presence of the constraint does not influence the outcome of our experiments. 
\newline\newline


\begin{acknowledgments}
This work was supported by the Swiss Natural Science Fundation through grant PBFRP2\_145872, by Ministerio de Econom\'\i a y Competitividad (Spain) through grant PRODIEVO, by the ERA-Net on Complexity through grant RESINEE, and by Comunidad de Madrid (Spain) through grant MODELICO-CM.
\end{acknowledgments}

\newpage
\newpage

\section*{SI Materials and Methods}

Here we give the details of the evolutionary dynamics that we consider in order to update the MCC behavioral parameters $\{q,p,r\}$. 
It is important to notice that, differently from the traditional approach used in game theory where players are simply described by their individual probabilities of cooperating and defecting, 
in our case players' strategies are defined by their three MCC parameters that determine such probabilities. 
Hence the strategy update rules traditionally employed in the literature have to be modified accordingly.
\newline\newline
For imitative rules, a given player $i$ adopts a new strategy by copying all the MCC parameters from a selected counterpart $j$, which is one of the $|k_i|$ neighbors of $i$. 

\emph{Proportional Imitation} --- $j$ is chosen randomly, but the probability that $i$ copies $j$'s parameters 
depends on the difference between the payoffs that they obtained in the previous round of the game through the expression
$$\mathcal P\left\{\{q_j,p_j,r_j\}^t\rightarrow\{q_i,p_i,r_i\}^{t+1}\right\}=\begin{cases}
	(\pi_j^t-\pi_i^t)/\Phi_{ij}&\mbox{if $\pi_j^t>\pi_i^t$}\\
	0&\mbox{otherwise}
	\end{cases}$$
with $\Phi_{ij}=\max(k_i,k_j)[\max(R,T)-\min(P,S)]$ to ensure $\mathcal P\{\cdot\}\in[0,1]$. 
This rule is well known in the literature as it brings---for a large, well-mixed population---to an evolutionary equation which is equal to that of replicator dynamics.

\emph{Fermi rule} --- as in Proportional imitation, $j$ is chosen randomly, but the probability that $i$ copies $j$'s parameters depends now on the payoff difference 
according to the Fermi distribution function:
$$\mathcal P\left\{\{q_j,p_j,r_j\}^t\rightarrow\{q_i,p_i,r_i\}^{t+1}\right\}=\frac{1}{1+\exp[-\beta\,(\pi_j^t-\pi_i^t)]}$$
Note that the Fermi Rules allows for mistakes: players can copy the parameters of others who are performing worse.
The Fermi rule has been widely used in the literature because of being analytically tractable. 

\emph{Death-Birth rule} --- player $i$ copies the parameters of one of her neighbors $j$, or herself's, with a probability proportional to the payoffs
$$\mathcal P\left\{\{q_j,p_j,r_j\}^t\rightarrow\{q_i,p_i,r_i\}^{t+1}\right\}=\frac{\pi_j^t-\psi}{\sum_{k\in \mathcal{N}_i^*}\pi_k^t-\psi}$$
where $\mathcal{N}_i^*$ is the set which includes $i$ and her neighbors and $\psi=\max_{j\in \mathcal{N}_i^*}(k_j)\min(0,S)$ to ensure $\mathcal P\{\cdot\}\in[0,1]$. 
Again with this rule a player can adopt, with low probability, the parameters of a neighbor that has done worse than herself.
  
\emph{Unconditional Imitation} or ``Imitate the Best'' --- this rule makes each player $i$ copy the parameters of the neighbor $j$ with the largest payoff, 
provided this payoff is greater than the player's:
$$\mathcal P\left\{\{q_j,p_j,r_j\}^t\rightarrow\{q_i,p_i,r_i\}^{t+1}\right\}=1\quad\mbox{ if }j:\pi_j^t=\max_{k\in \mathcal{N}_i^*}\pi_k^t$$
 
\emph{Voter model} --- $i$ simply copies the parameters of a neighbor $j$ selected at random. 
Such update rule represents an imitation mechanism of purely social nature, which thus incorporates the effect of the social pressure. 
\newline\newline
For innovative rules instead:

\emph{Best Response} --- Here every player chooses her MCC parameters as a best response to what her neighbors did in the last round. 
This means that each player $i$, given $x_i^t$ from the previous round $t$, computes the payoffs that she would have obtained by cooperating or defecting, respectively:
$$\pi_i^t(C)=R\,x_i^t+S\,(1-x_i^t)\qquad;\qquad\pi_i^t(D)=T\,x_i^t+P\,(1-x_i^t)$$
Then if in the previous round $i$ defected, she tries to increase the quantity
$$\pi_i^t(\cdot|D)=P^t(C|D)\,\pi_i^t(C)+P^t(D|D)\,\pi_i^t(D)=q_i^t\,\pi_i^t(C)+(1-q_i^t)\,\pi_i^t(D)$$
To do so, the players uses her current value of $q_i^t$ as well as two ``shifted'' values $q_i^t-\delta$ and $q_i^t+\delta$, 
and pick as her new $q_i^{t+1}$ the one that maximizes $\pi_i^t(\cdot|D)$ (and which satisfies $0\le q_i^{t+1}\le 1$).
Instead if in the previous round $i$ cooperated, the quantity that she tries to increase is
$$\pi_i^t(\cdot|C)=P^t(C|C)\,\pi_i^t(C)+P^t(D|C)\,\pi_i^t(D)=(p_i^t\,x_i^t+r_i^t)\,\pi_i^t(C)+(1-p_i^t\,x_i^t-r_i^t)\,\pi_i^t(D)$$
To do so, she uses the current values of $(p_i^t,r_i^t)$ as well as the four combinations $(p_i^t-\delta,r_i^t)$, $(p_i^t+\delta,r_i^t)$, $(p_i^t,r_i^t-\delta)$, $(p_i^t,r_i^t+\delta)$,
and chooses as new parameters $(p_i^{t+1},r_i^{t+1})$ the ones that maximize $\pi_i^t(\cdot|C)$ (and which satisfy $0\le p_i^{t+1}+r_i^{t+1}\le 1$).

Note that we do not use exhaustive Best Response (which consists in choosing the values of $q$ or $p,r$ that maximize $\pi_i^t(\cdot|D)$ or $\pi_i^t(\cdot|C)$, respectively) 
as it would lead immediately to $P(C|C)=P(C|D)=0$ (\emph{i.e.}, to $q=p=r=0$), which is the Nash equilibrium of PD games. 
Note also that Best Response belongs to a family of updating rules often referred to as \emph{Belief Learning} models. According to this family of rules, 
players update beliefs about what others will do based on history, and then use those beliefs to determine which strategies lead to the best outcome. 
Best Response is restrictive in considering only last round's choices to determine the strategies. While in principle more information can be used as well, 
we only restrict our attention to Best Response for three main reasons. Firstly, to have a fair comparison to the other dynamics, that only use last round's information 
(being it either actions or payoffs). Secondly because, in our non-exhaustive formulation of Best Response, history is contained in the current values of the MCC parameters. 
Thirdly because in PD the dominant strategy is to defect always, so that full defection becomes the final outcome whatever information is used to build beliefs about neighbors' actions.

\emph{Reinforcement Learning} --- When learning, players use only information about their own past choices and payoffs. 
Parameters updating takes place in three steps. First, after each round $t$ of the game each player $i$ calculates her \emph{stimulus} 
$s_i^t$ as $$s_i^t=\frac{\pi_i^t/k_i-A_i^t}{\max\{|T-A_i^t|,|R-A_i^t|,|P-A_i^t|,|S-A_i^t|\}}$$
where $A_i^t$ is the current \emph{aspiration level} of player $i$, and the normalization of the stimulus assures $|s_i^t|\le1$ $\forall i,t$. 
Second, each player updates her MCC parameters. Note however that, whereas Reinforcement Learning for game theory is usually modeled as a stochastic process, 
in our case we have to modify the classical algorithm to account for the two-step memory of moody conditional cooperators. 
Hence the parameters updating depends on the actions chosen at time steps $t$ and $t-1$. There are four cases:
\begin{enumerate}
 \item if $i$ defected at $t-1$ and cooperated at $t$, 
	$$q_i^{t+1}=
	\begin{cases}
	q_i^t+\lambda s_i^t(1-q_i^t)&\quad\mbox{if $s_i^t>0$}\\
	q_i^t+\lambda s_i^tq_i^t&\quad\mbox{if $s_i^t<0$}
	\end{cases}$$
  \item if $i$ defected at $t-1$ and defected at $t$, 
	$$q_i^{t+1}=
	\begin{cases}
	q_i^t-\lambda s_i^t(1-q_i^t)&\mbox{if $s_i^t<0$}\\
	q_i^t-\lambda s_i^tq_i^t&\mbox{if $s_i^t>0$}
	\end{cases}$$
 \item if $i$ cooperated at $t-1$ and cooperated at $t$, 
	$$p_i^{t+1}=
	\begin{cases}
	p_i^t+\lambda s_i^t(1-p_i^t)\\
	p_i^t+\lambda s_i^tp_i^t
	\end{cases}
	;\quad
	r_i^{t+1}=
	\begin{cases}
	r_i^t+\lambda s_i^t(1-r_i^t)&\quad\mbox{if $s_i^t>0$}\\
	r_i^t+\lambda s_i^tr_i^t&\quad\mbox{if $s_i^t<0$}
	\end{cases}$$
 \item if $i$ cooperated at $t-1$ and defected at $t$, 
	$$p_i^{t+1}=
	\begin{cases}
	p_i^t-\lambda s_i^t(1-p^t_i)\\
	p_i^t-\lambda s_i^tp_i^t
	\end{cases}
	;\quad
	r_i^{t+1}=
	\begin{cases}
	r_i^t-\lambda s_i^t(1-r_i^t)&\quad\mbox{if $s_i^t<0$}\\
	r_i^t-\lambda s_i^tr_i^t&\quad\mbox{if $s_i^t>0$}
	\end{cases}$$
\end{enumerate}
where $\lambda\in(0,1]$ represents the learning rate---accounting for different $\lambda$s covers for both the cases of slow and fast learning. 
Finally, player $i$ may adapt her aspiration level as $A_i^{t+1}=(1-h)A_i^t+h\pi_i^t/k_i$, where $h\in[0,1)$ is the adaptation (or habituation) rate.

\section*{SI Results} 

We now discuss the robustness of our results with respect to two additional features of our numerical study: 
spontaneous mutations of MCC parameters, as well as a different modeling of the MCC behavior.

\emph{Mutations} --- Spontaneous mutations of strategies (genotypes) represents an important aspect of evolutionary game theory, 
particularly in the modeling of evolving populations of life forms. In order to validate our findings against mutations, we introduce, for each player, 
a probability $\mu$ that (after every strategy updating takes place) one of her MCC behavioral parameters, chosen randomly, 
varies of a quantity drawn from a normal distribution $N[0,\sigma]$. Such variation is yet bounded to respected the constraints $p_D\in[0,1]$ and $p_C(x)\in[0,1]$. 
We observe that the introduction of spontaneous mutations does not have a significant impact on the system's behavior, unless the amount ($\sigma$) and frequency ($\mu$) 
of such mutations become dominant with respect to the changes of MCC parameters given by the strategy updating. In general, the major effect of mutations is 
a noisier and slower evolution (which also causes the stationary distributions of MCC parameters to remain broader and smoother). Moreover, the consequence of imposing a constraint 
on mutations is that the cooperation level increases when it was originally close to zero, and decreases when it was close to one. 
In particular, the fully defective state becomes inaccessible because of a minimal amount of cooperation, caused by mutations, which is proportional to $\sigma$ and $\mu$. 
The situation which is mostly affected by mutations is games played on lattices with the Death-Birth rule, where cooperation becomes unstable even for small values of $\mu$.
Notably, also in the presence of mutations, Reinforcement Learning remains the only evolutionary dynamics which can successfully reproduce all of the experimental outcomes.

\emph{Modified MCC} --- In this work, according to experimental inputs, we have modeled players as moody conditional cooperators. 
The reader should not be misled to think that this way of modeling distinguishes players from each other. In fact, $p_D$ does not represent the strategy of a defector, 
but just the probability of cooperating after having defected---with no reference to a hypothetical player kind. Equivalently, $p_C(x)$ is not the strategy of a cooperator, 
but just the probability of cooperating after having cooperated.
Aside from this consideration, it is indeed possible to think of introducing a dependence on observed cooperation in the probability of cooperating after having defected: 
$p_D(x)=s\,x+q$. While this choice does not agree well with experimental evidence, it brings to an elegant symmetry between $p_C(x)$ and $p_D(x)$; moreover, 
the original MCC behavior can be still recovered for $s\rightarrow0$. Therefore we extend our analysis to this modified MCC behavior, 
observing that the cooperation level $c$ does not change significantly with respect to what was obtained for $s=0$, irrespectively of the update rule employed. 
Naturally what changes is the stationary distributions of the parameter $q$ (as it is now coupled with $s$). 
Concerning imitative dynamics, as expected we do not observe a stationary distribution of $s$; additionally, full defection outcomes are still due to $q\rightarrow0$ 
(which, together with $x\rightarrow0$, makes $s$ irrelevant). For Best Response, still $c\rightarrow0$ together with $q$ and $r$; remarkably, 
in this case also $s\rightarrow0$ (differently from $p$, which remains broadly distributed). Finally, a stationary, non-trivial distribution of $s$ is again obtained only with 
Reinforcement Learning. Note that $s$ remaining finite (but small) in this case is probably due to the particular formulation of the learning algorithm proposed here.

\section*{SI EWA}

Here we present methods and results for the EWA (experience-weighted attraction) updating scheme. 
We recall that EWA is an evolutionary dynamics that combines Belief Learning (to which Best Response belongs) and Reinforcement Learning. 
While the original formulation of EWA cannot be trivially generalized to our MCC scenario---where multiple parameters and neighbors' actions regulate the strategies, 
we can still reproduce the key features of the EWA updating by a simple linear combination of Best Response and Reinforcement Learning. 
Indeed, such formulation updates the strategies exactly like the original EWA does. At the same time, since we do not aim at quantitatively reproducing experimental outcomes, 
we do not need to impose any initial attractors, nor any particular experience and growth rate. We thus implement the EWA dynamics as follows. 
For each player $i$, at each updating $t$ we compute the shift of MCC parameters given by Best Response and Reinforcement Learning, which we denote as 
$\{\delta q_{i(BR)}^t,\delta p_{i(BR)}^t,\delta r_{i(BR)}^t\}$ and $\{\delta q_{i(RL)}^t,\delta p_{i(RL)}^t,\delta r_{i(RL)}^t\}$, respectively. Then parameters are updated as:
\begin{eqnarray*}
q_i^{t+1}&=&q_i^t+\gamma\,\delta q_{i(RL)}^t+(1-\gamma)\,\delta q_{i(BR)}^t \nonumber \\
p_i^{t+1}&=&p_i^t+\gamma\,\delta p_{i(RL)}^t+(1-\gamma)\,\delta p_{i(BR)}^t \nonumber \\
r_i^{t+1}&=&r_i^t+\gamma\,\delta r_{i(RL)}^t+(1-\gamma)\,\delta r_{i(BR)}^t \nonumber
\end{eqnarray*}
where $\gamma\in(0,1)$ is the mixing parameter. 
Note that $\delta_{(BR)}=\delta$ and $\delta_{(RL)}\sim\lambda$ (as the stimulus is such that $|s_i^t|<1$), thus for the two terms to be comparable we usually choose $\delta=\lambda$.

\begin{figure*}
\includegraphics[width=0.24\textwidth]{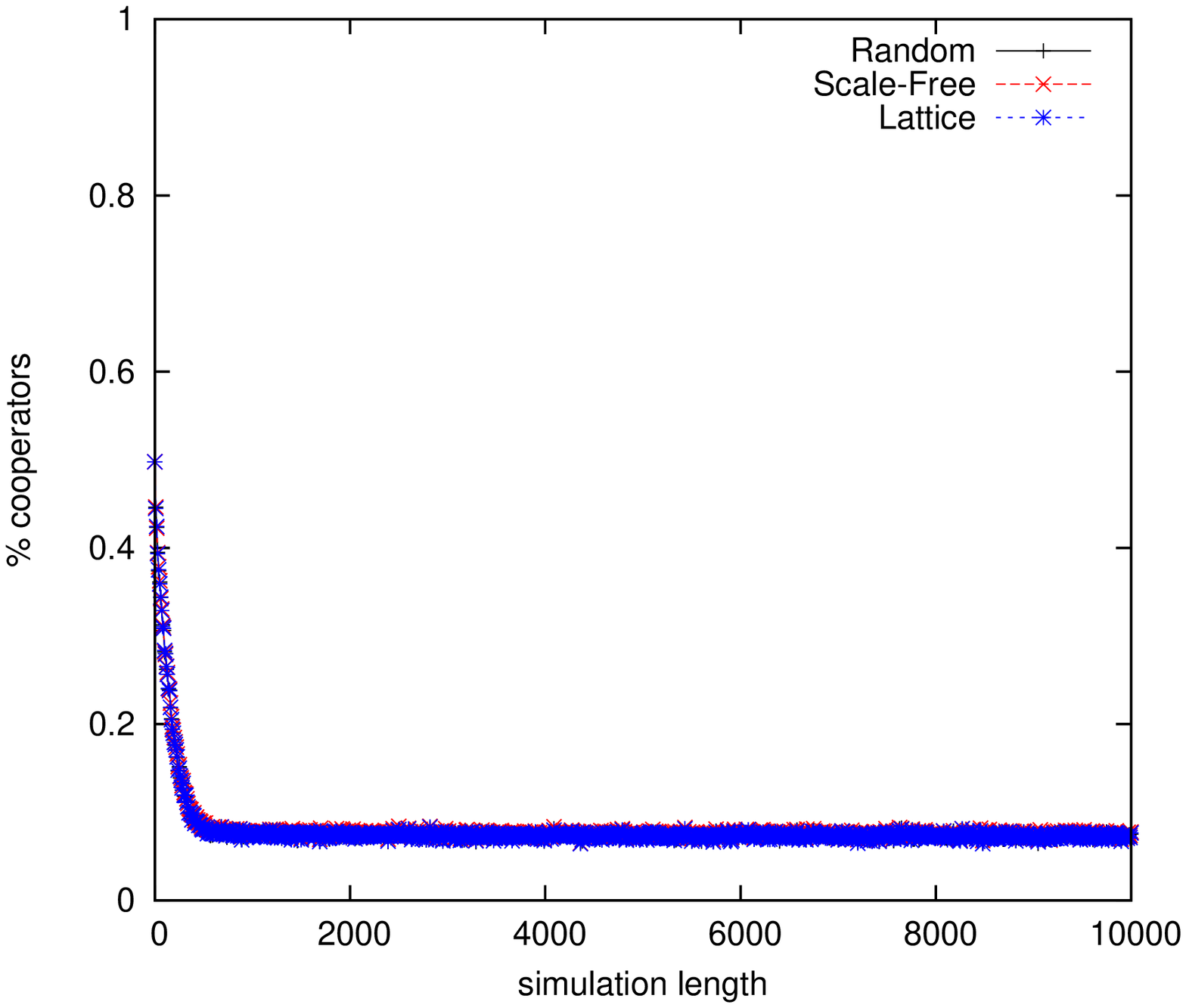}
\includegraphics[width=0.24\textwidth]{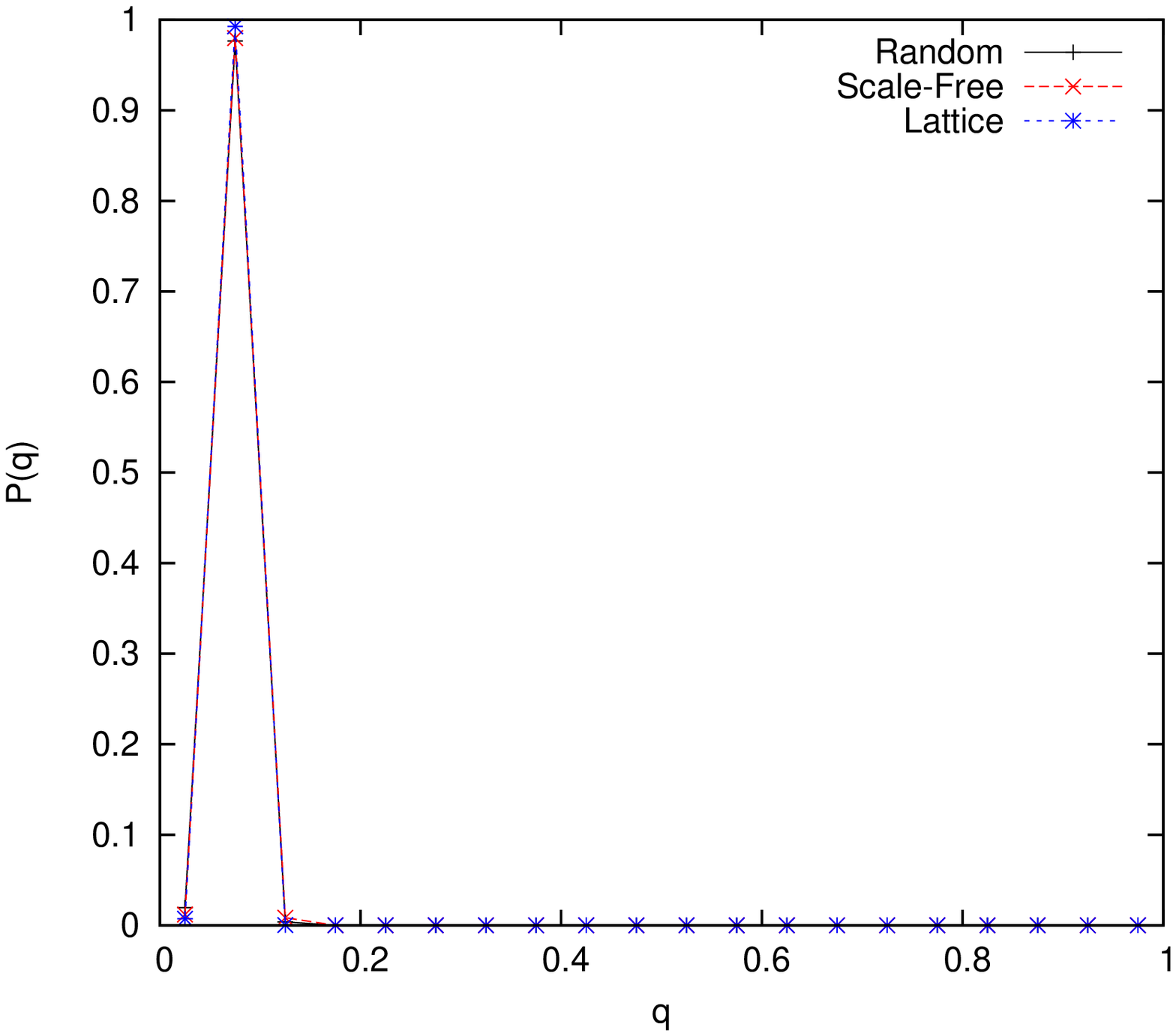}
\includegraphics[width=0.24\textwidth]{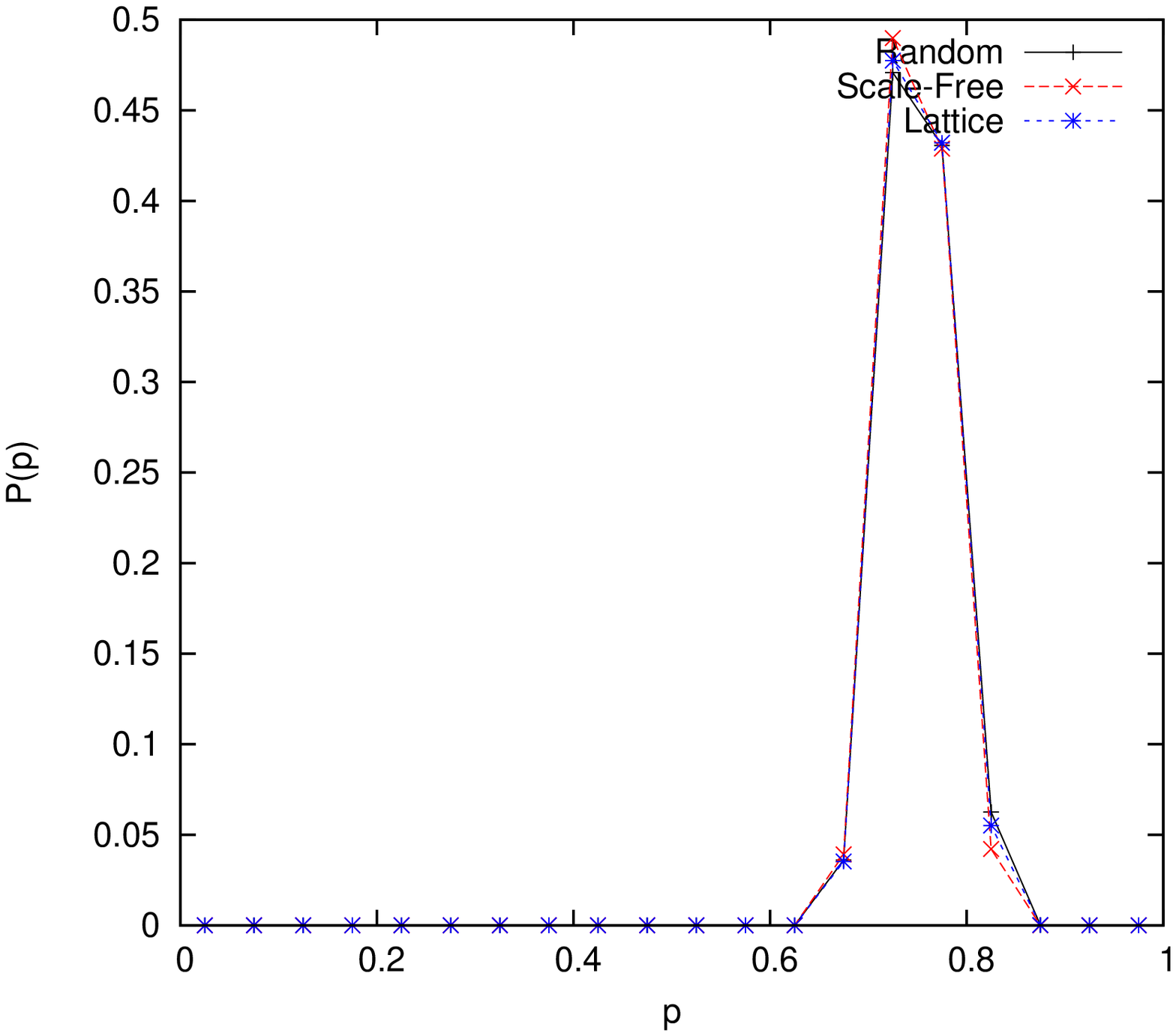}
\includegraphics[width=0.24\textwidth]{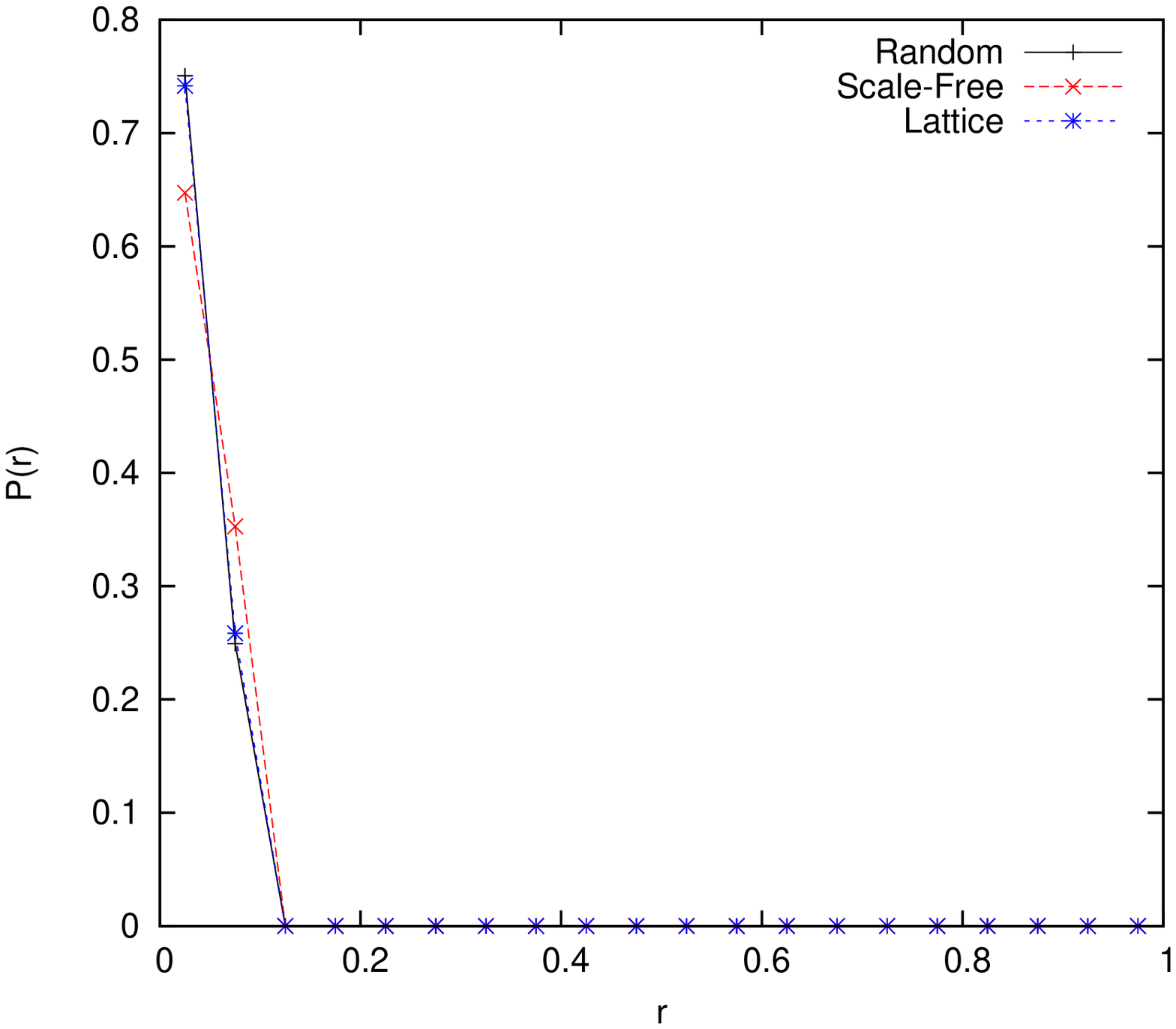}
\newline
\includegraphics[width=0.24\textwidth]{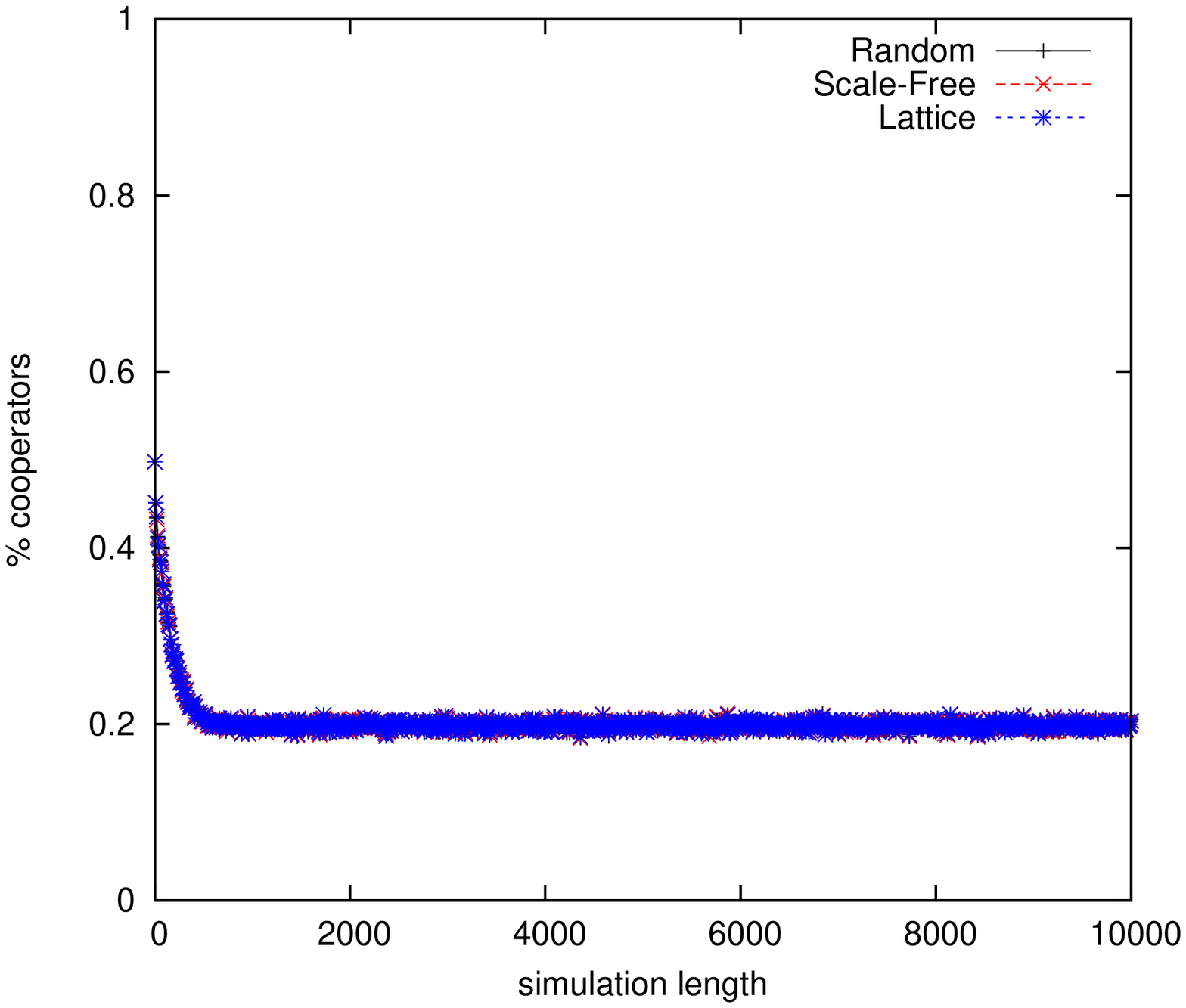}
\includegraphics[width=0.24\textwidth]{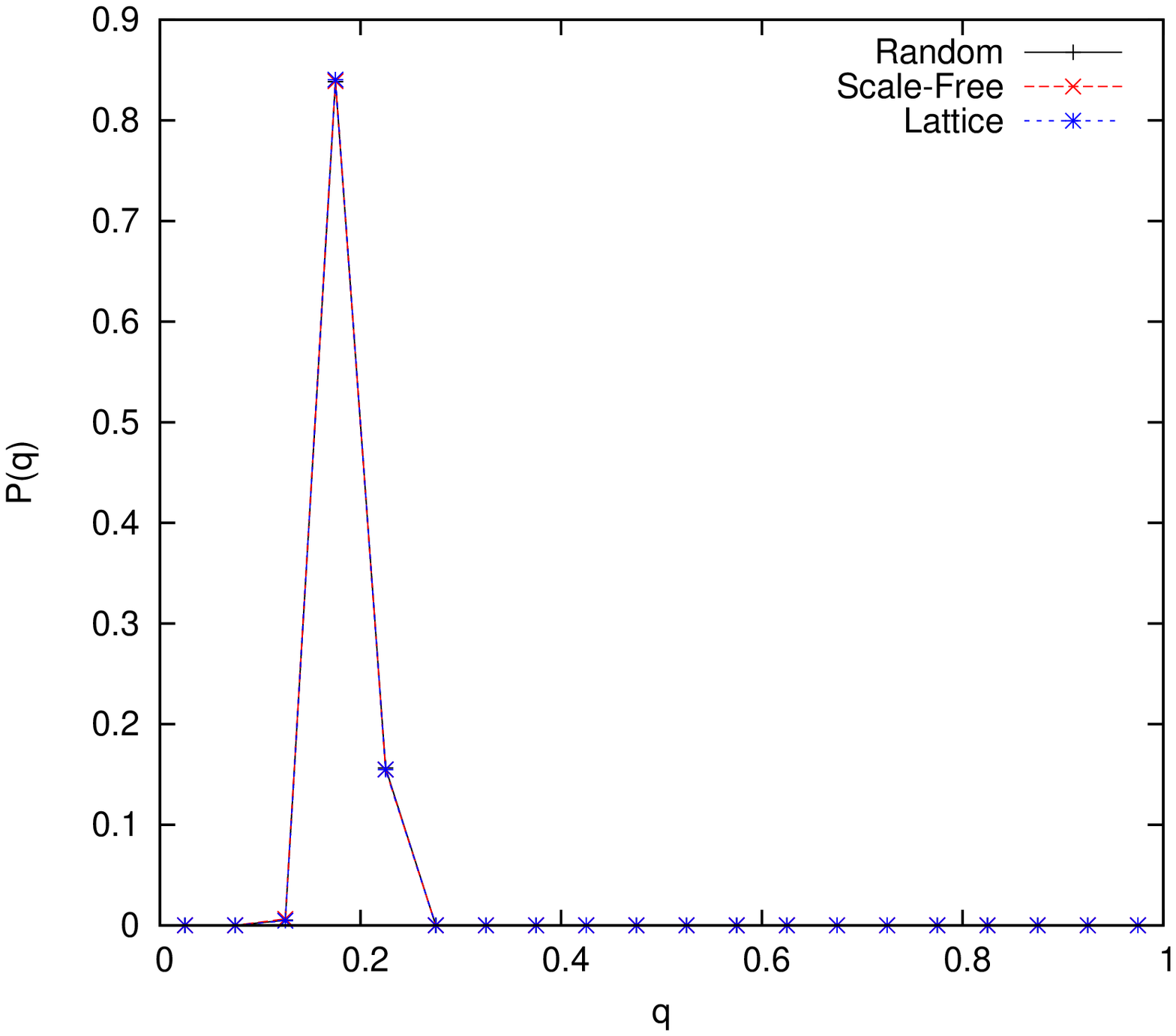}
\includegraphics[width=0.24\textwidth]{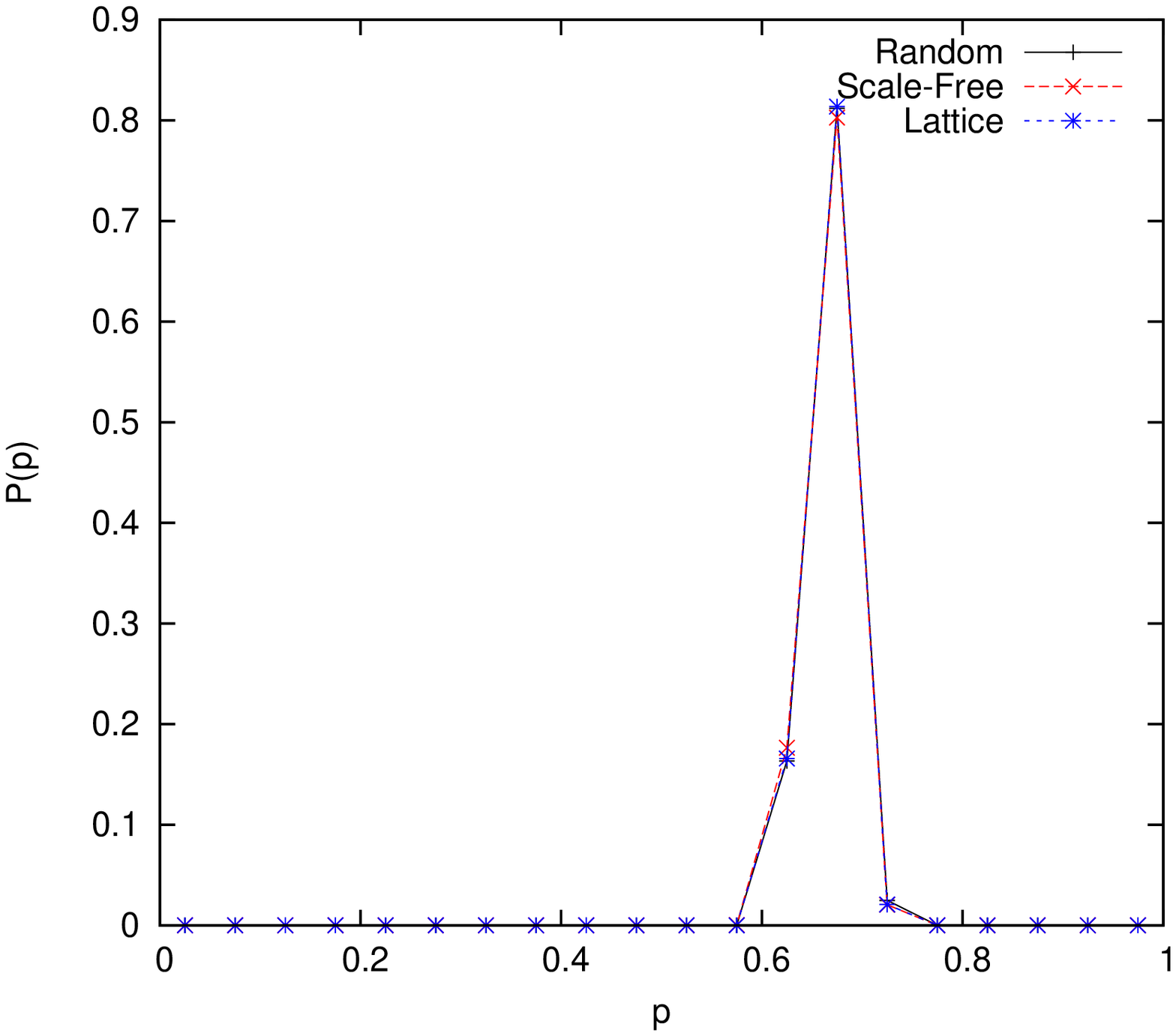}
\includegraphics[width=0.24\textwidth]{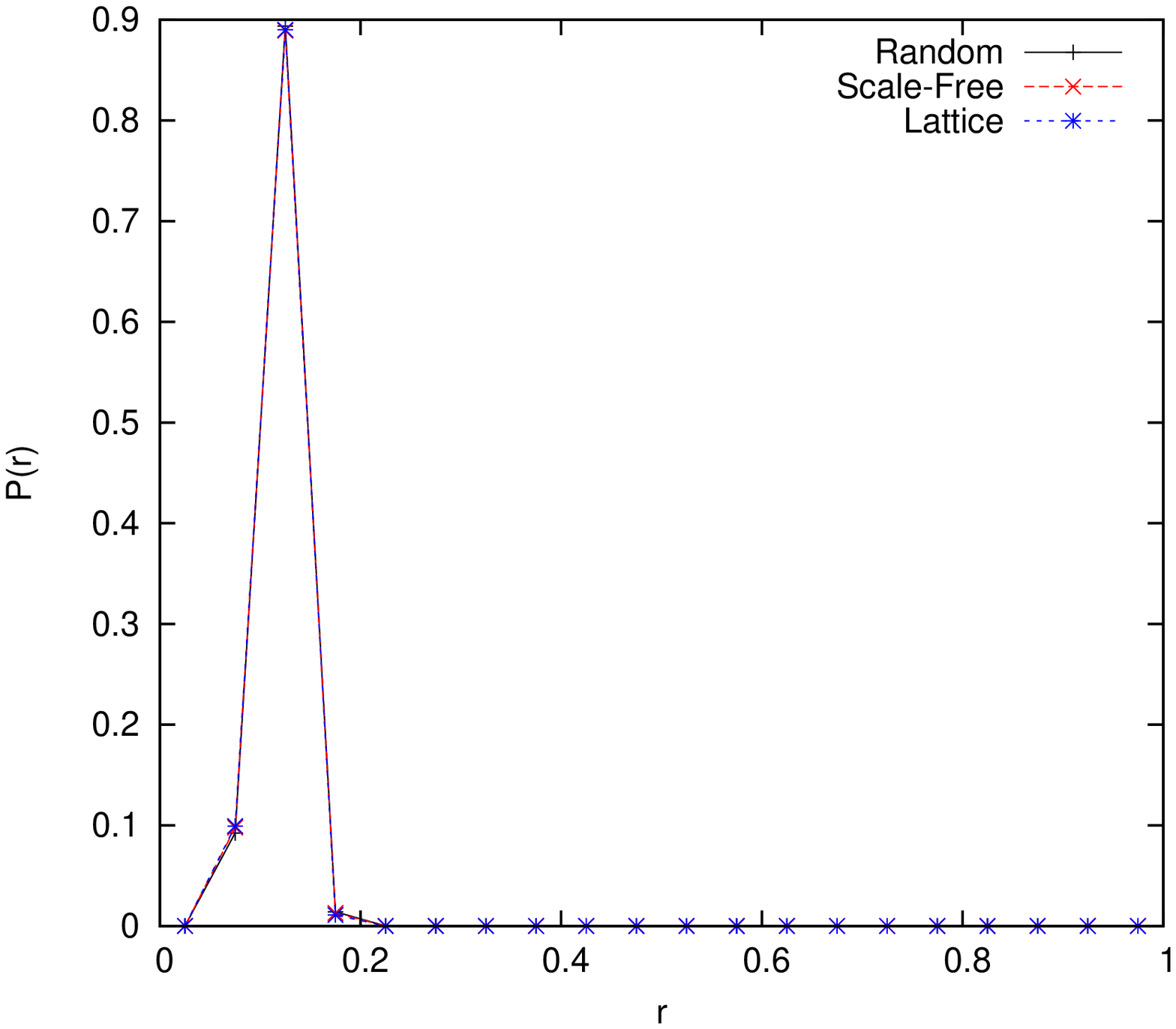}
\newline
\includegraphics[width=0.24\textwidth]{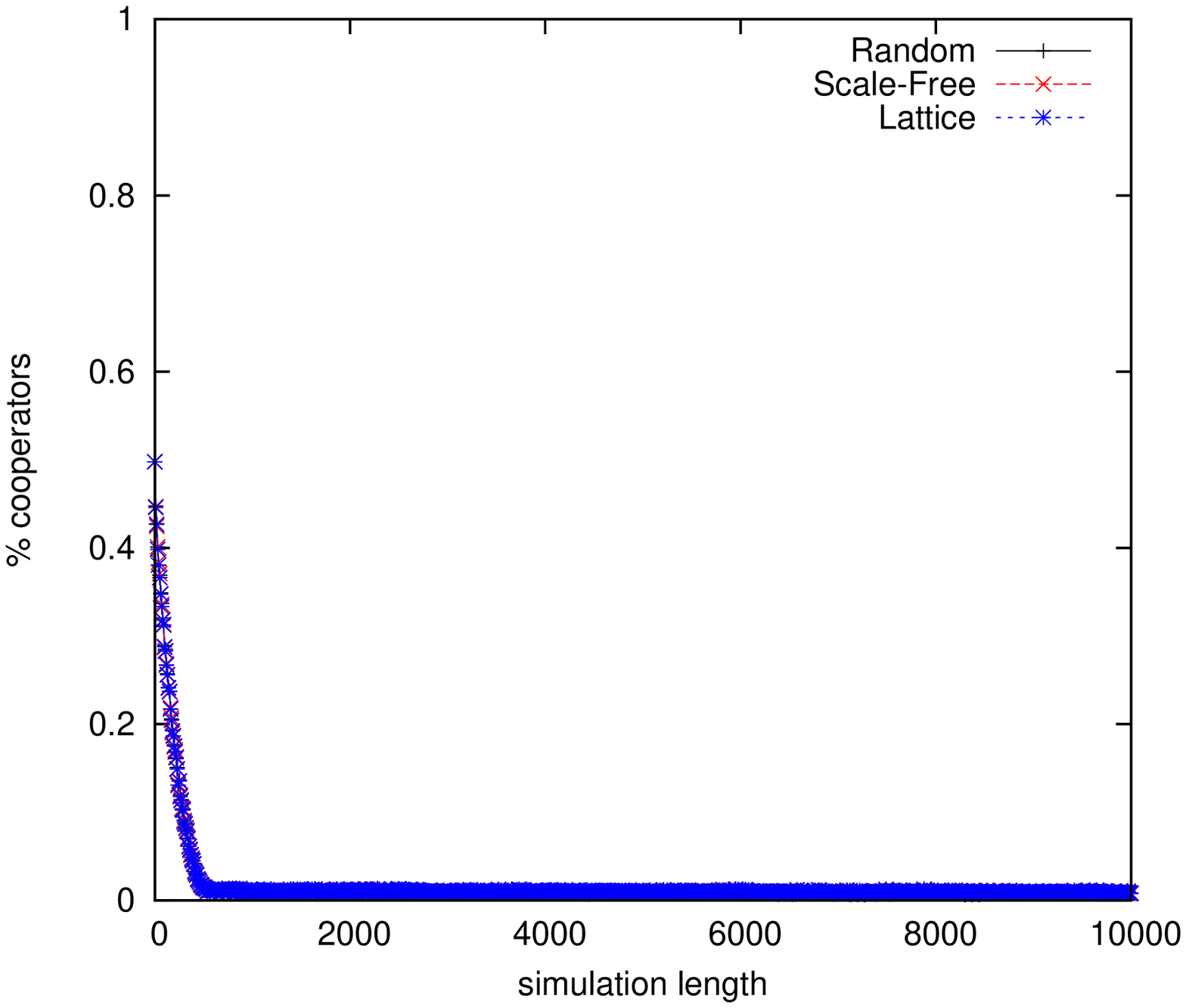}
\includegraphics[width=0.24\textwidth]{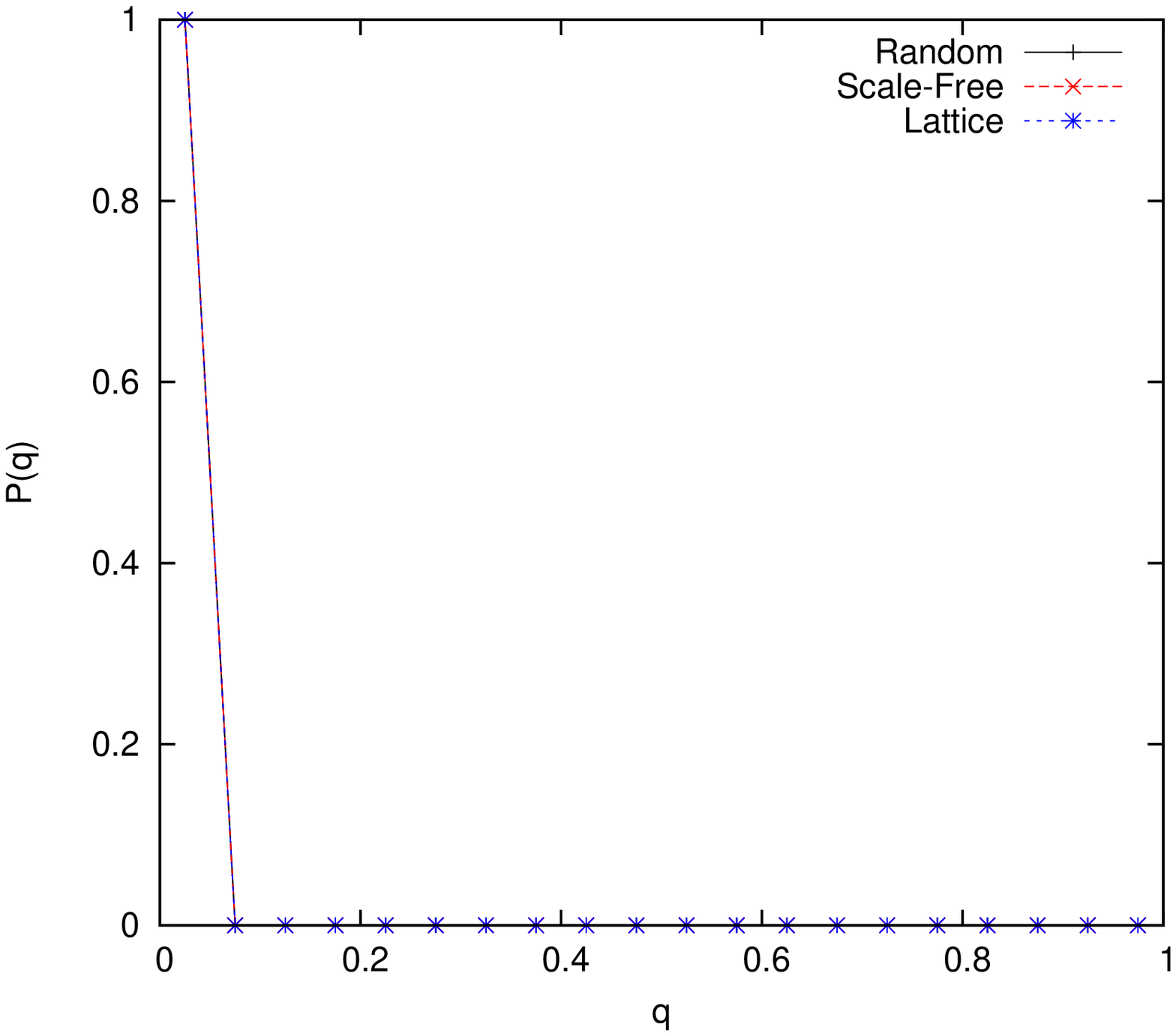}
\includegraphics[width=0.24\textwidth]{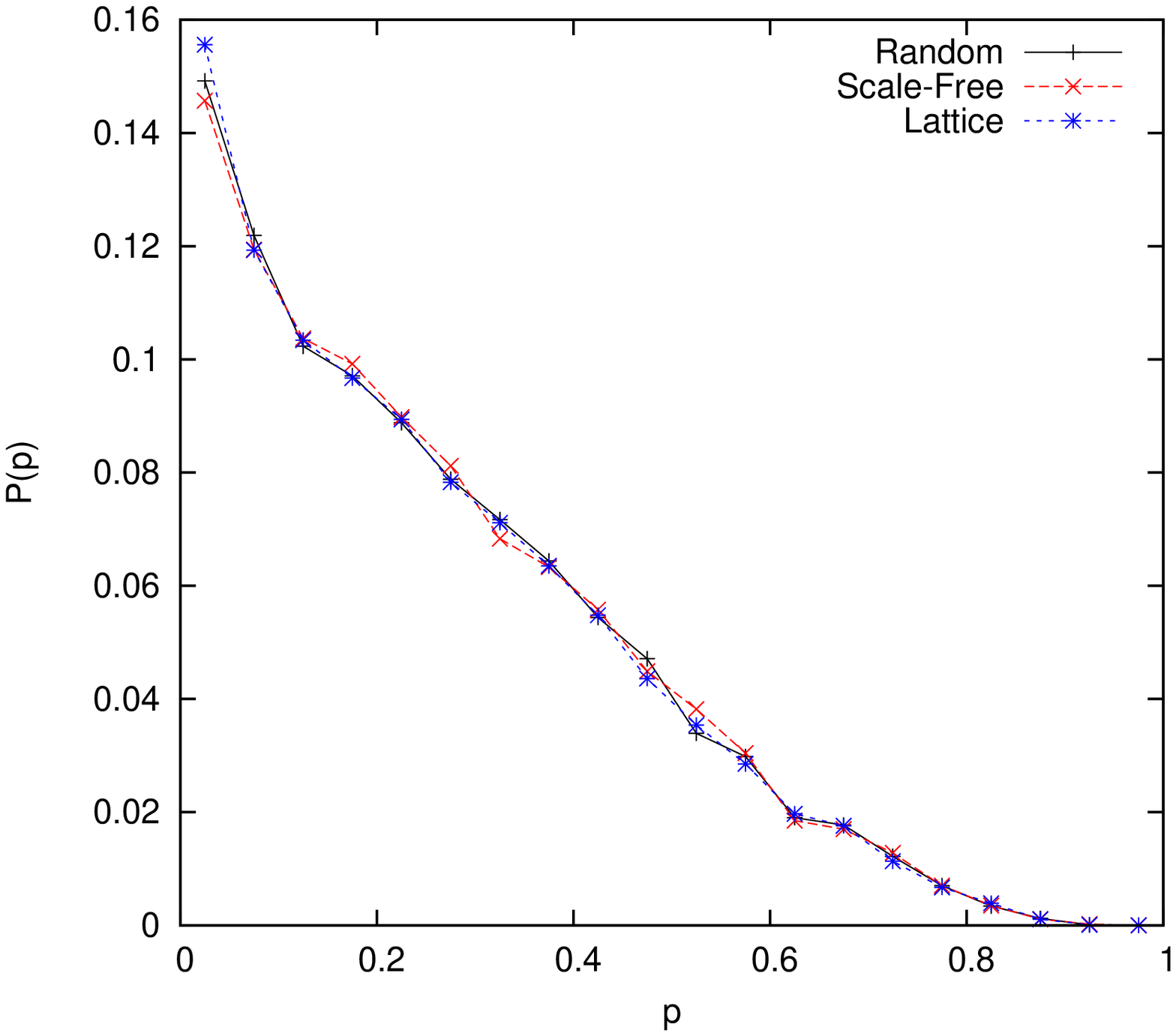}
\includegraphics[width=0.24\textwidth]{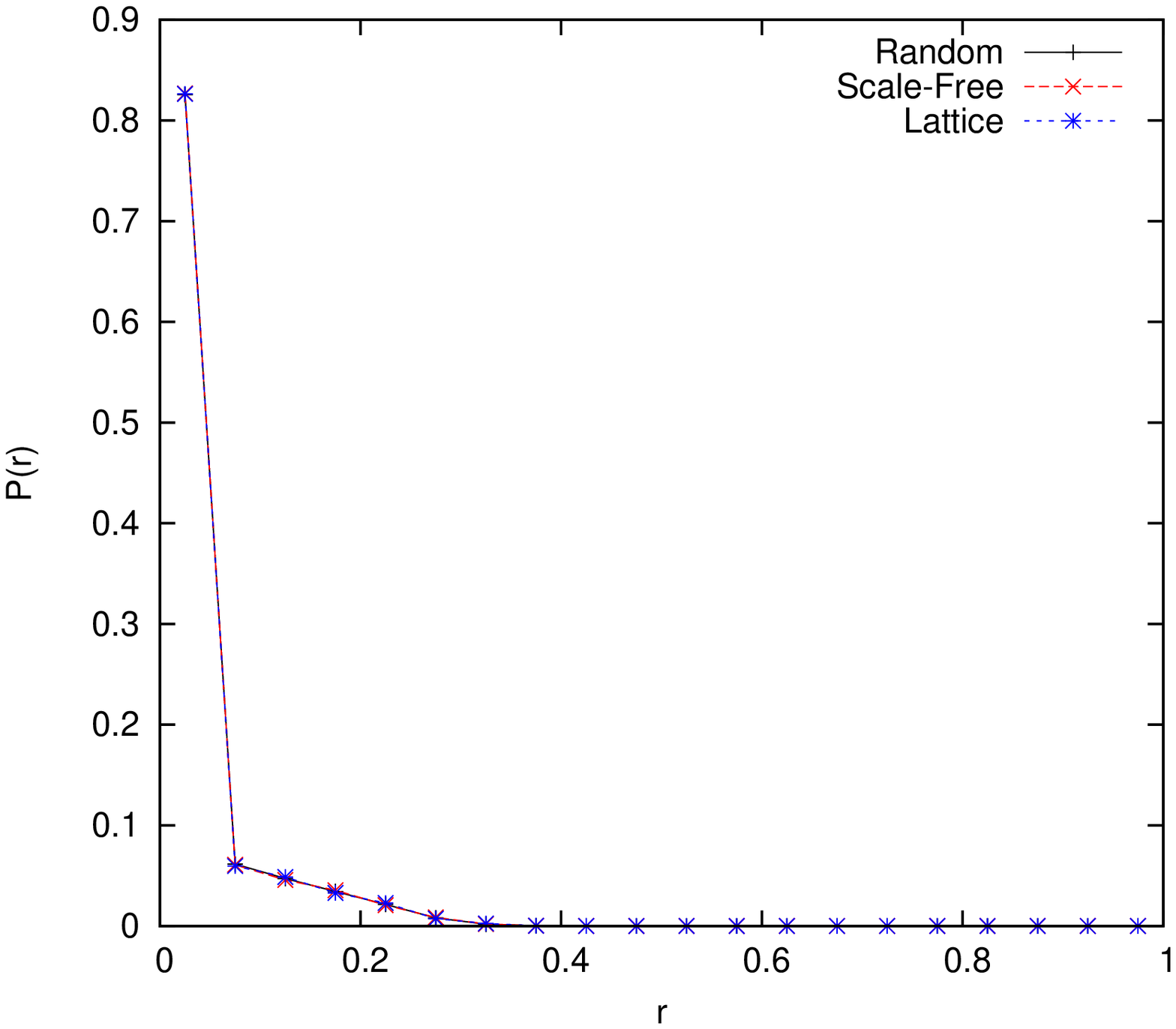}
\caption{Evolution of the level of cooperation (left column) and stationary distributions of MCC parameters (from left to right: $q$, $p$ and $r$) when the evolutionary dynamics is EWA with $\delta=\lambda=10^{-2}$ and $\gamma=3/4$. From top to bottom: $A=1/2$, $A=5/4$ and adaptive $A$ ($h=0.2$, $A^0=1/2$).}\label{fig.C_EWA}
\end{figure*}

Results for the EWA updating are shown in Fig.\ \ref{fig.C_EWA}. We observe that, when the drift to full defection due to Best Response is not dominant ($\gamma>1/2$), 
the cooperation level lies midway between the ones obtained separately by Best Response and Reinforcement Learning. Stationary distributions of MCC parameters do exist, 
but they are narrower and, except for the parameter $p$, concentrate on smaller values than with Reinforcement Learning alone. 
We can conclude that EWA updating brings to situations that are compatible with experimental outcomes, provided that the contribution from Reinforcement Learning remains dominant.

\begin{thebibliography}{38}
\bibitem{foot1} Note that the general MCC behavior includes full defectors and full cooperators as special cases: if for a given player $q=0$, then as soon as she defects she will keep defecting until $q$ changes; vice versa, if $r=1$ then as soon as she cooperates she will keep cooperating until $r$ changes.
\bibitem{foot4} We checked that our results are robust with respect to system size and link density.
\end{thebibliography}
\end{document}